\documentclass[twocolumn,iop, aps,superscriptaddress,showpacs]{revtex4-1}

\usepackage{mathptmx}
\usepackage{dcolumn}
\usepackage{amsmath,amssymb}
\usepackage{bm}
\usepackage{color}
\usepackage{latexsym}
\usepackage{epstopdf}
\usepackage{color}
\usepackage[english]{babel}
\usepackage{latexsym}

\usepackage{psfrag,graphicx} 
\usepackage{epsf} 
\usepackage[caption=false]{subfig}
\usepackage{graphicx}
\usepackage{amsmath} 
\usepackage{amssymb} 
\usepackage{amsfonts}
\usepackage{bm}
\usepackage{natbib}
\usepackage{epstopdf}
\usepackage{appendix}
\usepackage{changes}

\graphicspath{{./Figures/}}

\definecolor{mygrey}{gray}{0.35}
\definecolor{myblue}{rgb}{0.2,0.2,0.8}
\definecolor{myzard}{cmyk}{0,0,0.05,0}
\definecolor{mywhite}{rgb}{1,1,1}
\definecolor{myred}{rgb}{1,0.,0.3}

\usepackage[colorlinks=true,citecolor=myblue,linkcolor=myred]{hyperref}

\def\be{\begin{equation}}
\def\ee{\end{equation}}
\def\ba{\begin{align}}
\def\enda{\end{align}}
\def\bi{\begin{itemize}}
\def\ei{\end{itemize}}

 \def\ee{\mathord{\rm e}}

 \def\ee{\mathord{\rm e}}

\renewcommand{\ee}{{\rm e}}

\def\beq{\begin{equation}}
\def\beq{\begin{equation}}
\def\eeq{\end{equation}}

\begin{document}

\title[Short Title]{Nano-NMR based flow meter}

\author{D. Cohen}
\affiliation{Racah Institute of Physics, The Hebrew University of Jerusalem, Jerusalem 
91904, Givat Ram, Israel}
\author{R. Nigmatullin}
\affiliation{Complex Systems Research Group, Faculty of Engineering and IT, The
University of Sydney, Sydney, New South Wales 2006, Australia}
\author{O. Kenneth}
\affiliation{Dept. of Physics, Technion, Israel}
\author{F. Jelezko}
\affiliation{Institute for Quantum Optics, Ulm University, Albert-Einstein-Allee 11, Ulm 89081, Germany}
\author{M. Khodas}
\affiliation{Racah Institute of Physics, The Hebrew University of Jerusalem, Jerusalem 
91904, Givat Ram, Israel}
\author{A. Retzker}
\affiliation{Racah Institute of Physics, The Hebrew University of Jerusalem, Jerusalem 
91904, Givat Ram, Israel}
\date{\today}

\begin{abstract}
{Microfludic channels are now a well-established platform for many purposes, including bio-medical research and Lab on a Chip applications. However, the nature of flow within these channels is still unclear. There is evidence that the mean drift velocity in these channels deviates from the regular Navier-Stokes solution with 'no slip' boundary conditions. Understanding these effects, is not only of value for fundamental fluid mechanics interest, but it also has practical importance for the future development of microfluidic and nanofludic infrastructures. We propose a nano-NMR based setup for measuring the drift velocity near the surface of a microfludic channel in a non-intrusive fashion.  We discuss different possible protocols, and provide a detailed analysis of the measurement's sensitivity in each case. We show that the nano-NMR scheme outperforms current fluorescence based techniques.}
\end{abstract}
\maketitle

\section{INTRODUCTION}
In the past few decades there has been a surge in the use of microfludic devices. They are now a well-established platform, which has revolutionized bio-medical research, primarily as a result of the Lab-on-a-Chip approach to chemical and biological analysis \cite{cancer,LOC1,LOC2,LOC3,LOC4,OOCRev,CellEnvironment,MicroforBio}. Microfludic channels present rich physical phenomena which depend on a large number of parameters (geometry, fluid type, fabricated material, etc.). Understanding the physics of these channels is key to the future development of the field, not only for the fabrication of such devices, but also because their research applications demand both high temporal and spatial measurement resolution, which require an in-depth understanding of the flow characteristics \cite{MicroforBio,Review93}. One of the main interests is the flow profile near the boundary of microfludic channels, since these boundary effects will play a major role when attempting to downsize  these structures into the nanoscale regime. In recent years accumulating evidence has shown that the flow in such channels does not always obey the no-slip boundary condition \cite{Review93,Noslip1,Noslip2}, which is commonly used to solve the Navier-Stokes equations. Though there have been advances in the microscopic theory of this phenomenon \cite{Noslip2}, an experimental method to accurately measure the velocity near the surface has not yet been developed. 

Current methods for velocity measurement in microfludic channels are based mostly on fluorescent beads or dyes. Typically, fluorescent molecules are injected into the channel and their propagation is tracked using a confocal microscope \cite{Tillnow1,Tillnow2,Tillnow3,Tillnow4,Tillnow5,liron,Jorg2001}. Alternatively, the velocity can be determined by light scattering from a periodic array \cite{liron}. Using fluorescent molecules can also be used to measure the diffusion coefficient of these molecules within a liquid \cite{Tillnow5}. These methods lack in performance and only achieve an accuracy of about 5\% for the mean channel velocity due to technical issues which include laser beam focusing and the temporal resolution of the camera. Though progress in technology may help overcome these limitations in the future, these methods suffer from fundamental issues as well. First, the fluorescent molecules are often large, and in a small channel they can affect the flow profile. Second, the flow trajectories of these molecules generally pass through  the middle of the tube (in a Poiseuille settings) because of the velocity gradient, which supplies little information about the flow profile near the surface. Finally, current methods can only be used to measure the diffusion coefficient of the fluorescent molecule within the liquid, whereas the interesting  parameters are usually the self-diffusion and rotational diffusion coefficients of the liquid.

\begin{figure}[h]
\begin{center}
\includegraphics[width=0.48\textwidth]{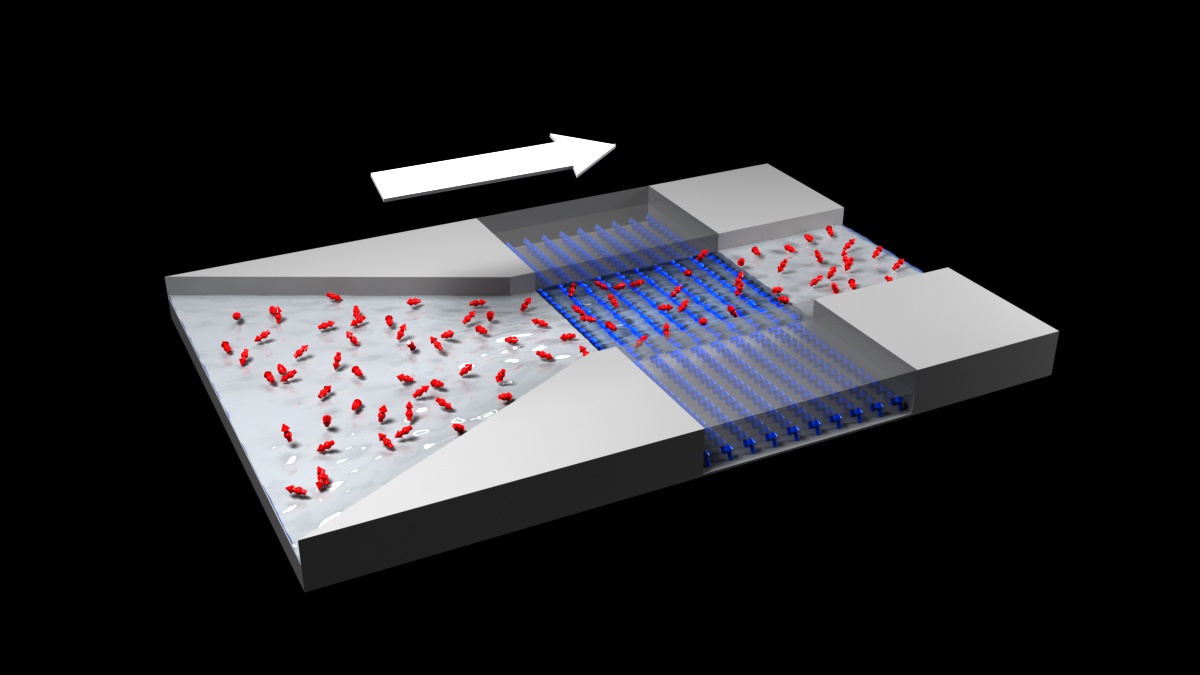}
\end{center}
\caption{{\em The scheme--}  The nuclear spins (red arrows), which are randomly oriented, generate a random magnetic field on an ensemble of NV centers (blue arrows), which are located next to the surface of the diamond. The NVs sense the magnetic field, whose randomness is amplified due to the flow. By optically measuring the state of the NV centers, the power spectrum of the magnetic noise can be estimated, 
from which the velocity can be deduced.}
\label{scheme}
\end{figure}

Here, we propose a non-intrusive measurement technique for the mean drift velocity and the self-diffusion coefficient, which is sensitive to surface effects, based on nano Nuclear Magnetic Resonance (nano-NMR).
Nano-NMR is an emerging field,  that attempts to adapt classic NMR techniques into the nano scale \cite{Glenn,Staudacher,NanoNMRstutgart,NanoNMRIBM}. The essential difference is that the core assumption of a macroscopic number of molecules is no longer valid, and as a result the NMR signal is below the signal-to-noise ratio of classic NMR measurement devices. Recent experiments have demonstrated that Nitrogen-Vacancy (NV) centers are a promising platform for nano-NMR \cite{NanoNMRHarward,NanoNMRIBM,NanoNMRMeriles,NanoNMRQdyne,NanoNMRstutgart,sar,qdyne,Glenn,Staudacher}. 
This method is based on the fact that the NV center is an excellent magnetometer at the nano scale  that can read the magnetic field created by the nuclear spins effectively and thus replace the role of the coils in the regular NMR setting. 
The main idea of our proposal is sketched in fig.\ref{scheme}, where a flow through a microfluidic channel is sensed by an NV center ensemble via a similar setup to the one in \cite{micro_Joerg}. The unpolarized spins motion induces a random magnetic field at the location of the NV centers. The power spectrum of the magnetic field noise can be estimated by optically probing the NV center. By analyzing the noise characteristics the flow properties can be deduced.
It should be noted that velocimetry methods using classic NMR techniques exist \cite{NMRvelocity1,NMRvelocity2,NMRvelocity3}. These, however, usually relay on magnetic field gradient spin-echo experiments and therefore will not work in the micro- and nano- fluidic regimes due to the aforementioned signal-to-noise problem of classic NMR devices. These methods can also be applied, in principle, in the nano-NMR settings. Implementation of such an experiment is challenging, since it requires efficient polarization of nuclear spins together with a strong and stable magnetic field gradient.

In this paper, we show that a nano-NMR approach for velocity estimation in microfludic channels outperforms current fluorescent based methods by orders of magnitude.  
The paper is constructed as follows. First, we discuss measurement protocols. Then, we estimate the sensitivity for the velocity measurement assuming a Lorentzian noise spectrum. We proceed by showing that in fact the spectrum deviates form the Lorentzian approximation due to the diffusion process and that the corrected spectrum leads to an improved sensitivity even when the diffusion process is more dominant than the drift. Finally, we present molecular dynamics simulations that support our analytic results.

\section{RESULTS}
\subsection{Measurement scheme}

We shall consider the NV center as a two level system (spin $\frac{1}{2}$) with an energy gap of $\gamma_e B_{ext}$, where $B_{ext}$ is the externally applied magnetic field \cite{Review_NV}.  The power spectrum generated by the motion can be read by a decoherence spectroscopy method \cite{avron,Biercuk,hall,Cole,Hall,Steinert} or spin relaxation \cite{Wood,de2007dangling}.
These methods exploit the fact that the transition rate or relaxation time in a two-level
system is proportional to the spectral density evaluated at the transition frequency or the Rabi frequency of an external drive. 
In both cases, the relaxation rate is given by $\Gamma = S(\omega)$, where $S(\omega)$ is the power spectrum at frequency $\omega$, which is either the external drive's Rabi frequency or the energy gap of the two level system.
The relaxation rate and, therefore, the power spectrum can be efficiently estimated from the population decay. 
Since all the effects of the velocity emerge from the  magnetic noise induced on the NV, polarization is not essential here and it can be measured in unpolarized samples in a way that is similar to diffusion measurements \cite{NV_depth}.

\begin{figure}[h]
\begin{center}
\includegraphics[width=0.48\textwidth]{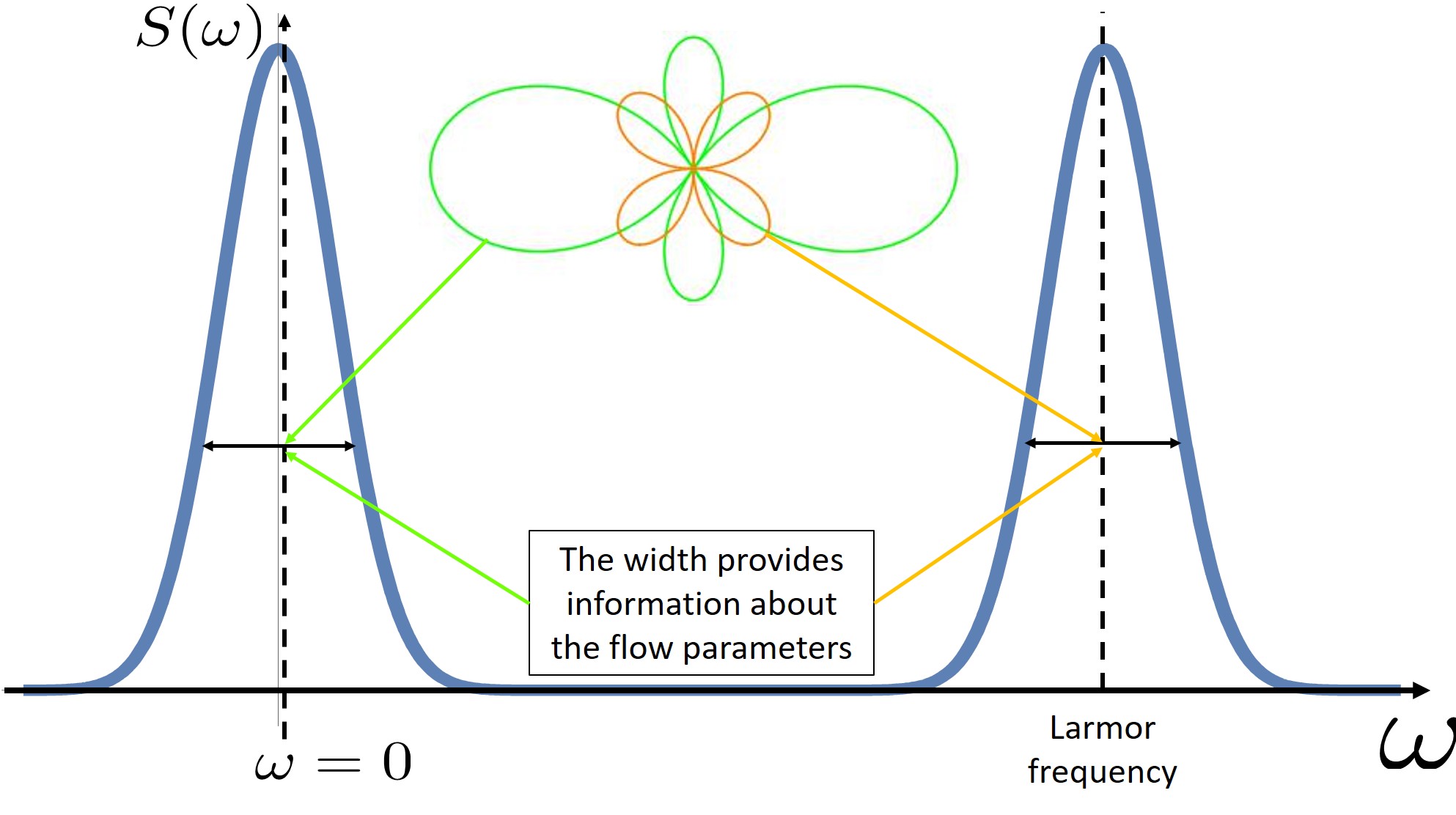}
\end{center}
\caption{A diagram of the total spectrum that can be probed by the two terms. The $T_0$ term (Green) probes the zero frequency part of the spectrum, and the $T_{\pm 1}$ terms (orange) probe the Larmor frequency part of the spectrum. The fluid drift velocity can be determined from the structure of the spectrum.}
\label{Total_Spect}
\end{figure}

The different components of the magnetic field can be sensed by probing the different parts of the dipole-dipole interaction separately. We denote $S_i\ (I_i)$ as the spin operator of the NV (nuclear spin) at the $i$ direction and $S_{\pm}\ (I_\pm)$ as the NV's (nuclear spin's) raising/lowering operators. The velocity can be sensed by the $T_0 = \left( 3 \cos(\theta)^2 -1 \right) S_z I_z$ term or the $T_{\pm 1} = \frac{3}{2} \sin \theta \cos \theta e^{\mp i \phi} \left( S_z I_{\pm} + S_{\pm} I_z \right)$ terms (see for example \cite{CohenTa}) as shown in fig. \ref{Total_Spect}. The $T_0$ is the classical term, deriving from the thermal polarization fluctuations of nuclei within a volume of $d^3,$ which  inflict a random field on the NV. In the case of a polarized fluid, the noise is decreased and this term vanishes. However, probing the dynamics is still possible if the nuclei are polarized in the $x-y$ plane \cite{Supp}. 

The second term, $T_{\pm 1}$, is a rotating term which manifests itself at high frequencies and is routinely used to probe the nano-NMR spectrum \cite{Glenn,Staudacher,NanoNMRstutgart,NanoNMRIBM}.
In order to probe this term, a rotation of the NV at the Larmor frequency must be introduced. 
The method of choice is to introduce driving by pulsed dynamical decoupling \cite{NanoNMRHarward,NanoNMRIBM,NanoNMRMeriles,NanoNMRQdyne,NanoNMRstutgart,sar,qdyne}, which yields the following effective Hamiltonian:
$H_1 = \frac{3}{2} \sin \theta \cos \theta S_z \left( I_x \cos(\delta t + \phi) + I_y \sin(\delta t +\phi) \right),$ where $\delta$ is the frequency difference between the Larmor frequency and the frequency of the pulses.


\begin{figure*}
\subfloat[]{\includegraphics[width=0.32\textwidth]{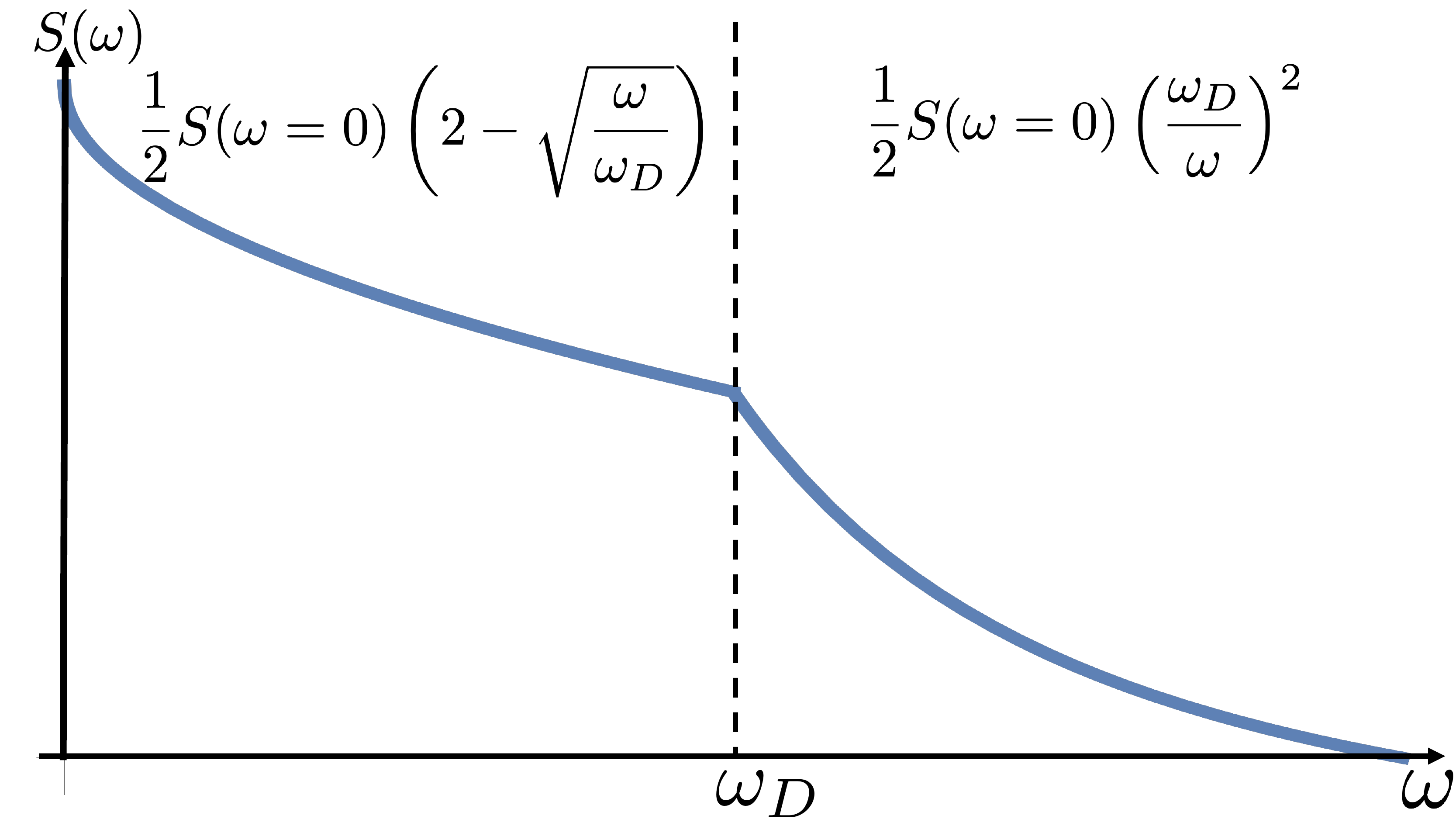} \label{power_diff}}
\hfill
\subfloat[]{\includegraphics[width=0.32\textwidth]{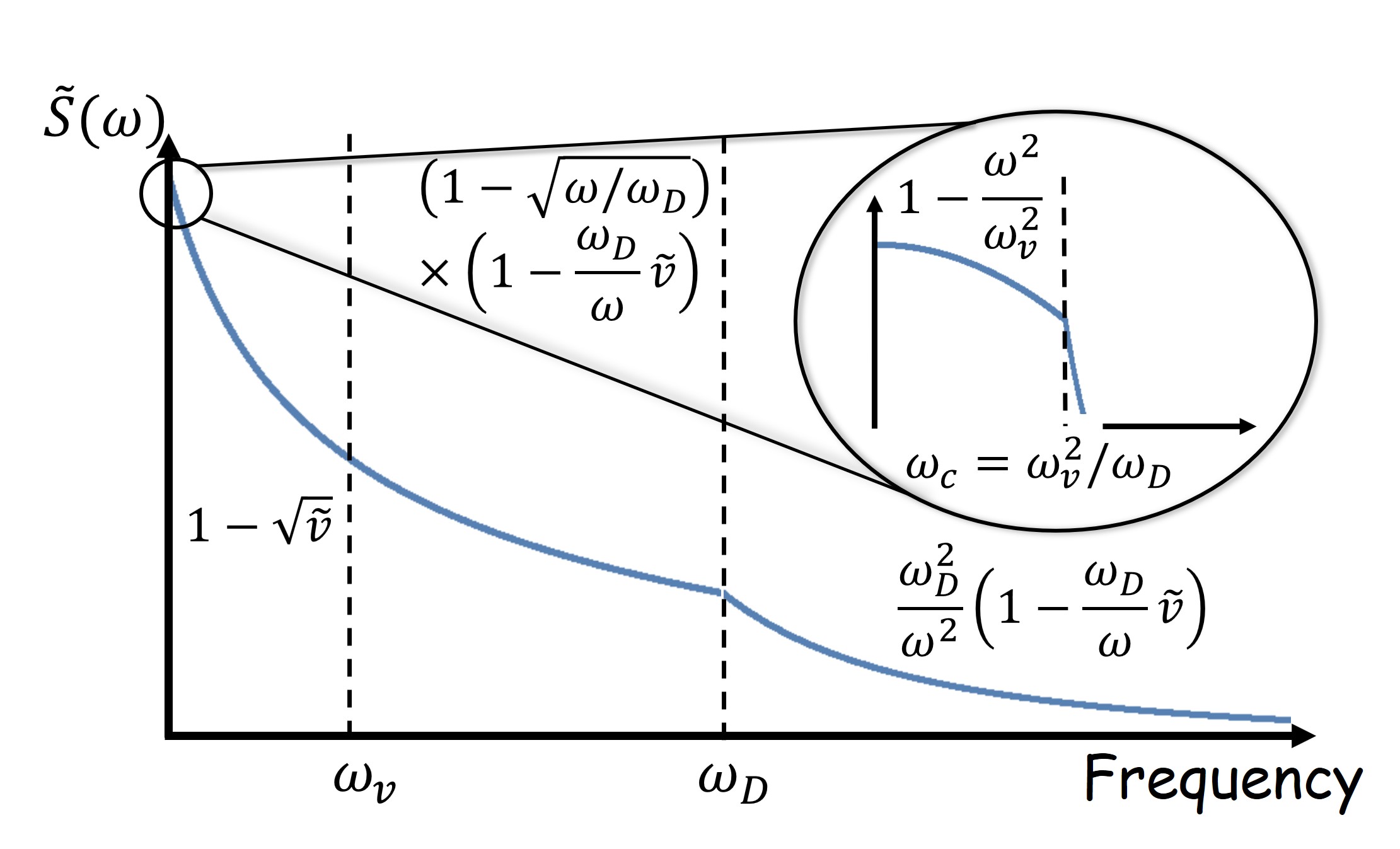} \label{power_drift}}
\hfill
\subfloat[]{\includegraphics[width=0.32\textwidth]{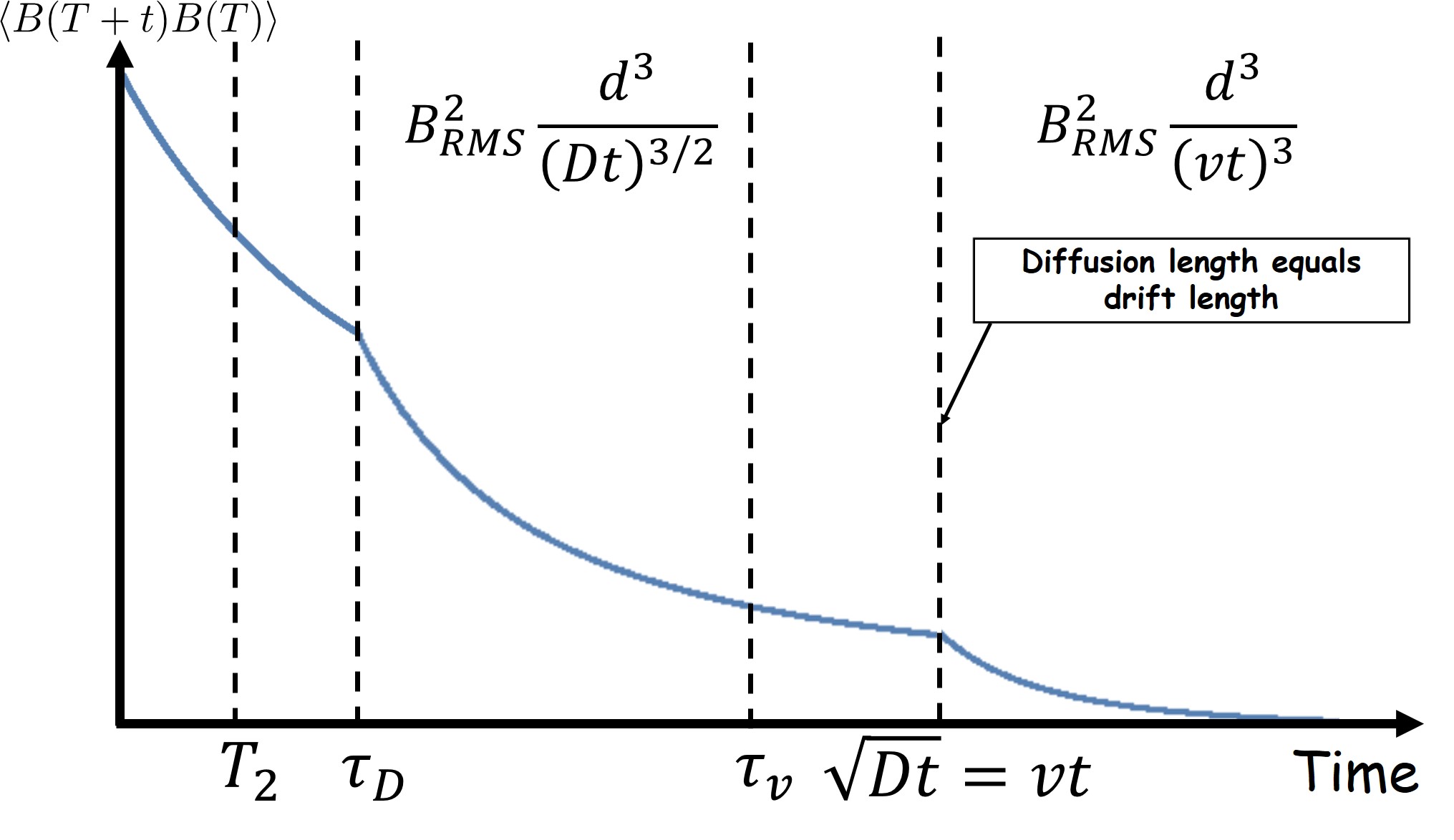} \label{corr_fun}}
\caption{(a) The structure of the power spectrum for diffusion only --- The power spectrum is divided into two regimes which are defined by $\omega_D.$ The regime $\omega \gg \omega_D$ is characterized by the expected Lorenzian type decay. The $\omega \ll \omega_D$ regime is described by a $\sqrt{\omega}$ decay. (b) The structure of the power spectrum for diffusion and drift in normalized units - The power spectrum has three different regimes which are divided by $\omega_v$ and $\omega_D$. The high frequency part $\omega > \omega_D$
is roughly equal to the power spectrum at zero velocity and damped by a factor of $\left(1 - \tilde{v}\right)$. This also holds for the intermediate region, $\omega_v<\omega<\omega_D$. The low frequency part is nearly a constant, which equals $S_{v=0}\left(\omega=\omega_v\right)$. Near zero the power spectrum decays quadraticaly as expected by the behavior of the correlation function.  (c) The structure of the correlation function --- The decay of correlations changes at the point where the diffusion distance equals the drift distance, $t=\frac{D}{v^2}$.  In the case that the correlation did not decay substantially within an NV coherence time, $T_2$, the correlations in the NV measurements are proportional to the magnetic correlations. }
\label{corr} 
\end{figure*}

\subsection{Lorentzian Model}
The velocity precision can be understood in a simplified model. In this nano-NMR setting there are two characteristic time scales: the diffusion time scale $\tau_D=\frac{d^2}{D}$ and the drift time scale $\tau_v=\frac{d}{v}$, where $d$ is the NV's depth below the diamond surface, $D$ is the self-diffusion coefficient and $v$ is the mean drift velocity. These time scales control the correlation time of the NMR signal if only one of the processes is dominant. Respectively, they are the inverse characteristic widths of the NMR spectra. In the case when both processes are significant, the natural expectation is that the characteristic width is $\sigma^2=\left(\tau_D^{-1}\right)^2+\left(\tau_v^{-1}\right)^2\equiv \omega_D^2+\omega_v^2.$ In order to estimate the accuracy we follow the commonly accepted assumption that the nano-NMR spectrum is a Lorentzian function \cite{NV_depth,Glenn}, $S(\omega)=\frac{\sigma^2 S(\omega=0)}{\omega^2+\sigma^2}$. The peak is given by $S(\omega=0) = \gamma_e^2 \frac{B_{RMS}^2}{\sigma}$, and $B_{RMS}^2=n\left(\frac{\mu_0\hbar\gamma_N}{4\pi}\right)^2\frac{1}{d^3}$, where $\gamma_e = 2 \pi \times 28 \ \textrm{GHz T}^{-1}$
is the electron gyromagnetic ratio, $\gamma_{N}$ is the nuclei gyro-magnetic ratio and $n$ is the nuclei number density \cite{NV_depth}, which results in a $B_{RMS}$ of $250\ \textrm{kHz}/\gamma_e$ for a 5 $\textrm{nm}$ deep NV and water density of $33\ \textrm{nm}^{-3}$. The sensitivity is best described by the dimensionless quantities:
\beq\label{units}
\tilde{S}(\omega)=\frac{S(\omega)}{\gamma_e^2B_{RMS}^2\tau_D},\ \tilde{v}=\frac{v}{d/\tau_D}.
\eeq    
The first denominator can be interpreted as a unit power spectrum taken as $S(\omega=0,v=0)$, whereas the second denominator is a unit velocity for which $\omega_v=\omega_D,$ i.e.,  $\tilde{v} = \frac{\omega_v}{\omega_D}$. We use these units henceforth in our estimations.  

The decay rate $\Gamma$ of the NV can be estimated with an accuracy of $\sim \Gamma$ for a single shot measurement within one experiment which lasts for $t \sim \Gamma^{-1}$. Repeating this measurement for a total duration of $T$ results in an accuracy of $\Delta \Gamma= \sqrt{\frac{\Gamma}{T}}$. Since most NV measurements are not single shot, the accuracy is reduced by an order of magnitude. Since the standard regime  is the one in which the velocity effect is dominated by the diffusion effect; i.e., $\omega_v \ll \omega_D$ the decay rate equals to $\tilde{S} \approx 1-\frac{\tilde{v}^2}{2},$ which is quadratic in the velocity; hence the sensitivity with respect to the velocity is substantially reduced in this regime.
This leads to the natural conclusion that whenever the effect of diffusion is larger than the velocity effect, it is very challenging to estimate the velocity. This can be intuitively understood in the following way. The velocity is estimated from changes in the magnetic field due to the drift of the fluid, as in fig. \ref{scheme}. Diffusion, however, causes the magnetic field to change as well. When the molecules move away from the vicinity of the NV due to diffusion rather than drift, the effect of the drift is suppressed.

Before we proceed, let us quantify this effect. The sensitivity of the velocity is optimal at $\omega=0$. 
Therefore, the uncertainty in the velocity, $\frac{\sqrt{S\left( \omega=0 \right)}}{\sqrt{T}} \frac{1}{\left \vert \frac{\partial S(\omega=0)}{\partial v} \right \vert}$, assuming $\omega_v \ll \omega_D$, is 
\beq\label{Lorentzian_Sensativity}
\Delta \tilde{v} = \frac{1}{\tilde{v}}\frac{1}{\gamma_e B_{RMS} \tau_D^{1/2}  \sqrt{T}} =  \frac{1}{\tilde{v}}\frac{1}{ \sqrt{S(\omega=0,v=0) T}}.
\eeq 
This can be also rewritten as $\Delta \tilde{v}  = \frac{1}{\tilde{v} \phi \sqrt{M}},$ where $\phi$ is the phase which is collected during a time of $\tau_{D}$ and $M$ is the number of measurements.
For the experimental parameters of water: $D\sim2.3 \cdot 10^3 \ \textrm{nm}^2\;\mu\textrm{s}^{-1},\ n\sim 33 \ \textrm{nm}^{-3}$ \cite{water_diff}  and $d\sim10 \ \textrm{nm}, v=1 \ \textrm{mm} \; \textrm{s}^{-1}$, the sensitivity is approximately: $\frac{\Delta v}{v}\approx3\frac{ 10^3 \ \textrm{s}^{-1/2}}{\sqrt{T}}$. Limited contrast and collection efficiency will reduce the sensitivity by a factor of 10, whereas collection from many NVs can increase it by a factor of $50 \sqrt{ \frac{A}{\mu \textrm{m}^2}}$, where $A$ is the area covered by the NV centers \cite{NV_ensemble}.
Thus in total $\frac{\Delta v}{v}\approx 600\ \frac{\textrm{s}^{-1/2}}{\sqrt{T}} \sqrt{\frac{(\mu\textrm{m})^{2}}{A}},$ ($A$ is the area in units of $(\mu m)^2$) which requires a large NV area $\left(150\ \mu\textrm{m}\right)^2$ to get reasonable accuracy. Differences in the distance from the surface in the NV ensemble result in negligible changes \cite{Supp}.

 In the following we show that the Lorentzian approximation of the power spectrum is very crude since in fact the power spectrum is linear or proportional to the square root of the velocity, depending on the frequency regime. These corrections enhance the sensitivity by $\tilde{v}^{3/2}$. For water flowing in a microfluidic channel $\tilde{v}\sim5\cdot10^{-3}$, which implies improvement by a factor of $5\cdot10^{-3}$  in sensitivity.

A heuristic sketch of the power spectrum without drift, based on Sec. VIII of \cite{Supp}, is shown in fig. \ref{power_diff}. While a Lorentzian type behavior is valid for large frequencies, the behavior for low frequencies deviates considerably from a Lorentzian and decays as $1-\sqrt{\omega/\omega_D}.$
This behavior derives from the fact that the correlation function at long times decays as $\frac{1}{t^{3/2}}$  due to diffusion dynamics which transforms to  $\sqrt{\omega}$  dependence in the power spectrum. This non-Lorentzian behavior been observed experimentally for diffusing Rb vapor in $\textrm{N}_2$ gas  \cite{NonLorentzian1,NonLorentzian2}. 
In the following we show that this non-analytic behavior has important implications for parameter estimation.

The velocity can be estimated either from the correlation function or from the power spectrum. Although these two quantities are related via the Fourier transform, 
the sensitivity obtained from each method is different, because of the modifications in the experimental procedure for probing them.
The optimal strategy corresponds to different regimes of the coherence time of the NV center.
In the case where the coherence time of the NV center is shorter than the correlation time of the signal, it is advantageous to probe the correlation function, since in this regime each measurement probes approximately a constant magnetic field. Therefore, the correlation between measurement results is proportional to the magnetic correlation function. In the opposite regime it is best to probe the power spectrum directly. Because an increased signal-to-noise ratio can be achieved by an ensemble measurement, a reasonable NV coherence time is $100 \; \mu \textrm{s}$ \cite{Glenn} after dynamical decoupling and should be compared to $\tau_D$ when deciding on a correlation function or power spectrum probing.

\subsection{Estimation via the correlation function}
The general structure of the correlation function, based on Secs. V and VI of \cite{Supp}, is shown in fig. \ref{corr_fun}. In the case where $\tau_D$ is longer than the coherence time of the NV, which is valid for viscous fluids such as oil and NVs deeper than $5\; \textrm{nm}$, each measurement result is a function of the instantaneous magnetic field, thus making it advantageous to probe the correlation function. The time separation that yields information about the velocity can be found in the regime in which the drift length is larger than the diffusion length ($v t > \sqrt{D t}$).  In this regime, the phase acquired by the NV during one measurement is $\phi \approx \gamma_e B(t) T_2,$ and thus the correlation function is $\langle P_\uparrow(t_1) P_\uparrow(t_1+t) \rangle \approx \langle \frac{1+\sin \phi(t_1)}{2} \frac{1+\sin \phi(t_1+t)}{2} \rangle  \approx \frac{1}{2} + \frac{1}{2} \langle \phi_\uparrow(t_1) \phi_\uparrow(t_1+t) \rangle \approx  \frac{1}{2} + \frac{1}{2} T_2^2 \gamma_e^2 B_{RMS}^2 \frac{d^3}{(v t)^3}.$ Assuming that $v t \approx d$ the uncertainty is $\frac{\Delta v}{v} = \frac{1}{3}\frac{1}{T_2^2 \gamma_e^2 B_{RMS}^2}.$ As the velocity in oil can reach  $10^{-2} \; \textrm{mm s}^{-1},$ yielding an optimal depth of $25 \; \textrm{nm}$ and taking into account a limited collection efficiency, a fractional uncertainty of $5\cdot10^{-2}$ is achieved. Thus, the total uncertainty is of the order of $\frac{\Delta v}{v} = 10^{-2}\frac{\sqrt{\textrm{s}}}{\sqrt{T}}\sqrt{\frac{(\mu\textrm{m})^{2}}{A}}.$
The limitation is that this method works as long as $T_2$ is shorter than $\tau_D$ and $\tau_v$, which only applies to very viscous fluids. For example, for an oil and NV depth of $25\; \textrm{nm}$, the diffusion time $\tau_D \approx \textrm{ms}$ \cite{Oded}.
For water on the other hand and for a shallower NV, $\tau_D = 0.02  \; \mu\textrm{s},$ which reduces the sensitivity by four orders of magnitude.

\subsection{Estimation via the power spectrum}
For lower viscosity fluids it is advantageous to estimate the velocity directly from the power spectrum. 
The structure of the power spectrum, based on Secs. II and IX of \cite{Supp}, is shown in fig.\ref{power_drift} and its behavior is divided into three regimes: the low frequency regime $\omega < \omega_v,$ the intermediate regime  $\omega_v<\omega<\omega_D$ and the high frequency regime $\omega>\omega_D$ .  In the high and intermediate frequency regimes the power spectrum is linear in the velocity and in the low frequency regime it has a square root dependence.  

In the intermediate frequency regime, for $\omega \approx \omega_D$, the uncertainty is 
\begin{equation}
\Delta \tilde{v} = \frac{1}{\sqrt{T}} \frac{1}{\sqrt{ \tau_D } \gamma_e B_{RMS}},
\label{sens1}
\end{equation}
this constitutes an improvement by a large factor of $\tilde{v}^{-1}=\frac{\omega_D}{\omega_v}$ in comparison with eq. \eqref{Lorentzian_Sensativity}. Substituting the parameters suiting water we estimate  $\Delta v= 10 \frac{\textrm{mm}}{\textrm{s}}\sqrt{\frac{\textrm{s}}{T}}$ for $15 \textrm{nm}$ deep NV, which is the optimal depth as  the power spectrum at this depth equals $T_2^{-1}$.
For an ensemble of NVs we achieve the enhancement    $\Delta v=  2 \frac{\textrm{mm}}{\textrm{s}}\sqrt{\frac{\textrm{s}}{T}} \sqrt{ \frac{(\mu \textrm{m})^2}{A}}$. To appreciate this accuracy, consider that for water the dimensionless parameter $\tilde v = 1$ corresponds to a drift velocity of $v=150\ \frac{\textrm{mm}}{\textrm{s}}.$

In the low frequency region the decay rate is equal to
\beq\label{lowfreq}
 \tilde{S}= 1- \sqrt{ \tilde{v}}
 \eeq 
 and the uncertainty is therefore,
 \beq\label{lowfreqsens}
 \Delta\tilde{v}=\frac{2}{\gamma_e B_{RMS}\sqrt{\tau_D}} \sqrt{\frac{\tilde{v}}{T}}
\eeq
  which is a substantial improvement by a factor of $\sqrt{\tilde{v}}$ compared to the intermediate range (eq. \ref{sens1}).
  This regime is applicable whenever $T_2 \gg \frac{1}{\omega_v}$, which is attainable for low viscosity fluids as this time scale is around $10 \mu \textrm{s}$ in that case.
    The accuracy in this regime,
  $\frac{\Delta{v}}{v}=\frac{2}{\gamma_e B_{RMS} \tau_{D}} \sqrt{\frac{d}{Tv}}$, is a decreasing function of $v$ as expected.
  Using the same parameters of water, and $d= 15 \textrm{nm}$ the accuracy in this is estimated by $\frac{\Delta{v}}{v}=0.3 \frac{\sqrt{\textrm{s}}}{\sqrt{T}} \sqrt{ \frac{(\mu m)^2}{A}}$.
This result represents an improvement by more then three orders of magnitude in fractional uncertainty over state of the art methods even when demanding high temporal (second) and spatial resolution (micrometer).


\subsection{Molecular dynamics}

\begin{figure}[h]
	\begin{center}
		\includegraphics[width=0.48\textwidth]{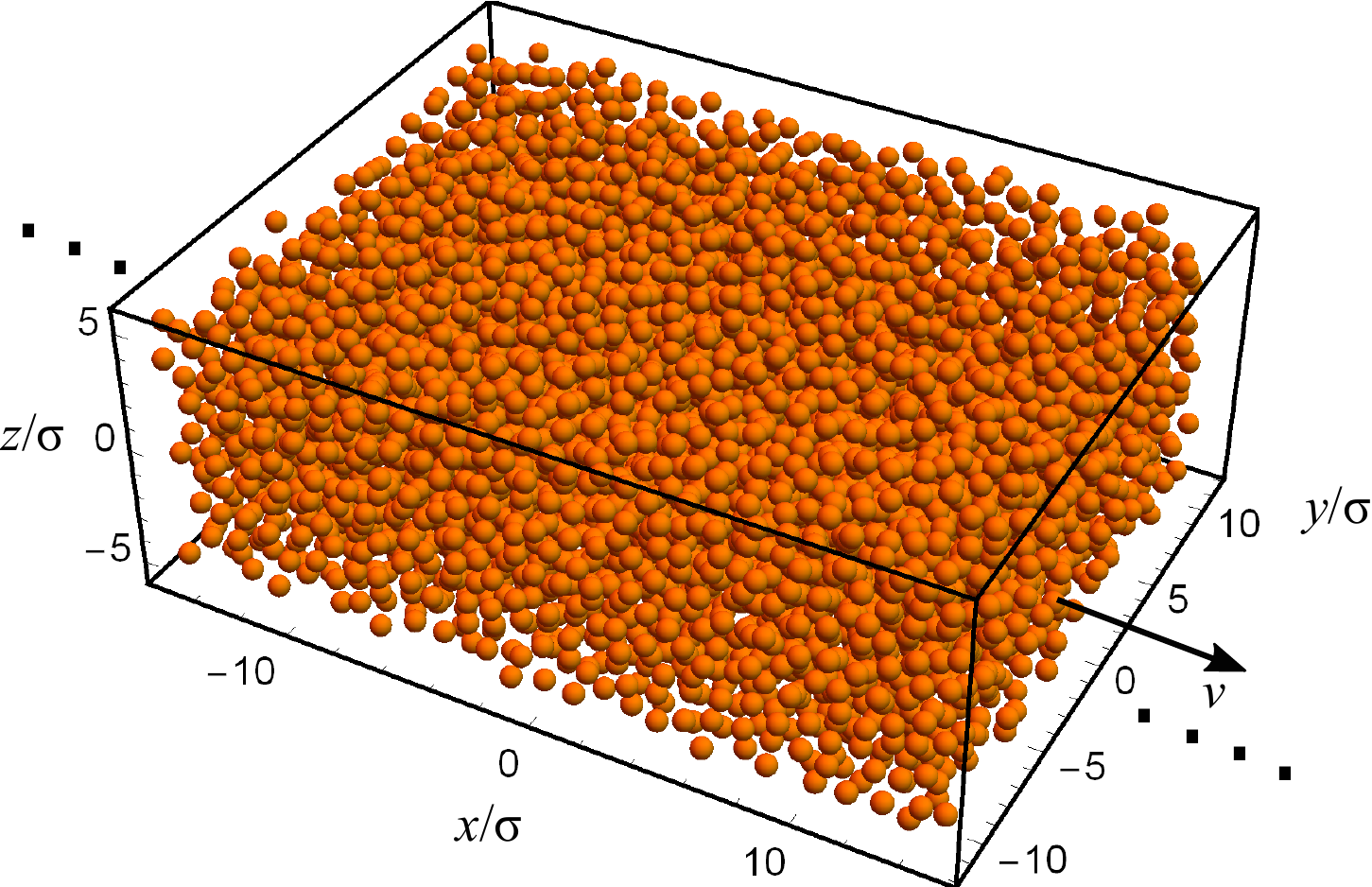}
	\end{center}
	\caption{The dipole bath is simulated using LJ fluid model. The thermal motion results in fluid self-diffusion. A drift velocity $v$ is added to every particle at the start of the simulation. The dimensions of the simulation box were $L_{x}=163.08\sigma, L_{y}=22.68\sigma$ and $L_{z}=10.8\sigma$. The number of particles was $31710$ giving the density of $\rho=0.79\sigma^{-3}$.}
	\label{fig:simbox}
\end{figure}
In order to verify our analytical results we preformed a molecular dynamics simulation.  
Fig. \ref{fig:simbox} illustrates the simulation setup. The system consists of N dipolar particles confined to a simulation box of volume $L_{x}L_{y}L_{z}$. Particles are interacting via the Lennard Jones (LJ) potential $4\epsilon\left[\left(\frac{\sigma}{r}\right)^{12}-\left(\frac{\sigma}{r}\right)^{6}\right]$ with an interaction cut-off distance of $r_{c}=2.5\sigma$. The system is initialized into a thermal state at temperature $T$ by applying a Langevin thermostat and each particle is assigned a random spin value $I_{z}\in\{-1,1\}$. After thermalization, the average particle velocity is subtracted from each particle, e.g. $\bar{v}_{j}\leftarrow\bar{v}_{j}-\left<\bar{v}\right>$. The initialization concludes with the addition of the drift velocity $v$ in the x-direction to each particle $v_{xj}\leftarrow v_{xj}+v$. The simulation dynamics is deterministic, following Newton's laws, and integrated using the Velocity-Verlet method with step size $\Delta t=0.00125\sqrt{\epsilon/(m\sigma^{2})}$. During the simulation the $z$ component of the magnetic field induced by the particles at the position of the NV center is computed using the equation
\beq
B(t)=\sum_{i=1}^{N}\frac{1}{r_{i}^{3}(t)}\left[3\cos^{2}(\theta_{i}(t))-1\right]I^{i}_{z},
\eeq
where $r_{i}$ is the distance between particle $i$ and the NV center, and $\theta_{i}$ is the angle between $r_i$ and the NV's quantization axis.  The NV is placed a distance $d$ below the simulation box in the center of the xy plane. Specular reflections are applied at the boundaries that are perpendicular to the z-axis and periodic boundary conditions are used in x and y directions. To emulate particles coming into and leaving the simulation box, we flip the spin of particles that reach one of the periodic boundary walls with probability 1/2. 

\begin{figure}[t!]
	\begin{center}
		\includegraphics[width=0.48\textwidth]{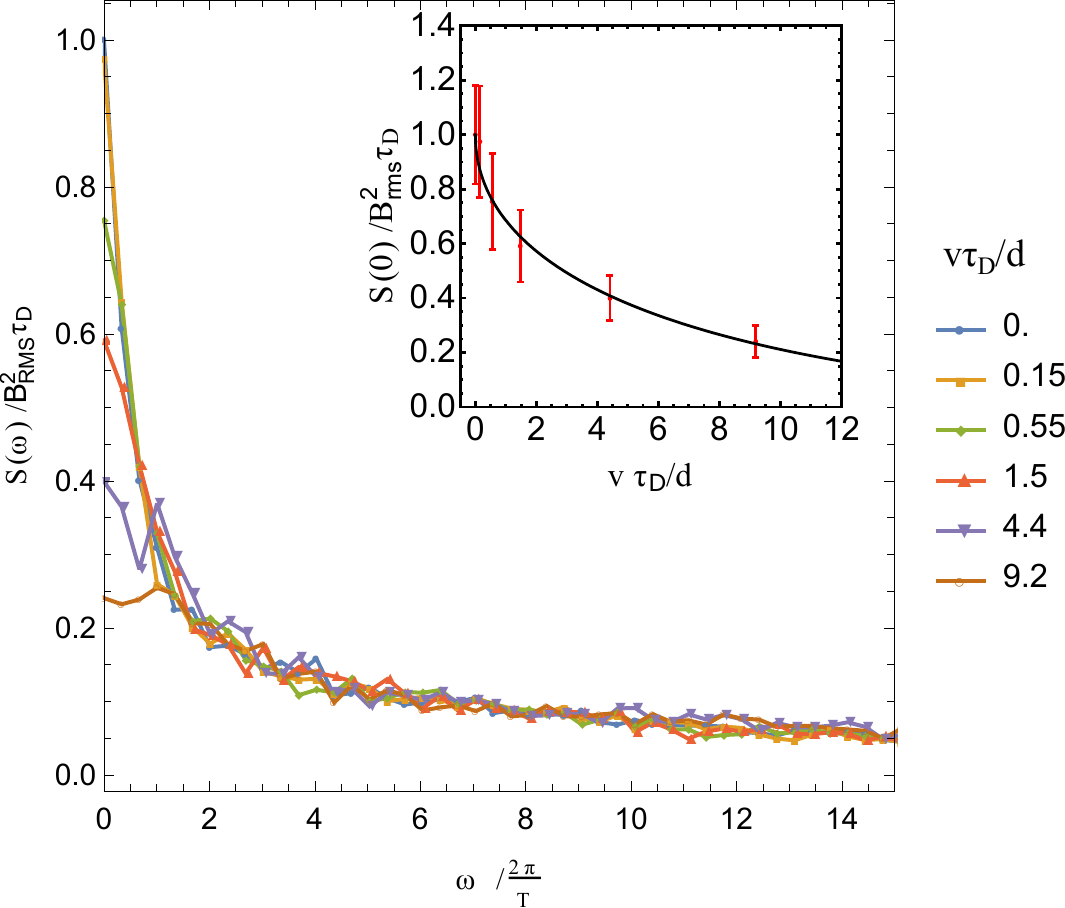}
	\end{center}
	\caption{Power spectra $S(\omega)$ obtained using LJ fluid model for six different values of drift velocity v. The simulation box dimensions were $L_{x}=163.08\sigma, L_{y}=22.68\sigma$ and $L_{z}=10.8\sigma$. The particle density was $\rho=0.79\sigma^{-3}$ and the NV center was placed at a distance $d=5\sigma$. LJ fluid parameters in reduced units were $\epsilon=1, \sigma=1$, and temperature $T=1$. The dynamics was integrated for time $18.7\tau_{D}$. The average $S(\omega)=\left\langle |\mathcal{F}[B(t)]|^{2}\right\rangle$  was taken over 130 simulation runs. The inset shows the sensitivity of the power spectrum at zero frequency on the drift velocity. The data was numerically fitted to a function $1+b\sqrt{v}+cv$, with $b=-0.42$ and $c=0.053$ which yielded $R^{2}=0.9926$. The error bars indicate the 95$\%$ confidence intervals. Each point corresponds to averaging of 130 runs.}
	\label{fig:Powerspectra-interactions}
\end{figure}

Fig. \ref{fig:Powerspectra-interactions} shows the power spectra computed using the LJ fluid model for $d=\sigma$ and $L_{x}=163.08\sigma$. The diffusion time scale $\tau_{D}$ is computed from the relation $S(\omega=0;v=0)=B_{\textrm{rms}}^{2}\tau_{D}$. The integration was carried out for $3.2\times10^{6}$ time steps, which resulted in total integration time of $18.7\tau_{D}$.
The value of the power spectrum at $\omega=0$ for different velocities, shown in the inset of the figure, was fitted using least-squares algorithm to the function $1+b\sqrt{v}+cv$, with $b=-0.42$ and $c=0.053$ which yielded $R^{2}=0.9926$, thus corroborating eq. \eqref{lowfreq} and the following accuracy estimation \eqref{lowfreqsens}. The difference $b<1$ between Eq. \eqref{lowfreq} and the simulation fit should be accounted for by the missing prefactor of the $\sqrt{{\tilde{v}}}$ term, which cannot be calculated using our analytic tools. 

\section{CONCLUSION}  We have proposed a novel setup to study flow properties in micro- and nano - fluidic structures which meets a crucial need in the fast growing field of microfluidics. 
The proposed setup is quantum inspired and is based on quantum sensing via color centers in solids. The proposal relies utterly on statistical polarization and therefore can be easily implemented. The evaluated sensitivity of the setup outperforms current velocimetry methods by orders of magnitude, even when diffusion effects are dominant.


\section*{ACKNOWLEDGMENTS}  This project received funding from the European Union Horizon 2020 research and innovation
programme ERC grant QRES under grant agreement No 770929 and the collaborative European project ASTERIQS. 
M.K. acknowledges the support by the Israel Science Foundation, Grant No. 1287/15.
We thank Carlos Meriles and Daniel Louzon for useful discussions.

\section*{AUTHOR CONTRIBUTIONS} 
A.R. conceived the idea for nano-NMR based velocimetry. 
M.K. made the general estimates of the power spectrum, which D.C. and M.K. verified for specific geometries in collaboration with O.K. A.R. and D.C. made the sensitivity estimation. R.N. made and analyzed the molecular dynamics simulation. A.R., M.K., D.C. and R.N. wrote the manuscript. 
F. J. has verified the experimental feasibility. All authors commented on the manuscript and it was revised accordingly. 


\end{document}


\title[Short Title]{Supplementary material for Nano-NMR based flow meter}
\author{D. Cohen}
\affiliation{Racah Institute of Physics, The Hebrew University of Jerusalem, Jerusalem 
91904, Givat Ram, Israel}
\author{R. Nigmatullin}
\affiliation{Complex Systems Research Group, Faculty of Engineering and IT, The
University of Sydney, Sydney, New South Wales 2006, Australia}
\author{O. Kenneth}
\affiliation{Dept. of Physics, Technion, Israel}
\author{F. Jelezko}
\affiliation{Institute for Quantum Optics, Ulm University, Albert-Einstein-Allee 11, Ulm 89081, Germany}
\author{M. Khodas}
\affiliation{Racah Institute of Physics, The Hebrew University of Jerusalem, Jerusalem 
91904, Givat Ram, Israel}
\author{A. Retzker}
\affiliation{Racah Institute of Physics, The Hebrew University of Jerusalem, Jerusalem 
91904, Givat Ram, Israel}
\date{\today}

\maketitle

\setcounter{figure}{0}
\renewcommand{\thefigure}{S\arabic{figure}}%

\section{SUMMARY}
In the following we derive the behavior of the power spectrum and correlation function presented in the main text.
We start in sec. \ref{Measure} with a detailed explanation about the measurement scheme. Then, in sec. \ref{universal}, we provide a general analysis of the expected universal behavior of the power spectrum. Afterwards, we calculate explicitly the behavior of the power spectrum for specific drift and geometry. In sec. \ref{Sphere} we calculate the power spectrum for spherical geometry with constant drift, which coincides with known results. We extend this calculation for the  power spectrum at $\omega=0$ for a more complex drift profile. In secs. \ref{Powernodrift} and \ref{Powerlowdrift} we provide approximations for the power spectrum of freely diffusing and drifting particles in a planar geometry. These new calculations, which coincide with our universal estimates, are our main result, as they lead to the enhanced sensitivity scaling presented in the main text. We also provide analytic solutions for the temporal correlation function of the magnetic field - the correlation function of freely diffusing particles in a planar geometry is presented in sec. \ref{timeplanar}, and the correlation function for diffusing and drifting particles in the whole space is found in sec. \ref{timevelocity}. Though these two results are geometry dependent, their asymptotic scaling is expected to be universal. 

\section{MEASUREMENT SCHEME}
\label{Measure}
The various parts of the dipole-dipole interaction could be used to sense the dynamics of the nuclei. A natural division of the different terms in the interaction is via the $T_{0,\pm 1, \pm2}$ parts \cite{cohen}. In the following, we mainly concentrate on the $T_0$ and $T_{\pm 1}$ and analyze their efficiency to probe the dynamical quantities of polarized and unpolarized fluids. 
We will show that in complete contrast to all other NMR examples, polarization does not help in this case and all quantities could be estimated also in the unpolarized case with a similar signal-to-noise ratio. 

\subsection{The term $T_{0}=\left(3\cos^2(\theta)-1\right)S_{z}I_{z}$ }

This term could probe both the polarized and the unpolarized dynamics.
The two cases will, however, result in the same efficiency.

\subsubsection{Unpolarized nuclear spin ensemble}
The Master equation in this case originates from the following stochastic
Hamiltonian:
\beq\label{ME0}
H \propto S_{z}\sum_{i=1}^N \frac{1}{r^3_i(t)} \left(3\cos^{2}(\theta_{i}(t))-1\right)I_{z}^{i}\equiv S_{z}\sum_{i=1}^Nf(\vec{r_{i}}(t))I_{z}^{i},
\eeq
where $\vec{r}_i(t)$ is the vector connecting the NV center and the $i$th nucleus,  
$\theta_i$ is the angle between between $\vec{r}_i$ and the quantization axis,
$S_z$ and $I_z^i$ are the components of the spin operators of  NV and $i$th nucleus respectively along the quantization axis of the NV.
The total number of nuclei is denoted by $N$.

As the dynamics of the nuclei is probed via a noise measurement and is manifested in the dephasing of the NV 
we apply the Master equation starting with the equation of motion for the joint density matrix of the NV and the nuclei,
\beq\label{ME1}
\dot{\chi}=-i\left[H(t),\chi(t)\right]\, , \quad \chi(t)=\chi(0)-i\int_{0}^{t}\left[H(t'),\chi(t')\right]dt'\ .
\eeq
It follows from Eqs.~\eqref{ME0} and \eqref{ME1}  
\begin{align}\label{ME2}
& \dot{\rho}_{NV}=-\int_{0}^{t}\left \langle\left[H(t),\left[H(t'),\chi(t')\right]\right] \right \rangle dt'\\ \nonumber
&\dot{\rho}_{NV}=\int_{0}^{t} \left \langle\left[S_{z}\sum_{i=1}^Nf(\vec{r_{i}}(t))I_{z}^{i},\left[S_{z}\sum_{j=1}^Nf(\vec{r_{j}}(t'))I_{z}^{j},\chi(t')\right]\right] \right \rangle dt' \ ,
\end{align}
where $\left<\cdot\right>$ stands for average over realizations of nuclear spin polarization and their trajectories.
Namely, $\left<\cdot\right>$ denotes tracing over the spin degrees of freedom, and averaging of the spatial ones.    
The density matrix $\chi$ of the nuclear spins and NV is assumed to be in the form $\chi=\rho_{NV}\otimes\rho_{Nuclei} $, where $\rho_{Nuclei}$ is approximately constant in time. Hereon, we denote the NV's density matrix as $\rho$ for brevity. 
As all the spin operators commute, the nuclei spin operators, $I_z^i$ can be assumed to be classical uncorrelated random variables taking values 
$x_{i}=\pm1$ in the non-interacting limit.
Equation \eqref{ME2} in the weak coupling regime gives
\beq\label{ME3}
\dot{\rho}=\int_{0}^{t}\langle B(t)B(t')\rangle dt'\left(S_{z}\rho S_{z}-\rho\right),
\eeq
where
\beq\label{ME4}
\langle B(t)B(t')\rangle=\left \langle\left(\sum_{i=1}^{N}f(\vec{r_{i}}(t))x_{i}\right)\left(\sum_{j=1}^{N}f(\vec{r_{j}}(t'))x_{j}\right) \right \rangle\, .
\eeq
The master equation \eqref{ME3} describes a dephasing process due to the random magnetic field induced by the nuclear spins at the NV's location. 
The dephasing is the strongest when the nuclei are immobile since both diffusion and drift, tend to 
smear the magnetic field thereby reducing the fluctuations.

The correlation function, \eqref{ME4} could be further simplified as follows:
\begin{equation}\label{ME5}
\langle B(t)B(t')\rangle = \left\langle   \sum_{i,j=1}^{N}f(\vec{r_{i}}(t))x_{i}f(\vec{r_{j}}(t'))x_{j}  \right\rangle = \left\langle  \sum_{i=1}^{N} f(\vec{r_{i}}(t))f(\vec{r_{i}}(t')) \right\rangle 
\end{equation}
where the last equality holds as different nuclear spins are uncorrelated in the non-interacting limit.
We rewrite the expression \eqref{ME5} in terms of the nuclear density, $n$ and the conditional probability 
$P(\vec{r},t|\vec{r}_0,t_0)d^3 r$ for the nucleus located at $\vec{r}_0$ at the time $t_0$ to be found in the volume element $d^3 r$ centered at $\vec{r}$ at a later time instant $t > t_0$ \cite{Abragam1961},
\begin{equation}\label{PS1}
\langle B(t)B(t')\rangle = n \int d^3r d^3r'  f\left(\vec{r}\right)f\left(\vec{r}'\right) P\left(\vec{r},t|\vec{r}',t'\right ),
\end{equation}
where since the correlation function \eqref{ME5} is symmetric with respect to the interchange of $t$ and $t'$, we have defined 
$P\left(\vec{r},t'|\vec{r}',t\right ) =P\left(\vec{r},t|\vec{r}',t'\right )$ for $t'<t$.
In addition, we have assumed the steady state with the nuclear density being constant in space and time.

By introducing the dynamical decoupling we obtain
\beq\label{ME6}
\dot{\rho}=\int_{0}^{t}e^{i\omega(t-t')}\langle B(t)B(t')\rangle dt'\left(S_{z}\rho S_{z}-\rho\right)\approx S_{BB}\left(\omega\right)\left(S_{z}\rho S_{z}-\rho\right),
\eeq
where $\omega$ is the dynamical decoupling frequency. This result is valid in the limit in which the correlation time of the noise is much shorter than the coherence time of the NV centre and should be valid for low viscosity fluids.
Equation \eqref{ME6} expresses the dephasing rate via the power spectrum, $S_{BB}(\omega)$.
In this way the measurement of the dephasing rate allows one to study the effect of the flow on the power spectrum. 

\subsubsection{Statistically polarized nuclear spin ensemble}
In the case of finite polarization, the polarization $x_i$ takes the value $\pm 1$ with the probability $p$ and $1 - p$, respectively.
We have $\langle x_i \rangle =1 - 2 p$ for the mean  and  $\langle (x_i - \langle x_i \rangle )^2 \rangle = \alpha = 2 \sqrt{p(1-p)}$ for the STD.
The finite mean average leads to an Overhauser field which is not important for us here, since it does not depend on the parameters $v$ and $D$, characterizing the flow. Therefore, only the random fluctuations affect the noise. 
This will add a multiplicative prefactor of $4p(1-p)$ in Eq.~\eqref{ME5}. 
It ensures that the dephasing is absent when the spins are fully polarized as expected.

\subsubsection{Fully polarized nuclear spin ensemble}\label{Sec.Pol.}
In the case of full polarization the calculation above produces a null result ($p=0$
or $p=1$) as in that case the NV will only feel the Overhauser field with no noise at all. 
This means that there is no classical noise.  
Noise, however,  could still be induced quantum mechanically.

\begin{figure}
\begin{center}
\includegraphics[width=0.98\textwidth]{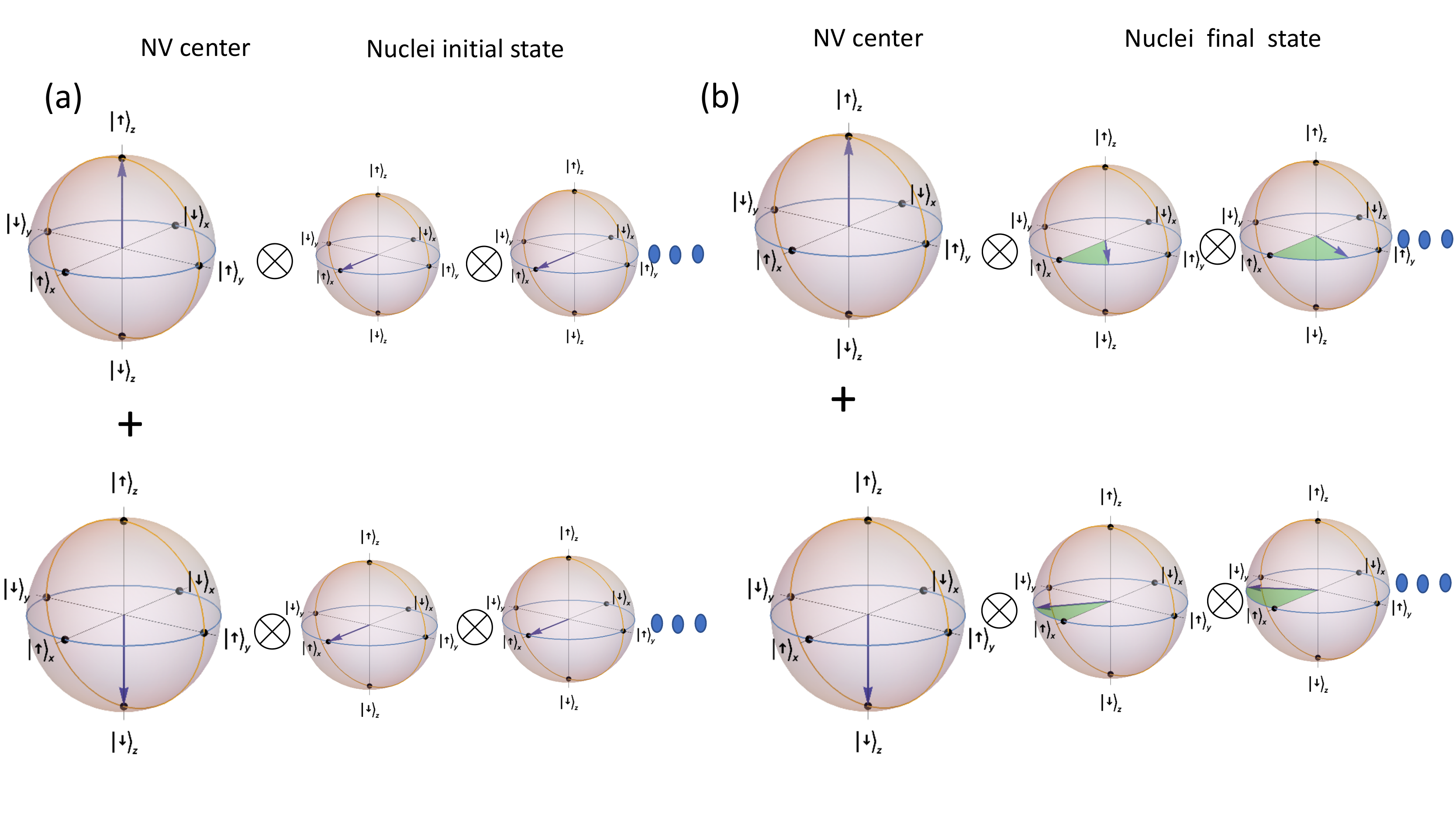}
\end{center}
\caption{\textbf{Bloch sphere representation of the decoherence sensing in the polarized case.}  \textbf{(A)} The NV starts its dynamics in the $\ket \uparrow_x  = \frac{1}{2}\left( \ket \uparrow_z + \ket \downarrow_z \right).$ In the polarized case all the nuclei start in a definite direction in the $x-y$ plane, for example the $x$ direction. \textbf{(B)} The interaction between the NV and the nuclei creates an entangled states of the GHZ type where the $\ket \uparrow_z$ is entangled with the nuclei which rotate clockwise and $\ket \downarrow_z$ is entangled with the nuclei which rotate counter-clockwise; although each nucleus rotates by a different angle which is dictated by the individual interaction strength. This entanglement is detected via the NV by a measuring its coherence (purity). The faster the flow the weaker the effective interaction between the NV the the nuclei, which can be estimated the by amount of dephasing. The same effect is detected for polarized and unpolarized  nuclei.}
\label{pol_dec_1}
\end{figure}

This can be done by generating the nuclear spin polarization 
in the $x-y$ plane and let the NV-nuclei interaction decohere the NV.  This can be seen in the following way. Starting with the NV in the $\ket{\uparrow_x}$ and the nuclei in the $x$ direction (see fig. \ref{pol_dec_1}(A)), the state of the NV and the nuclei is

\beq\label{ME16}
\vert\psi\rangle=\vert\uparrow_{x}\rangle_{NV}\vert\uparrow_{x}\uparrow_{x}...\uparrow_{x}\rangle=\frac{1}{\sqrt{2}}\left(\vert\uparrow_{z}\rangle_{NV}+\vert\downarrow_{z}\rangle_{NV}\right)\vert\uparrow_{x}\uparrow_{x}...\uparrow_{x}\rangle.
\eeq
Under the $T_{0}$ term this state evolves to (see fig. \ref{pol_dec_1}(B))
\beq\label{ME17}
\vert\psi\rangle=\frac{1}{\sqrt{2}}\left(\vert\uparrow_{z}\rangle_{NV}\vert\uparrow_{\theta_{1}}\uparrow_{\theta_{2}}...\uparrow_{\theta_{3}}\rangle+\vert\downarrow_{z}\rangle_{NV}\vert\uparrow_{-\theta_{1}}\uparrow_{-\theta_{2}}...\uparrow_{-\theta_{3}}\rangle\right)\equiv\frac{1}{\sqrt{2}}\left(\vert\uparrow_{z}\rangle_{NV}\vert\psi_{1}\rangle+\vert\downarrow_{z}\rangle_{NV}\vert\psi_{2}\rangle\right),
\eeq
where the $\theta_{i}$'s are the rotation angles from the $x$ axis in the $x-y$ plane due to the NV - nuclei interaction.
The NV's density matrix is 
\beq\label{ME18}
\rho=\frac{1}{2}\left(\begin{array}{cc}
1 & \langle\psi_{1}\vert\psi_{2}\rangle\\
\langle\psi_{2}\vert\psi_{1}\rangle & 1
\end{array}\right).
\eeq
Measuring in the $x$ basis we get: 
\beq\label{ME19}
\textrm{Tr}\left(\rho\cdot\sigma_{x}\right)=\frac{1}{2}\left(\langle\psi_{1}\vert\psi_{2}\rangle+\langle\psi_{2}\vert\psi_{1}\rangle\right)=Re\langle\psi_{1}\vert\psi_{2}\rangle,
\eeq
as
\beq\label{ME20}
\langle\psi_{1}\vert\psi_{2}\rangle=\prod_{i=1}^{N}\cos\theta_{i}.
\eeq
The expectation value of the x-component of the NV spin polarization,
\beq\label{ME21}
\textrm{Tr}\left(\rho\cdot\sigma_{x}\right)=\prod_{i=1}^{N}\cos \theta_{i},
\eeq
which holds for immobile nuclei.
The nuclei's motion, however, decreases the dephasing rate due to dynamical averaging, which means
that instead of $\theta_{i} = f(\vec{r}_i) T$ we have $\theta_{i}=\int_0^T d t  f(\vec{r}_{i}(t))$, where $T$ is a single measurement time.  
Assuming the Gaussian distribution of the $\theta_i$s we have 
$\langle\cos\theta_i\rangle= \exp\left( -\frac{1}{2}\langle\theta_i^{2}\rangle\right)$. 
We then have from Eq.~\eqref{ME21}, in the limit of short correlation time, $\textrm{Tr}\left(\rho\cdot\sigma_{x}\right)= \exp\left( - S(\omega=0)T/2\right)$.
The same can be of course rigorously derived from the master equation as in Eqs. \eqref{ME2} and \eqref{ME3}.


We shall now show that this result also applies for an unpolarized ensemble of nuclear spins. First, lets examine a single nuclear spin polarized to $\ket{\downarrow_x}$. The joint state of the NV and the nucleus is $\ket{\uparrow_x}_{NV}\ket{\downarrow_x}$, and it propagates in time to $\frac{1}{\sqrt{2}}\left(\ket{\uparrow_z}_{NV}\ket{\downarrow_\theta}+\ket{\downarrow_z}_{NV}\ket{\downarrow_{-\theta}}\right)$. It follows that the inner product \eqref{ME20} is not affected by the different initial state of the nuclear spin. For many nuclear spins, as $\ket{\psi_{1,2}}$ are tensor product states, \eqref{ME20} still remain unchanged if each spin's polarization is chosen at random to be $\ket{\uparrow_{x}}$ or $\ket{\downarrow_x}$. This can be seen clearly by fig. \ref{pol_dec_1},  with simple modifications - initially each nuclear spin starts in $\ket{\uparrow_{x}/\downarrow_{x}}$, after some time, due to the interaction, the $i'$th nucleus is found rotated by an angle $\theta_i$ from it's initial position. Since the direction of the rotation is dependent on the state of the NV, and the value of $\theta_i$ is dependent on the position of the nuclear spin we arrive at the same dephasing dynamics as before.

Finally, we note that since the unpolarized ensemble is invariant under spin rotations, the direction chosen to represent the individual polarization of each nucleus is irrelevant. 
This scenario is, therefore, equivalent to the unpolarized ensemble presented in the previous sections.

\subsection{The flip-flop term \emph{$T_{0}'=-\frac{1}{4} \left( 3 \cos^2 \theta -1\right) \left(S_+ I_- + S_- I_+\right)$ }}
This term could also be used for sensing the parameters $v$ (kinematic viscosity) and $D$. In order to turn this interaction resonant, a very specific magnetic field \cite{Wood} or dynamical decoupling pulse sequence \cite{stark,joas} has to be applied. This term allows for an exchange of one energy quanta between the NV and a nuclear spin.
The optimal exchange is achieved for immobile spin ensemble, since the nuclei motion, due to diffusion or drift, shifts the interaction from resonance, and therefore decreases the efficiency of the process. By measuring the state of the NV center, the rate of the  process can be estimated and the flow parameters $v$ and $D$ can be deduced. 

The master equation for the NV's density matrix in this case is
\begin{equation}\label{ME22}
\dot{\rho} = -S_{BB}\left(\omega \right)\left(S_{+}\rho S_{-} + S_{-}\rho S_{+} - 2 \rho \right).
\end{equation}
This results in a dephasing dynamics with a dephasing rate which is equal to the power spectrum at the Larmor frequency. Namely, this result is the same as in the previous section, where the peak of the power spectrum is at the energy difference between the nuclei and the NV, which is approximately the NV's Larmor frequency.
Eq. \eqref{ME22} holds both for polarized and unpolarized nuclear spin ensembles, as in both cases the NV's initial state can be chosen to allow this type of energy exchange.

\subsection{The term \emph{$T_{\pm1}=\frac{3}{2}\sin\theta\cos\theta S_{z}\left(I_{x}\pm iI_{y}\right)$ }}

Unlike the $T_0$ term that contributes to the low frequency part of the power spectrum, this term affect the power spectrum at the vicinity of the nuclear spins Larmor frequency as in \cite{NV_depth,Kong,Peruncic,NanoNMRstutgart}. Therefore, dynamical decoupling is always required for this interaction to be the dominant one. After the dynamical decoupling this term can effectively be written as $T_{\pm1}=\frac{3}{2}\sin\theta\cos\theta S_{z}\left(I_{x}\cos\left(\delta t+\phi\right)+I_{y}\sin\left(\delta t+\phi\right)\right)$, where $\delta$ is the detuning of the dynamical decoupling frequency from the nuclear Larmor frequency.  
At first sight it seems that this term is very much different from the previous ones, however, we will see that the results are very much similar.

\begin{figure}[h]
\begin{center}
\includegraphics[width=0.95\textwidth]{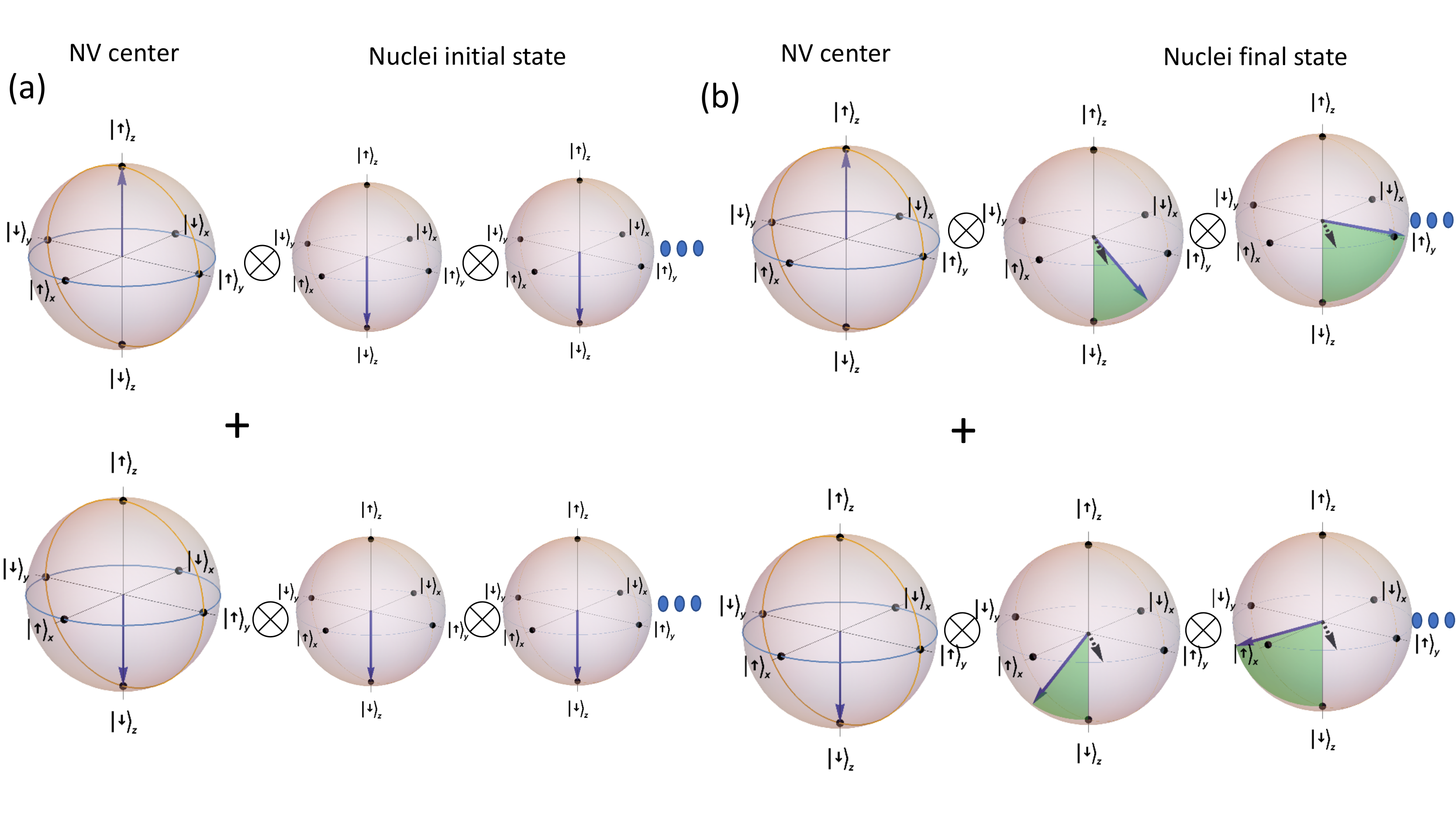}
\end{center}
\caption{ \textbf{Bloch sphere representation of the decoherence sensing in the polarized case in the interaction picture.}  \textbf{(A)} The NV start its dynamics in the $\ket \uparrow_x  = \frac{1}{2}\left( \ket \uparrow_z + \ket \downarrow_z \right).$ In the polarized case all the nuclei start in a $\ket \downarrow_z$ state. \textbf{(B)} The interaction between the NV and the nuclei creates an entangled states of the GHZ type where the $\ket {\uparrow_z}_{NV}$ is entangled with the nuclei which rotates in the counter-clockwise direction and $\ket {\downarrow_z}_{NV}$ is entangled with the nuclei which rotates in clockwise direction with respect to an axis in the $x-y$ plane (black arrow). This entanglement is detected via the NV by a decoherence measurement.}
\label{pol_dec_2}
\end{figure}

First, we shall examine a polarized nuclear spin ensemble.
The NV is initialized at the $\ket{\uparrow_x}_{NV}$ state and the nuclear spins are all initially oriented in the $\vert\downarrow_{z}\rangle$ state, see fig. \ref{pol_dec_2}. The $T_{\pm 1}$ terms causes them to rotate around an axis in the $x-y$ plane.
The axis will be time dependent, but for a propagation time $\tau$ in which $\delta \tau\ll1$ the rotation axis is approximately a constant vector in the x-y plane.

The exact axis of rotation is not important, as entanglement between the NV state and the nuclei state will be generated for any rotation axis in the x-y plane. Thus, the same dephasing dynamics as in the previous sections is achieved. This can also be seen by comparing fig. \ref{pol_dec_1} to fig. 
\ref{pol_dec_2} - the two processes are the same as in both cases we get a GHZ type state in different bases.
It can also be shown that quantitatively the result is similar.

\begin{figure}[h]
\begin{center}
\includegraphics[width=0.95\textwidth]{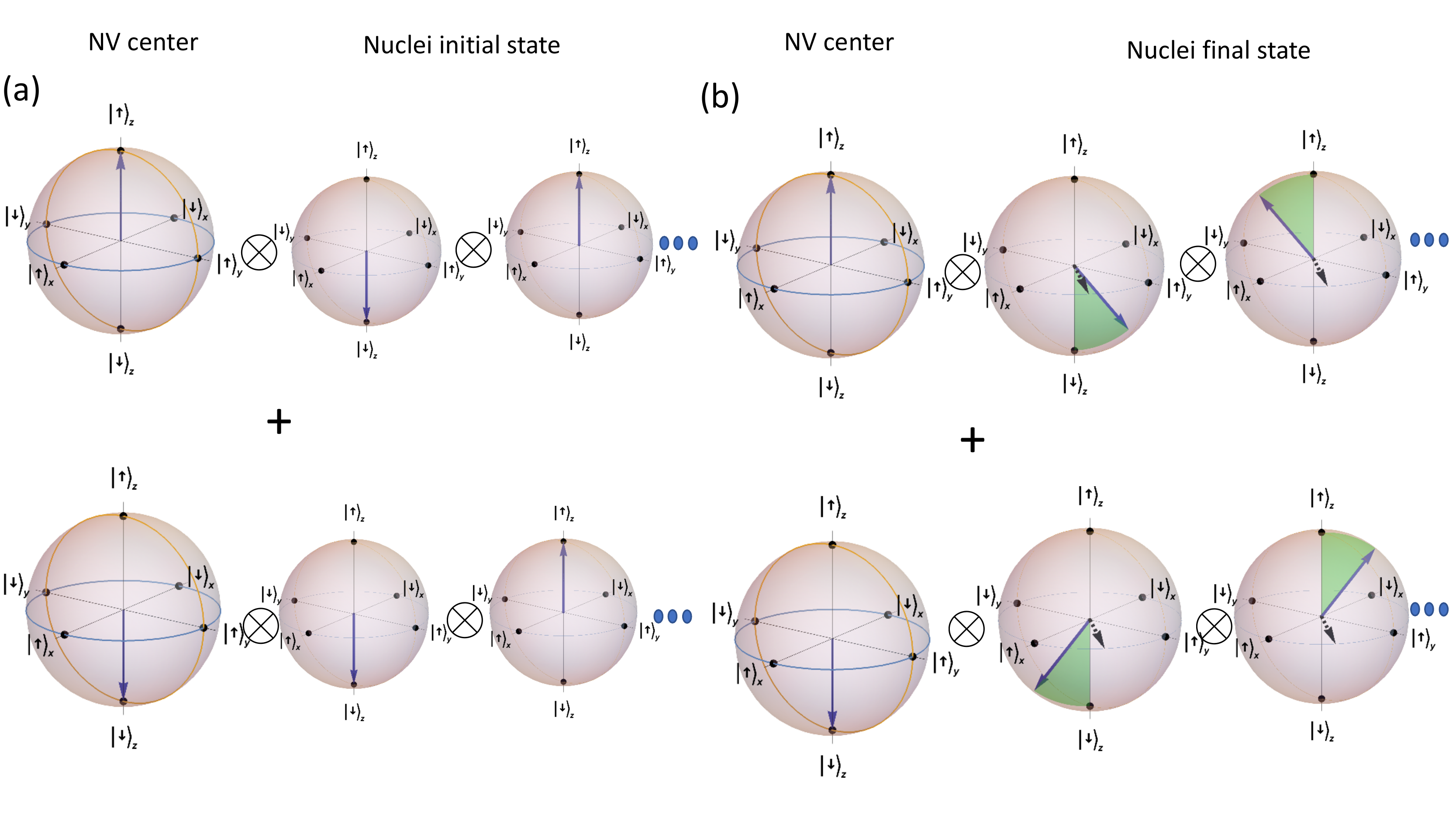}
\end{center}
\caption{\textbf{Bloch sphere representation of the decoherence sensing in the unpolarized case in the interaction picture.} \textbf{(A)} The NV starts its dynamics in the $\ket \uparrow_x  = \frac{1}{2}\left( \ket \uparrow_z + \ket \downarrow_z \right)$ state. In the unpolarized case half of  the nuclei start in a $\ket \downarrow_z$ state and half in $\ket \uparrow_z$. \textbf{(B)} The interaction between the NV and the nuclei creates an entangled states of the GHZ type where the $\ket {\uparrow_z}_{NV}$ is entangled with the nuclei which rotate in the counter-clockwise direction, upper row.   $\ket {\downarrow_z}_{NV}$ is entangled with the nuclei which rotates in clockwise direction with respect to an axis in the $x-y$ plane (black arrow), lower row. This entanglement is detected via the NV by a decoherence measurement.}
\label{unpol_dec_2}
\end{figure}



By following the same reasoning as in part \ref{Sec.Pol.}, the same dephasing dynamics can be shown to apply for an unpolarized spin ensemble. 
This can be seen by comparing fig. \ref{pol_dec_2} and fig. \ref{unpol_dec_2} - the important parameter is the angle difference between the nuclei which are entangled to the $\ket \uparrow_z$ and the ones which are entangled to the $\ket \downarrow_z$ as in Eq. \eqref{ME20}. This is left unaffected by the initial polarization of each nuclues.

\section{QUALITATIVE ESTIMATES OF THE POWER SPECTRUM BY DIMENSIONAL ANALYSIS}
\label{universal}
In the previous section we showed how the dephasing rate, which is proportional to the power spectrum could be measured, and we argued that the dynamical properties of the liquid can be estimated from this rate. Therefore, calculating the power spectrum and the temporal correlation function in terms of properties of the liquid is required for all our sensitivity estimates. In the following section we present crude estimation for the power spectrum based on dimensional analysis. We chose to present them first, since they are simple, independent of geometry, and most importantly they reproduce our main results while capturing their intuition.

\subsection{Conditional probability as a Green function of the drift-diffusion equation}
The noise caused by the drift/diffusion is characterized by the conditional probability $P(\bar{r},t|\bar{r}',t')$ of a particle being injected at time $t'$ at the location $\bar{r}'$ to be found at location $\bar{r}$ at subsequent times, $t>t'$.
At times exceeding the inter-particle collision time the motion becomes diffusive and for these times $P(\bar{r},t|\bar{r}',t')$ can be found by solving the differential equation it satisfies.
This equation at $t > t'$ follows from the Fick's law for the probability current, 
\beq\label{FL}
\bar{j} = - D \bar{\nabla} P(\bar{r},t|\bar{r}',t') + \bar{v}(\bar{r}) P(\bar{r},t|\bar{r}',t'),
\eeq
where $\bar{v}(\bar{r})$ is the drift velocity and $D$ is the diffusion coefficient.
The drift velocity $\bar{v}(\bar{r})$ is assumed to be time independent and given.
The probability conservation is given by
\beq\label{cont}
\partial_t P(\bar{r},t|\bar{r}',t') + \bar{\nabla} \cdot \bar{j} = 0.
\eeq
Combining \eqref{FL} and \eqref{cont} we readily obtain,
\beq\label{cont1}
\partial_t P(\bar{r},t|\bar{r}',t') - D \bar{\nabla}^2 P(\bar{r},t|\bar{r}',t') + \bar{\nabla} [\bar{v}(\bar{r}) P(\bar{r},t|\bar{r}',t')]= 0,\quad t >t'
\eeq
Assuming that the flow has no sources or drains in the region of interest, and that the fluid is incompressible we have,
\beq\label{assump}
\bar{\nabla} \cdot \bar{v} = 0
\eeq
and eq. \eqref{cont1} simplifies to
\beq\label{cont2}
\left[ \partial_t - D \bar{\nabla}^2 + \bar{v} \cdot \bar{\nabla} \right] P(\bar{r},t|\bar{r}',t')= 0,\quad t >t' \ .
\eeq
By definition, the conditional probability complies with the initial condition,
\beq\label{cont3}
\lim_{t \rightarrow t' + 0}  P(\bar{r},t|\bar{r}',t') = \delta ( \bar{r} - \bar{r}') \ .
\eeq
\eqref{cont2} and \eqref{cont3} can be combined into the single equation,
\beq\label{cont4}
\left[ \partial_t - D \bar{\nabla}^2 + \bar{v} \cdot \bar{\nabla} \right] P(\bar{r},t|\bar{r}',t')=  \delta ( \bar{r} - \bar{r}') \delta(t - t'), \qquad P(\bar{r},t<t' |\bar{r}',t')= 0
\eeq
It can be shown that the power spectrum of the noise is a weighted Fourier transformation of the conditional probability \cite{Abragam1961}, 
\beq\label{cont5}
\Gamma(\omega) = \int' d^3r \int' d^3r'  P_{\omega}(\bar{r},\bar{r}') f_1(\bar{r}) f_2(\bar{r}')= 2\mathrm{Re} \int_{- \infty}^{\infty} d t e^{ i \omega t }  \int' d^3r \int' d^3r'  P(\bar{r},t|\bar{r}',t') f_1(\bar{r}) f_2(\bar{r}'),
\eeq
where the weighting functions $f_{1,2}(\bar{r})$ depend on the known form of the dipole-dipole interaction, presented in the previous section. The first equality is shown in the previous section \eqref{PS1}. The second equality is due to the fact that the diffusion propagator is taken to be causal and coincides with \cite{Hwang1975}.
The  notation $\int'$ means integration over the volume occupied by fluid.
The Fourier transform of the conditional probability defined in \eqref{cont5} satisfies according to \eqref{cont4}
\beq\label{cont6}
\left[ - i \omega - D \bar{\nabla}^2 + \bar{v} \cdot \bar{\nabla} \right] P_{\omega}(\bar{r},\bar{r}')=  \delta ( \bar{r} - \bar{r}')
\eeq
with the standard requirements on analyticity in the space of a complex frequency.

\subsection{Universal features of the diffusion induced noise}
In general, the geometrical constraints provide us with the typical distance $d$ of the NV from the rest of the diffusing/drifting  spins. 
The diffusion frequency scale is
\beq\label{omega_D}
\omega_D = \frac{ D }{ d^2 }
\eeq
and we expect the power spectrum to depend on $\omega/\omega_D$.

\subsubsection{Static limit: $\omega = 0$}
In the static limit, $\omega = 0$, the problem is equivalent to electrostatics, and \eqref{cont6} becomes the Poisson equation.
We therefore have 
\beq\label{stat1}
P_{\omega}(\bar{r},\bar{r}')  \sim \frac{ 1}{ D |\bar{r} - \bar{r}'|} 
\eeq
where we omitted the contribution of ``image'' charges, which are necessary in order to satisfy the boundary condition.
This will be readily explained, while we first proceed with the approximation \eqref{stat1}.
Substituting \eqref{stat1} into \eqref{cont5} we obtain from dimensional analysis the following estimate,
\beq\label{stat2}
\Gamma(\omega=0) =  \int' d^3r \int' d^3r'  \frac{ 1}{ D |\bar{r} - \bar{r}'|} \frac{1}{r^3} \frac{1}{r'^3} \propto \frac{1}{d D}
\eeq
which means that
\beq\label{stat3}
\Gamma(\omega=0) = \frac{C_1}{d D},
\eeq
where $C_1$ is a non-universal constant.
The contribution of ``image'' charges may only affect the value of $C_1$, because the typical distances fixing their location are again $\sim d$. Therefore the ``images'' contribution can be eliminated.
This is the same result obtained in \cite{NV_depth} with an appropriate choice of $C_1$. This is used in the main text for $S(\omega=0)$ of the Lorentzian model, and therefore as the normalization factor of the power spectrum in Eq. (1).  
\subsubsection{Low frequencies, $\omega \ll \omega_D$}
Finite frequency yields a finite length scale, $l_{\omega} = \sqrt{D/\omega}$  .
Therefore, in the regime $l_{\omega} \gg d$ which is the same as $\omega \ll \omega_D$ we may estimate roughly the suppression of the power spectrum by cutting off the distances that are larger than this characteristic length:
\begin{align}\label{stat5}
\Gamma(\omega  \ll \omega_D) & \approx   \int' d^3 r \int' d^3 r'  \frac{ 1}{ D |\bar{r} - \bar{r}'|} \frac{1}{r^3} \frac{1}{r'^3}  - 
\int_{r \gtrsim \sqrt{D/\omega} }  d^3r' \int_{r' \gtrsim \sqrt{D/\omega} }d^3 r'  \frac{ 1}{ D |\bar{r} - \bar{r}'|} \frac{1}{r^3} \frac{1}{r'^3} 
\notag \\
& 
\approx  \frac{C_1 }{d D}   - \frac{ C'_2 }{ D \sqrt{D/\omega} } = \frac{ 1}{ d D } 
\left[ C_1 - C_2 \sqrt{ \frac{\omega}{\omega_D} } \right].
\end{align}
The constants $C_{1,2}$ depend on geometry but other than that the result \eqref{stat5} is universal.
This argument shows the origin of the nature of the convex of the power spectrum which is responsible of the enhanced sensitivity.

\subsubsection{Large frequencies, $\omega \gg \omega_D$}
\label{sec33}
Lets turn to \eqref{cont6}, substituting $v=0$, 
\beq\label{dyn12}
\left[ - i \omega - D \bar{\nabla}^2  \right] P_{\omega}(\bar{r},\bar{r}')=  \delta ( \bar{r} - \bar{r}').
\eeq
We estimate the spatial Fourier transform of \eqref{dyn12},
\beq\label{dyn14}
P(q,\omega) \propto \frac{1}{ - i \omega + D q^2 },
\eeq
by concluding that the important values of $q$ are of the order $d$. Therefore in the real space,
\beq\label{dyn16}
P(\bar{r},\bar{r}',\omega) \propto \frac{d^{-3}}{ - i \omega + D /d^2},
\eeq
which is the same as approximating \eqref{dyn12} directly by considering the $D \bar{\nabla}^2 \approx  - D/d^2$ as a perturbation on top of the large frequency, $\omega \gg D /d^2 $.
Returning to \eqref{dyn16}, since the typical distances are given by $d$,
\beq\label{dyn21}
\Gamma(\omega \gg \omega_D ) \approx  \mathrm{Re} \frac{d^{-3}}{ - i \omega + D /d^2} \approx
d^{-3} \frac{D}{d^2} \frac{1}{ \omega^2 } \propto \frac{ 1 }{ d D } \frac{ \omega_D^2 }{ \omega^2 },
\eeq
which is a typical damped oscillator behavior.

We could obtain this behavior starting from the Lorentzian curve,
\beq\label{dyn22}
\Gamma(\omega \gg \omega_D) \approx \Gamma(\omega =0 )\mathrm{Re} \frac{ \omega_D }{ - i \omega  + \omega_D}  \approx \Gamma(\omega = 0 ) \frac{ \omega_D^2}{ \omega^2} = \frac{1}{ d D} \frac{ \omega_D^2}{\omega^2}.
\eeq
This crude dimensional analysis is validated later on in the spherical geometry. 
Equations \eqref{stat5} and \eqref{dyn21} correspond to the behavior depicted in Fig. 3a in the main text.

\subsection{Estimates of the effect of the drift on the power spectrum}
We now make a crude but useful estimates of how the drift affects the power spectrum at different frequencies.
We characterize the drift by the parameter
\beq\label{omega_v}
\omega_v = \frac{v}{d},
\eeq
similar to \eqref{omega_D}.
To determine the change in the power spectrum we apply the idea of "Doppler shift" presented in \cite{Wang}.
Lets assume that the potential of the interaction between the NV and the spins in the fluid has a simple form,
\beq
V(x) = V_q \cos (q x)
\eeq
where the geometry is encoded in the scale $d$ such that the typical $q \sim 1/d$.
For a harmonic wave in principle there are two contributions, $\cos (q x) =\frac{1}{2}e^{ i q x } + \frac{1}{2}e^{ - i q x}$, which gives rise to the Doppler shifts 
$\pm \Delta_{Doppler}$, that enter the propagator  
\beq\label{Doppler1}
P_{v}\left(\omega\right) = \frac{1}{2} \left[P_{v=0}\left(\omega + \Delta_{Doppler}\right)  + P_{v=0}\left(\omega - \Delta_{Doppler}\right) \right]\ , \qquad  \Delta_{Doppler} = v q \sim \omega_v.
\eeq
In the following, we expand \eqref{Doppler1} for "small" Doppler shift. We derive the exact requirement post-priori. 
There is an obvious concern about the cancellation of the linear terms in the expansion of \eqref{Doppler1}.
The cancellation appears because the two terms $e^{\pm i q x}$ are present in \eqref{Doppler1} with equal weights.
This cancellation will not occur if the two components were with different weights.
This means that the magnetic fields $f_{1}(\bar{r}) \propto Y_{l=2}^{m'}$ and $f_{2}(\bar{r}') \propto Y_{l=2}^{m}$ enter \eqref{cont5} with different values of $m$, $m \neq m'$.
This is the case when the rotational symmetry is broken, i.e., when the NV's magnetization axis points in an arbitrary direction (exact definition is geometry dependent).  
This would bring about the breaking of the symmetry between the two senses of propagation.
We note that even in the symmetric case, cancellation does not occur for $\omega =0$ since 
\beq\label{Doppler2}
P_{v}\left(\omega=0\right) \approx \frac{1}{2} \left[P_{v=0}\left(\Delta_{Doppler}\right)  + P_{v=0}\left(-\Delta_{Doppler}\right)\right] = P_{v=0}\left(\Delta_{Doppler}\right).
\eeq
Hence, instead of \eqref{Doppler1} we proceed with
\beq\label{Doppler3}
P_{v}\left(\omega\right) = P_{v=0}\left(\omega + \Delta_{Doppler}\right) \ , \qquad  \Delta_{Doppler} = v q \sim \omega_v.
\eeq
We then Taylor expand \eqref{Doppler3} for sufficiently  weak Doppler shift,
\beq\label{Doppler4}
P_{v}\left(\omega\right)  \approx P_{v=0}\left(\omega\right) + \frac{ \partial P_{v=0}}{ \partial \omega } \Delta_{Doppler} .
\eeq
In every given range of frequencies it implies different limitations on the parameters as we discuss below.
By dimensional arguments \eqref{Doppler4} also implies the same relation for the power spectrum,
\beq\label{Doppler5}
\Gamma_v\left(\omega\right) \sim \Gamma_{v=0}\left(\omega\right) + \frac{ \partial \Gamma_{v=0}\left(\omega\right)}{ \partial \omega } \omega_v.
\eeq
Our claim is that all the results obtained in later sections are consistent with \eqref{Doppler5}.
We now consider different frequency regimes in turn.
\subsubsection{High Frequencies: $ \omega \gg \omega_D$} 
In this case the application of the Doppler shift equation \eqref{Doppler5} to the asymptotic expression \eqref{dyn21} gives
\beq\label{Doppler7}
\Gamma_v\left(\omega\right) =  \frac{ 1 }{ d D } \frac{ \omega_D^2 }{ \omega^2 } - \frac{ 2 }{ d D } \frac{ \omega_D^2 }{ \omega^2 }  \left( \frac{\omega_v}{\omega} \right).
\eeq
We can also deduce the relative change as
\beq\label{Doppler9}
\Gamma_v\left(\omega\right)  -\Gamma_{v=0}\left(\omega\right)  \approx -2\Gamma_{v=0}\left(\omega\right)  \left( \frac{\omega_v}{\omega} \right),
\eeq
which tells us that the expansion works in the range,
\beq\label{Doppler11}
\omega \gg \max\{\omega_D, \omega_v\}.
\eeq
Eq. \eqref{Doppler7} is used for the high frequency domain in Fig. 3b of the main text.
\subsubsection{Low frequencies: $\omega \ll \omega_D$}
In this case the application of the Doppler shift equation \eqref{Doppler5} to the asymptotic form \eqref{stat5} gives
\beq\label{Doppler31}
\Gamma_v\left(\omega\right)  -\Gamma_{v=0}\left(\omega\right)  \approx -\frac{\Gamma(\omega = 0)}{2} \left[  \sqrt{ \frac{ \omega_D }{\omega} } \left( \frac{\omega_v}{\omega_D} \right) \right].
\eeq
Lets understand when \eqref{Doppler31} holds.
First, the correction has to be small relative to the first term. Therefore, we have to impose the inequality,
\beq\label{Doppler33}
\omega \gg \frac{\omega_v^2}{\omega_D}.
\eeq
Since we are already in the regime  $\omega \ll \omega_D$, we must have $\omega_v \ll \omega_D$.
These conditions can be written in a compact manner together with \eqref{Doppler33}, 
\beq\label{Doppler34}
\frac{\omega_v^2}{\omega_D} \ll \omega \ll \omega_D\, , \qquad \omega_v \ll \omega_D.
\eeq 
Yet, if we want to stop our series expansion, we must ensure that the next order contribution is negligible. Consider including one more term in the expansion \eqref{Doppler31}
\beq\label{Doppler35}
\Gamma_v\left(\omega\right)  -\Gamma_{v=0}\left(\omega\right)  \approx -\frac{\Gamma(\omega = 0)}{2}\left[  \sqrt{ \frac{ \omega_D }{\omega} } \left( \frac{\omega_v}{\omega_D} \right)
-\frac{1}{2}
\left( \frac{ \omega_D }{\omega} \right)^{3/2} \left( \frac{\omega_v}{\omega_D} \right)^2 + \ldots
\right].
\eeq
The absolute value of the ratio of the third and the second term is given by $\frac{\omega_v}{2\omega}$, therefore, we have instead of \eqref{Doppler34} the following limitation, 
\beq\label{Doppler37}
\omega_v \ll \omega \ll \omega_D .
\eeq

Eq. \eqref{Doppler31} is later verified for the planar geometry, and is used for the intermediate regime of Fig. 3b of the main text.
\subsubsection{Intermediate range $\omega \sim \omega_D$}
In this case, the conditions \eqref{Doppler11} and \eqref{Doppler37} both become $\omega_v \ll \omega_D$.
This means that our estimates should agree in the intermediate range.
Both expressions \eqref{Doppler9} and \eqref{Doppler31} match as expected (up to a prefactor) and give
\beq\label{Doppler41}
\Gamma_v\left(\omega\right)  -\Gamma_{v=0}\left(\omega\right)  \sim \Gamma(\omega=0)  \left( \frac{\omega_v}{\omega_D} \right)
\eeq
\subsubsection{Low frequencies: $\omega \lesssim \omega_v$}
To get an estimate in this case we assume that at low frequencies the only relevant scale is Doppler shift itself, $\omega_v$.
Therefore, we can simply take the expression \eqref{Doppler31} and substitute in it $\omega = \omega_v$, 
\beq\label{Doppler43}
\Gamma_v\left(\omega\right)  -\Gamma_{v=0}\left(\omega\right)  \approx -\frac{\Gamma(\omega = 0)}{2}\sqrt{ \frac{ \omega_v }{\omega_D} }  
\eeq
Although this is a very crude estimation, it agrees with the results of the molecular dynamics simulation presented in the main text, and it reproduces the results obtained in the spherical geometry shown in the next section. This result also seems to be universal for sufficiently slow varying flow profile. 
Eq. \eqref{Doppler43} coincides with Eq. 4 of the main text, and is used in Fig. 3b of the main text for the low frequency region. 

\section{SPHERICAL GEOMETRY}
\label{Sphere}
After obtaining the generic behavior of the power spectrum we wish to consider a specific model. We start with a simple spherical geometry as a "toy model", which can be solved analytically and help develop intuition. 
Consider the fluid in the region outside a sphere of radius $d$ with the NV placed in it's center. Moreover, we introduce the drift as a rotation of the fluid as a whole at the angular frequency $\bar{\Omega} = \Omega \hat{z}$. 
We will first derive the propagator by solving Eq. \eqref{cont6} with $\bar{v} = \bar{\Omega} \times \bar{r}$. The drift term in this case,
\be\label{sp1}
\bar{v} \cdot \bar{\nabla} P_{\omega}(\bar{r},\bar{r}') = (\bar{\Omega} \times \bar{r} ) \cdot \vec{\nabla} P_{\omega}(\bar{r},\bar{r}') =\bar{\Omega} \cdot  (\bar{r} \times \bar{\nabla} P_{\omega}(\bar{r},\bar{r}') ) .
\ee
If $v$ is the velocity at the radius $d$ in equatorial plane, we simply have
\be
\omega_v = \frac{v}{d} = \Omega,
\ee
as previously defined in \eqref{omega_v}. Henceforth we use the notation $\omega_v$ instead of $\Omega$.

Using the definition of the angular momentum, $\bar{L} = \frac{\hbar}{i}\ \bar{r} \times \bar{\nabla}$ in spherical coordinates,
\be
L_z = \frac{\hbar}{i} \partial_{\phi},
\ee
we obtain for the drift term \eqref{sp1},
\be
\bar{v} \cdot \bar{\nabla} P_{\omega}(\bar{r},\bar{r}') = \frac{ i }{ \hbar} \omega_v L_z P_{\omega}(\bar{r},\bar{r}') = \omega_v \partial_{\phi} P_{\omega}(\bar{r},\bar{r}'),
\ee
and \eqref{cont6} takes the form
\begin{align}\label{sp2}
	( - i \omega - D \nabla^2 + \omega_v \partial_{\phi} ) P_{\omega}(\bar{r},\bar{r}') = \delta (\bar{r} - \bar{r}')
\end{align}
Recall that in spherical coordinates  
\be\label{sp3}
\nabla^2 P_{\omega}(\bar{r},\bar{r}')=  \frac{1}{r} \frac{\partial^2 }{\partial r^2 } (r P_{\omega}(\bar{r},\bar{r}')) - \frac{L^2}{ r^2 } P_{\omega}(\bar{r},\bar{r}'),
\ee
where the $L$-operator is divided by $\hbar$. Substituting \eqref{sp3} in \eqref{sp2} and using the spherical harmonics completeness relation,
\begin{align}\label{sp4}
	&- i \omega P_{\omega}(\bar{r},\bar{r}') -D \frac{1}{r} \frac{\partial^2 }{\partial r^2 } (r P_{\omega}(\bar{r},\bar{r}')) + D\frac{L^2}{ r^2 } P_{\omega}(\bar{r},\bar{r}') + \omega_v \frac{\partial }{\partial \phi} P_{\omega}(\bar{r},\bar{r}') 
	=\delta(\bar{r} - \bar{r}') = \frac{1}{r^2} \delta(r - r') \delta(\phi - \phi') \delta(\cos \theta - \cos \theta') 
	\notag \\
	& =\frac{1}{r^2} \delta(r - r') \sum_{lm} Y^*_{lm}(\theta',\phi') Y_{lm}(\theta,\phi). 
\end{align}
We shall look for a solution in the form
\be\label{form1}
P_{\omega}(\bar{r},\bar{r}') = \sum_{lm} g_{lm}(r,r')  Y^*_{lm}(\theta',\phi') Y_{lm}(\theta,\phi) .
\ee
Substituting \eqref{form1} into \eqref{sp4} results in
\be\label{target_0}
\left[- i \omega  -D \frac{1}{r} \frac{\partial^2 }{\partial r^2 } r  + D\frac{l(l+1)}{ r^2 }  - i m  \omega_v  \right] g_{lm}(r,r') = \frac{1}{r^2} \delta(r - r').
\ee
Let us introduce the Doppler shifted frequency,
\be\label{tilde_omega}
\tilde{\omega} = \omega + m \omega_v.
\ee
\eqref{target_0} can then be rewritten as
\begin{align}\label{sp5}
	\left[
	r^2  \frac{\partial^2 }{\partial r^2 }  + 2 r   \frac{\partial }{\partial r } - l(l+1) + i r^2 \frac{\tilde{\omega}}{D}  
	\right] g_{lm}(r,r')
	= - \frac{1}{D}\delta(r - r') 
\end{align}
Defining the dimensionless distances
\be\label{rescale}
x = r \sqrt{ \frac{|\tilde{\omega}| }{D}} = \frac{r}{d} \sqrt{\frac{|\tilde{\omega}|}{\omega_D}} ; \qquad x' = r' \sqrt{ \frac{|\tilde{\omega}| }{D}} =   \frac{r'}{d} \sqrt{\frac{|\tilde{\omega}|}{\omega_D}}
\, , \quad \bar{d} = d \sqrt{ \frac{|\tilde{\omega}| }{D}} =  \sqrt{\frac{|\tilde{\omega}|}{\omega_D}},
\ee
we can write \eqref{target_0} as follows,
\begin{align}\label{target}
	\left[
	x^2  \frac{\partial^2 }{\partial x^2 }  + 2 x   \frac{\partial }{\partial x} - l(l+1) +  (\pm i) x^2
	\right] g_{lm}(x,x')
	= - \frac{1}{D}  \sqrt{ \frac{|\tilde{\omega}| }{D}}    \delta(x - x') =- \frac{1}{ d D}  \sqrt{\frac{|\tilde{\omega}|}{\omega_D}} \delta(x - x'),
\end{align}
where 
the sign of $(\pm i)$ is the same as the sign of $\tilde{\omega}$.
For convenience we define,
\be\label{sq}
(+i)^{1/2}= \frac{1 + i }{ \sqrt{2} } \, , \qquad (-i)^{1/2}= \frac{-1 + i }{ \sqrt{2} } 
\ee
with the convention that the imaginary part is positive in both cases.
Note that for any solution, $\mathcal{F}(z)$ of  the spherical Bessel equation,
\be\label{sphB}
z^2 \frac{ d^2\mathcal{F}(z) }{ d z^2} + 2 z \frac{ d \mathcal{F}(z)}{ d z} - l(l+1) \mathcal{F}(z) + z^2 \mathcal{F}(z) = 0
\ee
the functions $\mathcal{F}( (\pm i)^{1/2} x)$  is the solution of the 
Eq.~\eqref{target} with $0$ on the right hand side.
The solutions of \eqref{sphB} are spherical Bessel functions.
The pair of solutions that we chose to consider are the spherical Hankel functions of first and second kinds, $h_l^{(1)}(z)$ and $h_l^{(2)}(z)$.
The two functions scales differently at infinity,  $h_l^{(1)}(z) \sim e^{i z} $, $h_l^{(2)}(z) \sim e^{-i z} $.
As long as the Hankel functions solve \eqref{sphB} the two functions 
$h_l^{(1)}((\pm i)^{1/2}x)$ and $h_l^{(2)}((\pm i)^{1/2}x)$ form a pair of independent solutions of \eqref{target} for $x \neq x'$.
Because of the convention \eqref{sq} the physically acceptable solutions will always be $h_l^{(1)}((\pm i)^{1/2}x)$ at large $x$.
The continuity at $x =x'$ and the convergence at $x \rightarrow \infty$ tells us that we have a solution of the form,
\be\label{form}
g_{lm}(x,x') = h_l^{(1)}[(\pm i)^{1/2}x_>] \left( A_1 h_l^{(1)}[(\pm i)^{1/2}x_<]  + A_2 h_l^{(2)}[(\pm i)^{1/2}x_<] \right)
\ee
To fix the two constant $A_{1,2}$ we need two conditions, one is vanishing of the flux at $r = d$ and the second is the jump of the derivative at $r = r'$.
Vanishing of the flux across the hard surface gives
\be\label{sp_cond1}
0 = \frac{ \partial g_{lm}(x,x') }{\partial x }\bigg\vert_{x = \bar{d}} = A_1 \frac{ d h_l^{(1)} [(\pm i)^{1/2}x] }{d x} \bigg\vert_{x = \bar{d}} + A_2 \frac{ d h_l^{(2)}[(\pm i)^{1/2}x ]}{d x} \bigg\vert_{x = \bar{d}} = 0\, .
\ee
Lets introduce the notation,
\be
h_l^{\prime(1)}[(\pm i)^{1/2} \bar{d}] =  \frac{ d h_l^{(1)} [(\pm i)^{1/2}x] }{d x} \bigg\vert_{x = \bar{d}}
\ee
with a similar notation for $h_l^{\prime(2)}[(\pm i)^{1/2} \bar{d}]$.
Then by \eqref{sp_cond1} we have the relation,
\be\label{A12}
A_1 = -  A_2 \frac{   h_l^{\prime(2)} \left[(\pm i)^{1/2} \bar{d}\right] }{h_l^{\prime(1)}\left[(\pm i)^{1/2} \bar{d}\right] } .
\ee
To find $A_2$ we multiply \eqref{target_0} by $(- \frac{r}{D})$ and apply the operation $\lim_{\epsilon \rightarrow 0} \int_{r'- \epsilon}^{r' + \epsilon} d r$ to both sides to get,  
\be
\frac{ d }{ d r} ( r g_{lm} (r,r') ) \bigg\vert_{r = r' + \epsilon} - \frac{ d }{ d r} ( r g_{lm} (r,r') ) \bigg\vert_{r = r' - \epsilon} = - \frac{ 1 }{ r' D},
\ee
which in terms of rescaled variables \eqref{rescale} translates into,
\be
\frac{ d }{ d x} ( x g_{lm} (x,x') ) \bigg\vert_{x = x' + \epsilon} - \frac{ d }{ d x} ( x g_{lm} (x,x') ) \bigg\vert_{x = x' - \epsilon} = - \frac{ 1 }{ x' } \sqrt{ \frac{| \tilde{\omega}| }{ D^3 }}.
\ee
Using \eqref{form} we write the last equation as
\begin{align}
	& \frac{ d }{ d x} ( x h_l^{(1)}[(\pm i)^{1/2}x] )\bigg\vert_{x = x' }  \left( A_1 h_l^{(1)}[(\pm i)^{1/2}x']  + A_2 h_l^{(2)}[(\pm i)^{1/2}x'] \right)  
	\notag \\
	& -
	h_l^{(1)}[(\pm i)^{1/2}x'] \frac{ d }{ d x} ( x \left( A_1 h_l^{(1)}[(\pm i)^{1/2}x]  + A_2 h^{(2)}_l[(\pm i)^{1/2}x] \right)\bigg\vert_{x = x' }
	= - \frac{ 1 }{ x' } \sqrt{ \frac{ |\tilde{\omega} |}{ D^3 }}
\end{align}
Simplifying the last equation we are left with the following equation for $A_2$,
\begin{align}
	A_2 \left\{ \frac{ d }{ d x} (  h_l^{(1)}[(\pm i)^{1/2}x] )\bigg\vert_{x = x' }  h_l^{(2)}[(\pm i)^{1/2}x'] 
	-
	h_l^{(1)}[(\pm i)^{1/2}x'] \frac{ d }{ d x}  h^{(2)}_l[(\pm i)^{1/2}x] \bigg\vert_{x = x' } \right\}
	= - \frac{ 1 }{ x^{' 2} } \sqrt{ \frac{ |\tilde{\omega} |}{ D^3 }}.
\end{align}
Notice that the Wronskian of the two independent spherical Hankel functions is 
\be
W[ h_l^{(1)}(z),h_l^{(2)}(z) ] \equiv  h_l^{(1)}(z) [h_l^{(2)}(z) ]' - [h_l^{(1)}(z)]' h_l^{(2)}(z) =- 2 i /z^2
\ee
which gives 
\be
A_2 (\pm i)^{1/2} \frac{2 i }{ (\pm i) x^{'2} } = - \frac{ 1}{ x^{'2} } \sqrt{ \frac{ |\tilde{\omega}| }{ D^3 }}\, , \qquad
A_2 = - \frac{ (\pm i)^{1/2}}{ 2 i} \sqrt{ \frac{ |\tilde{\omega}| }{ D^3 }} = \frac{1}{2} i (\pm i)^{1/2} \sqrt{ \frac{ |\tilde{\omega} |}{ D^3 }} 
\ee
Then from \eqref{A12}
\be\label{A12_a}
A_1 = -  \frac{1}{2} i (\pm i)^{1/2} \sqrt{ \frac{ |\tilde{\omega} |}{ D^3 }}  \frac{   h_l^{\prime(2)} \left[(\pm i)^{1/2} \bar{d}\right] }{h_l^{\prime(1)}\left[(\pm i)^{1/2} \bar{d}\right] } 
\ee
And from \eqref{form}
\be\label{form_1}
g_{lm}(x,x') =- \frac{ (\pm i)^{1/2}}{ 2 i} \sqrt{ \frac{ |\tilde{\omega}| }{ D^3 }} h_l^{(1)}[(\pm i)^{1/2}x_>] \left( - \frac{   h_l^{\prime(2)} \left[(\pm i)^{1/2} \bar{d}\right] }{h_l^{\prime(1)}\left[(\pm i)^{1/2} \bar{d}\right] }  h_l^{(1)}[(\pm i)^{1/2}x_<]  + h_l^{(2)}[(\pm i)^{1/2}x_<] \right)
\ee
Alternatively,
\be\label{form_2}
g_{lm}(x,x') = \frac{ (\pm i)^{1/2}}{ 2 i} \sqrt{ \frac{ |\tilde{\omega}| }{ D^3 }} \left(  \frac{   h_l^{\prime(2)} \left[(\pm i)^{1/2} \bar{d}\right] }{h_l^{\prime(1)}\left[(\pm i)^{1/2} \bar{d}\right] } h_l^{(1)}[(\pm i)^{1/2}x_>]  h_l^{(1)}[(\pm i)^{1/2}x_<]  - h_l^{(1)}[(\pm i)^{1/2}x_>] h_l^{(2)}[(\pm i)^{1/2}x_<] \right)
\ee
It will be more convenient to express the result in terms of the dimensionless frequency, according to \eqref{rescale},
\be\label{bar_o}
\bar{d} = d \sqrt{ \frac{|\tilde{\omega}| }{D}} = \sqrt{|\bar{\omega}|}\, , \qquad \bar{\omega} = \tilde{\omega} \frac{d^2}{D} = \frac{\tilde{\omega}}{\omega_D}.   
\ee
With \eqref{bar_o}, \eqref{form_2} can be written as
\be\label{form_3}
g_{lm}(x,x') = \frac{ (\pm i)^{1/2}}{ 2 i} \frac{1}{d D} \sqrt{ |\bar{\omega}| } \left(  \frac{   h_l^{\prime(2)} \left[(\pm i)^{1/2} \sqrt{|\bar{\omega}|}\right] }{h_l^{\prime(1)}\left[(\pm i)^{1/2} \sqrt{|\bar{\omega}|}\right] } h_l^{(1)}[(\pm i)^{1/2}x_>]  h_l^{(1)}[(\pm i)^{1/2}x_<]  - h_l^{(1)}[(\pm i)^{1/2}x_>] h_l^{(2)}[(\pm i)^{1/2}x_<] \right)
\ee

\begin{figure}
	\includegraphics[width=0.6\textwidth]{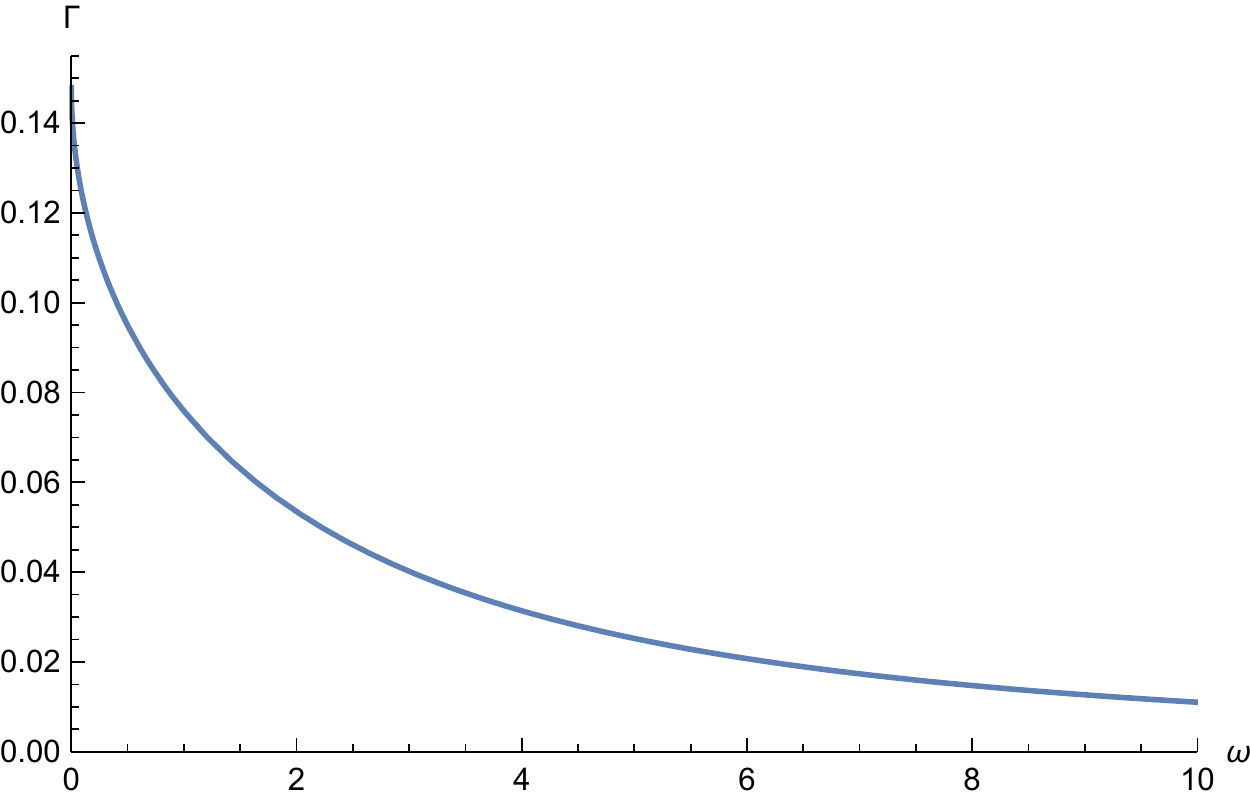} 
	\caption{\textbf{The exact power spectrum, $\Gamma(\omega)$ as given by \eqref{exact}.} 
		The units of power spectrum are $1/(d D)$ and the power spectrum is plotted as a function of the dimensionless frequency, $\omega/\omega_D$.\label{fig:res}}
\end{figure}

\subsection{Exact expression for the power spectrum}
In accordance with \eqref{cont5} the power spectrum in given by
\be
\Gamma(\omega) = 2\mathrm{Re} \int_{r>d}  \frac{d \bar{r} }{ r^3} \int_{r'>d}  \frac{d \bar{r}' }{r'^3} Y_{2,m_1}^*(\theta \phi) Y_{2,m_2}(\theta' \phi') P_{\omega}(\bar{r},\bar{r}').
\ee 
After the rescaling \eqref{rescale}, substituting the solution \eqref{form} together with \eqref{form_2} and integrating over the angular coordinates, the previous equation will read
\begin{align}\label{target_1}
	\Gamma(\omega) & = 2 \mathrm{Re}  \frac{ (\pm i)^{1/2}}{ 2 i} \frac{1}{d D} \sqrt{ |\bar{\omega}| } \int_{\sqrt{\bar{\omega}}}^{\infty} \frac{d x}{x} \int_{\sqrt{\bar{\omega}}}^{\infty} \frac{d x'}{x'} 
	\notag \\
	\times & 
	\left(  \frac{   h_l^{\prime(2)} \left[(\pm i)^{1/2} \sqrt{|\bar{\omega}|}\right] }{h_l^{\prime(1)}\left[(\pm i)^{1/2} \sqrt{|\bar{\omega}|}\right] } h_l^{(1)}[(\pm i)^{1/2}x_>]  h_l^{(1)}[(\pm i)^{1/2}x_<]  - h_l^{(1)}[(\pm i)^{1/2}x_>] h_l^{(2)}[(\pm i)^{1/2}x_<] \right),
\end{align}
where we used the spherical harmonics orthonormality.
With the definitions 
\be\label{def_F}
F(\alpha) = 1 + \frac{   h_l^{\prime(2)} (\alpha) }{h_l^{\prime(1)}(\alpha) },
\ee
\be\label{alpha}
\alpha = (\pm i)^{1/2} \sqrt{ |\bar{\omega}| } \equiv  (\pm i)^{1/2} \sqrt{ |\tilde{\omega}|/\omega_D },
\ee
and changing variables, $y = (\pm i)^{1/2} x$,  $y' = (\pm i)^{1/2} x'$ we rewrite \eqref{target_1},
\be\label{beforelast}
\Gamma(\alpha) = - 2\frac{1}{d D} \mathrm{Im} \left\{    
\frac{ \alpha}{2} F(\alpha) \left[ \int_{\alpha}^{\infty} \frac{ d y}{y}  h_l^{(1)}[y] \right]^2
- 
\alpha \int_{\alpha}^{\infty} \frac{ d y'}{y'}  h_l^{(1)}[y']  \int_{\alpha}^{y'}   \frac{ d y}{y}   \left(   h_l^{(1)}[y]  + h_l^{(2)}[y] \right)
\right\}.
\ee
Using the result of the exact integration for $l=2$, we find
\be\label{res1}
\alpha \int_{\alpha}^{\infty} \frac{ d y'}{y'}  h_l^{(1)}[y']  \int_{\alpha}^{y'}   \frac{ d y}{y}   \left(   h_l^{(1)}[y]  + h_l^{(2)}[y] \right) = \frac{(3+2 i \alpha ) \alpha ^2+3 e^{2 i \alpha } (\alpha +i)^2+3}{6 \alpha ^5},
\ee
\be\label{res2}
(\alpha/2)( \int_{\alpha}^{\infty} \frac{ d y}{y}  h_l^{(1)}[y] )^2= \frac{e^{2 i \alpha } (\alpha +i)^2}{2 \alpha ^5}.
\ee
Substituting
 \eqref{res1} and \eqref{res2} into \eqref{beforelast} we obtain the result in the closed form, recalling that $l=2$,
\be\label{exact}
\Gamma(\omega) = - 2\frac{1}{d D}\mathrm{Im} \left\{  
	\left( 1 + \frac{   h_2^{\prime(2)} \left[\alpha\right] }{h_2^{\prime(1)}\left[\alpha\right] } \right)  \frac{e^{2 i \alpha } (\alpha +i)^2}{2 \alpha ^5} 
	+
	\frac{(3+2 i \alpha ) \alpha ^2+3 e^{2 i \alpha } (\alpha +i)^2+3}{6 \alpha ^5}
	\right\}.
\ee
Given the definitions \eqref{tilde_omega}, \eqref{bar_o} and \eqref{alpha},
the expression \eqref{exact} simplifies to 
\begin{align}\label{exact_A}
	\Gamma(\omega) &= - \frac{2}{3 d D}\mathrm{Im} \left\{ 
	\frac{1}{ |\bar{\omega}| +\frac{27}{i^{1/2} |\bar{\omega}|^{1/2}+4 i}+9 i }
	\right\}=
	- \frac{2}{3 d D}\mathrm{Im} \left\{ 
	\frac{i^{1/2} |\bar{\omega}|^{1/2}+4 i}{ (|\bar{\omega}|  +9 i)(i^{1/2} |\bar{\omega}|^{1/2}+4 i)   +27} 
	\right\}
	\notag \\
	= & - \frac{2}{3 d D}\mathrm{Re} \left\{ 
	\frac{i^{-1/2} |\bar{\omega}|^{1/2}+4 }{ 
		|\bar{\omega}|^{3/2} i^{1/2} + 9 i^{3/2} |\bar{\omega}|^{1/2} + 4 i |\bar{\omega}| - 9} 
	\right\}=
	\frac{2}{3 d D} \frac{4}{9}\mathrm{Re} \left\{ 
	\frac{i^{-1/2} |\bar{\omega}|^{1/2}/4 +1 }{ 
		-|\bar{\omega}|^{3/2} i^{1/2}/9 -  i^{3/2} |\bar{\omega}|^{1/2} - 4 i |\bar{\omega}| /9+ 1} 
	\right\}
\end{align}.
We can replace the complex number in the curly brackets in the last equation \eqref{exact_A} by its complex conjugate to write it as
\begin{align}\label{exact_B}
	\Gamma(\omega) &=
	\frac{8}{27 d D} \mathrm{Re} \left\{ 
	\frac{i^{1/2} |\bar{\omega}|^{1/2}/4 +1 }{ 
		-|\bar{\omega}|^{3/2} i^{-1/2}/9 -  i^{-3/2} |\bar{\omega}|^{1/2} + 4 i |\bar{\omega}| /9+ 1} 
	\right\}
	=
	\frac{8}{27 d D} \mathrm{Re} \left\{ 
	\frac{i^{1/2} |\bar{\omega}|^{1/2}/4 +1 }{ 
		|\bar{\omega}|^{3/2} i^{3/2}/9 +  i^{1/2} |\bar{\omega}|^{1/2} + 4 i |\bar{\omega}| /9+ 1} 
	\right\}
\end{align}
Our result, \eqref{exact} in the form \eqref{exact_B} coincides exactly with the equation $(A13)$ of \cite{Hwang1975}.
The result \eqref{exact} or equivalently \eqref{exact_B} is shown in Fig.~\ref{fig:res}.
\subsection{Asymptotic behavior in the limit of large and small frequencies}
At lower frequencies the Taylor expansion of \eqref{exact} yields
\be\label{as1}
\Gamma(\omega \ll \omega_D) \approx \frac{1}{d D} \left[ \frac{4}{27} - \frac{1}{ 9 \sqrt{2} } \sqrt{  \frac{|\omega - m \omega_v|}{\omega_D} } \right] ,
\ee
in accordance with the general expectation \eqref{stat5}.
For finite fluid rotation frequency $\omega_v \neq 0$, \eqref{as1} describes  the Doppler shift of the diffusion smeared power spectrum. 
The Doppler shift holds of course for all frequencies as well.
Note that at zero frequency we have a non-analytic square root behavior of the power spectrum,
\be\label{as1a}
\Gamma(\omega=0) \approx \mathrm{const} - \frac{1}{dD }\sqrt{\frac{\omega_v}{\omega_D}}.
\ee
At large frequencies  and at $\omega_v =0$, the result \eqref{exact} reduces to
\be
\Gamma(\omega \gg \omega_D)  \approx \frac{3}{ d D} \left( \frac{ \omega_D }{ \omega }\right)^2
\ee
as again expected from \eqref{dyn22}.

The question arises whether the singular suppression of the power spectrum as given by \eqref{as1a} is generic.
The answer is unfortunately negative.
In the next section we present a model with the angular frequency decaying to zero at large distances, for which the singular square root is eliminated by the combined action of the diffusion and frequency variations across the fluid volume, see Fig.~\ref{fig:zero}.

\subsection{Suppression of the power spectrum at $\omega=0$ caused by the flow generated by fluid rotation with the angular velocity $\Omega \propto r^{-2}$.}

\begin{figure}
	\includegraphics[width=0.6\textwidth]{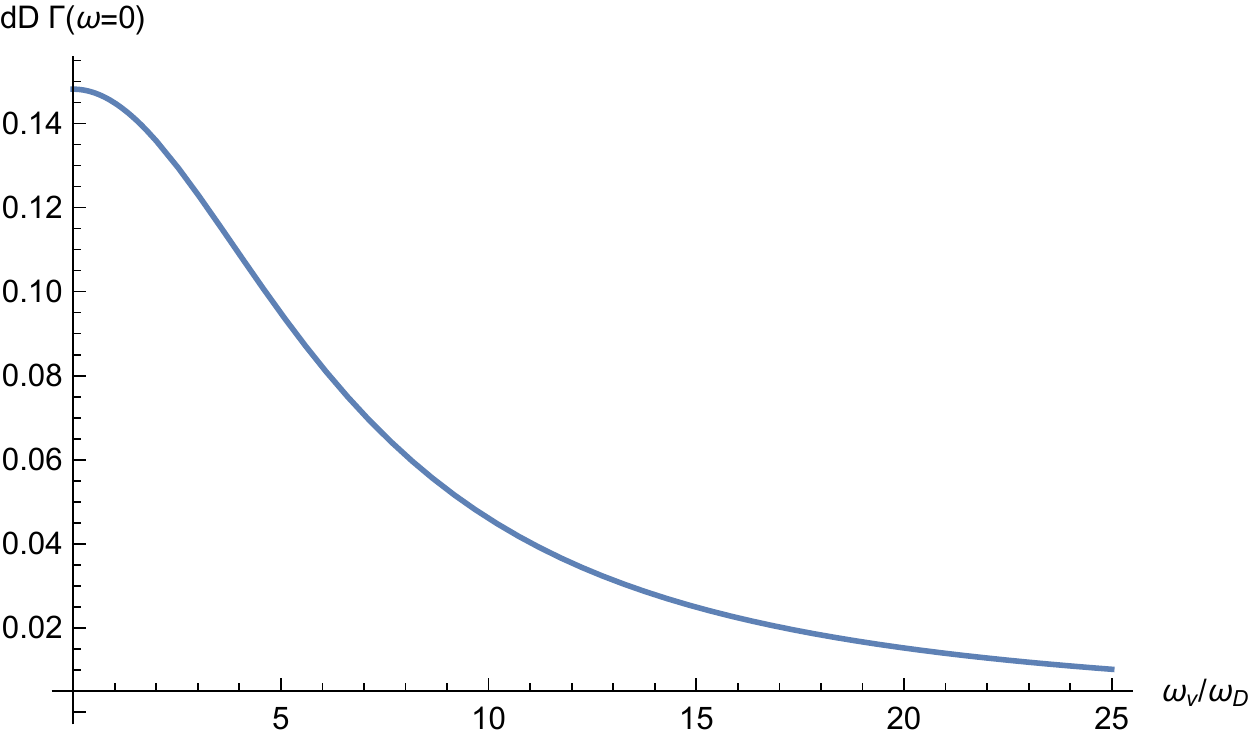} 
	\caption{\textbf{The exact power spectrum \eqref{Gamma_res} at $\omega =0$.} The spectrum is displayed in dimensionless units, for the model specified by the drift-diffusion equation \eqref{target_G0} with the angular velocity of the flow \eqref{Omega_r}.
		The spectrum is plotted as a function of the parameter $\omega_v/\omega_D$. 
		Note that this function is analytic in contrast to the singular square root decay of the power spectrum with $\omega_v$ in the model with a drift at a constant angular frequency, \eqref{as1a} as shown in Fig.~\ref{fig:res}.
		\label{fig:zero}}
\end{figure}

The model we previously presented allows for exact solution at finite drifts.
It is, however, artificial inasmuch the drift gives rise to only a single Doppler shift, rather than a spectrum of many such frequencies shifts.
Here we are try to overcome this limitation by looking at a different velocity profile in which the frequency decays with the distance,
\be\label{Omega_r}
\Omega(r) = \omega_v (d/r)^2, \omega_v = v_d/d
\ee
Repeating all the steps leading to \eqref{target_0}, eq. \eqref{form} is still valid, but instead of \eqref{target_0} we now get 
\be\label{target_G0}
\left[- i \omega  -D \frac{1}{r} \frac{\partial^2 }{\partial r^2 } r  + D\frac{l(l+1)}{ r^2 }  - i m  \omega_v (\frac{ d}{r} )^2 \right] g_{lm}(r,r') = \frac{1}{r^2} \delta(r - r') 
\ee
We will focus on $\omega=0$ as otherwise the solution is not given in terms of known special functions.
With $x = r/d$, and $x' = r'/d$, Eq. \eqref{target_G0} implies that the function $g_{lm}(r,r')$ in units of $1/(d D)$ satisfies 
\be\label{D_1}
\left[  - \frac{1}{x} \frac{\partial^2 }{\partial x^2 } x  + \frac{l(l+1)}{ x^2 }  - i \beta' \frac{ 1}{x^2} \right] g_{lm}(x,x') = \frac{1}{x^2} \delta(x - x') ,
\ee
where 
\be\label{beta'}
\beta'  = m  \omega_v/\omega_D.
\ee
We can rewrite \eqref{D_1},
\be\label{D_2}
\left[  - \frac{1}{x} \frac{\partial^2 }{\partial x^2 } x  + \frac{\lambda(\lambda+1)}{ x^2 }  \right] g_{lm}(x,x') = \frac{1}{x^2} \delta(x - x').
\ee
To determine $\lambda$ we may introduce the notation,
$(l+1/2)^2 \beta = \beta'$, which for $l=2$ gives $\beta' = \beta (25/4)$
\be\label{D_4}
l(l+1) = (l + 1/2)^2 - 1/4, \qquad \lambda(\lambda+1) = (\lambda + 1/2)^2 - 1/4.
\ee
We have the relation,
\be\label{D_6}
(l + 1/2)^2 - 1/4 + (l + 1/2)^2 i \beta = (\lambda + 1/2)^2 - 1/4
\ee
\be\label{D_8}
(\lambda + 1/2)^2 = (l + 1/2)^2 ( 1 + i \beta )\, , \qquad
\lambda = - 1/2 + (l+1/2) \sqrt{1 + i \beta } ,
\ee
where we assume $\beta >0$ for definiteness, and that the square root has a positive imaginary part.
The solution of \eqref{D_2} for $x\neq x'$ then reads,
\be\label{D_10}
g_{lm}(x,x') = x_>^{\lambda_-} ( A x_<^{\lambda_-} + B x_<^{\lambda_+} )\, , \qquad
\lambda_{\pm} = - 1/2 \pm (l+1/2) \sqrt{1 + i \beta } \, .
\ee
The boundary condition at the hard surface, $x=1$, is
\be\label{D_12}
0 = \frac{ \partial g_{lm}(x,x') }{ \partial x } \bigg\vert_{x=1} =
A \lambda_- + B\lambda_+ =0.
\ee
Another boundary condition is 
\be\label{D_14}
\frac{\partial }{\partial x } \left( x g_{lm}(x,x') \right)\bigg\vert_{x=x' + \epsilon}  - \frac{\partial }{\partial x } \left( x g_{lm}(x,x') \right)\bigg\vert_{x=x' - \epsilon}  = - \frac{1}{x'},
\ee
which gives
\be\label{D_16}
\frac{\partial }{\partial x } x^{\lambda_- + 1} \bigg\vert_{x=x' + \epsilon}  ( A x^{'\lambda_-} + B x^{'\lambda_+} ) -
x^{'\lambda_- }\frac{\partial }{\partial x } ( A x^{\lambda_- + 1 } + B x^{'\lambda_+ + 1} )\bigg\vert_{x=x' - \epsilon} = - \frac{1}{x'}.
\ee
The previous equation can be solved for $B$,
\be\label{D_18}
B (\lambda_- + 1) x^{'(\lambda_+ + \lambda_-)} - B  (\lambda_+ + 1) x^{'(\lambda_+ + \lambda_-)} = - \frac{1}{x'},
\ee
which finally gives,
\be\label{D_20}
B = \frac{ 1 }{ \lambda_+ - \lambda_-} = \frac{ 1 }{( 2 l + 1) \sqrt{ 1 + i \beta } }
\ee
The coefficient $A$ can then be calculated from \eqref{D_12} by substituting \eqref{D_20},
\be\label{D_23}
A = - B \frac{ \lambda_+}{\lambda_-} = -  \frac{ 1 }{( 2 l + 1) \sqrt{ 1 + i \beta } } \frac{  - 1/2 + (l+1/2) \sqrt{1 + i \beta } }{ - 1/2 - (l+1/2) \sqrt{1 + i \beta } }
=
-  \frac{ 1 }{( 2 l + 1) \sqrt{ 1 + i \beta } } \frac{  1 -  (2l+1) \sqrt{1 + i \beta } }{ 1 + (2l+1) \sqrt{1 + i \beta } }.
\ee
The solution \eqref{D_10} can now be explicitly written using \eqref{D_20} and \eqref{D_23},
\be\label{D_27}
g_{lm}(x,x') = \frac{ 1 }{( 2 l + 1) \sqrt{ 1 + i \beta } } 
\left( 
- \frac{\lambda_+}{\lambda_-} x_>^{\lambda_-} x_<^{\lambda_-} + 
x_<^{\lambda_+} x_>^{\lambda_-}  \right).
\ee
The power spectrum can now be written as
\begin{align}\label{D_31}
	\Gamma(\omega=0) = \int_{1}^{\infty} \frac{ dx }{ x } \int_{1}^{\infty} \frac{ dx' }{ x' } g_{lm}(x,x')=
	\frac{ 1 }{( 2 l + 1) \sqrt{ 1 + i \beta } } 
	\int_{1}^{\infty} \frac{ dx }{ x } \int_{1}^{\infty} \frac{ dx' }{ x' }   \left( 
	- \frac{\lambda_+}{\lambda_-} x_>^{\lambda_-} x_<^{\lambda_-} + 
	x_<^{\lambda_+} x_>^{\lambda_-}  \right).
\end{align}
The integrals are easily evaluated,
\begin{align}\label{D_33}
	\int_{1}^{\infty} \frac{ dx }{ x } \int_{1}^{\infty} \frac{ dx' }{ x' }   
	x_>^{\lambda_-} x_<^{\lambda_-}  & = \frac{1 }{\lambda_-^2 },
	\notag \\
	\int_{1}^{\infty} \frac{ dx }{ x } \int_{1}^{\infty} \frac{ dx' }{ x' }   
	x_>^{\lambda_-} x_<^{\lambda_+}  & = \frac{2 }{\lambda_- (\lambda_+ + \lambda_-) }.
\end{align}
Substituting \eqref{D_33} into \eqref{D_31} we obtain 
\begin{align}\label{Gamma_res}
\Gamma(\omega =0) &= \mathrm{Re} \left\{ \frac{ 1 }{\lambda_+ - \lambda_- } 
\left( - \frac{\lambda_+}{\lambda_-} \frac{1 }{\lambda_-^2 }  + \frac{2 }{\lambda_- (\lambda_+ + \lambda_-) } \right) \right\}\\\nonumber
&=16\text{Re}\left\{\frac{\left(2 l+1\right) \sqrt{4+\frac{25 i m \omega_v}{\omega_D}}+6}{\left[\left(2 l +1\right)\sqrt{4+\frac{25 i m \omega_v}{\omega_D}}+2\right]^3}\right\}=16\textrm{Re}\left\{\frac{6+5\sqrt{4+\frac{25 i m \omega_v}{\omega_D}}}{\left(2+5 \sqrt{4+\frac{25 i m \omega_v}{\omega_D}}\right)^3}\right\},
\end{align}
where in the last equality we substituted $l=2$.
This result is plotted in Fig.~\ref{fig:zero} as a function of the ratio $\omega_v/\omega_D$ with $m=1$.

\section{SPECIAL NOTATIONS}
\label{Notations}
In the next few sections we calculate integrals using series expansions of special functions. Here we clarify all the notations we use
\begin{enumerate} 
	\item  $Y_l^{(m)}(\Omega)$ - The spherical harmonics, $\Omega$ is the solid angle.
	\item  $Y_l^{(m)}(\theta)$ - The spherical harmonics without their polar angle dependence.
	\item $P_n(x)$ - The n'th Legendre polynomial.
	\item $J_n(x)$ - The n'th Bessel function of the first kind.
	\item $K_n(x)$ - The n'th modified Bessel function of the second kind
	\item $j_n(x)$ - The n'th spherical Bessel function of the first kind. 
	\item $\textrm{ber}_n\left(x\right)/\textrm{bei}_n\left(x\right)$ - The Kelvin functions - the real/imaginary parts of $J_n\left(x e^{\frac{3\pi i}{4}}\right)$.
	\item $\textrm{ker}_n\left(x\right)/\textrm{kei}_n\left(x\right)$ - The Kelvin functions - the real/imaginary parts of $K_n\left(x e^{\frac{3\pi i}{4}}\right)$.
	\item $h^{(1)}_n(x)$ - The n'th spherical Hankel function of the first kind.
	\item $Si(x))$ - The sine integral - $\int\limits_{0}^{x}\frac{\sin(t)}{t}dt$
	\item $Ci(x)$ - The cosine integral - $\int\limits_{0}^{x}\frac{\cos(t)}{t}dt$
	\item $Ei(x)$ - The exponential integral - $\int\limits_{-x}^{\infty}\frac{e^{-t}}{t}dt$
	\item $Erf(x)$ - The Gaussian integral - $\frac{2}{\sqrt{(\pi)}}\int\limits_0^x dt e^{-t^2}$
	\item $Erfc(x)$ - The complementary Gaussian integral - $1-Erf(x)$ 
	\item $\Gamma(x)$ - The Gamma function $\int\limits_0^{\infty}t^{x-1}e^{-t}dt$
	\item  $\Gamma(x,z)$ -The (upper) incomplete Gamma function - $\int\limits_z^{\infty}t^{x-1}e^{-t}dt$
	\item $\,_{p}F_{q}\left(\left\{a_1,\cdots,a_p\right\};\left\{b_1,\cdots,b_q\right\};z\right)$ -  The generalized hypergeometric function.
	\item $\gamma$ - The Euler constant.
	\item $D_{m_1,m_2}^l\left[\mathcal{R}\right]$ - The Wigner matrices.
\end{enumerate}

\section{TIME CORRELATION CALCULATION - PLANAR GEOMETRY WITHOUT DRIFT}
\label{timeplanar}
In this section we derive a closed expression for the temporal auto-correlation function of the magnetic field, for a planar boundary condition, meaning, we assume that the NV is situated in a depth $d$ within a planar diamond surface, and that the liquid fills the half space above the diamond surface.
\subsection{  Calculating the autocorrelation function in the time domain}
\label{Time_correlation_no_drift_main}
{\it Solution to the diffusion equation with a planar boundary condition ---} As in the previous sections, we are interested in correlation functions of the type \eqref{cont5},
\beq\label{corrform}
G^{\left(m_1,m_2\right)}\left(t\right)=\int\int f^{\left(m_1\right)*}\left(\bar{r}\right)f^{\left(m_2\right)}\left(\bar{r}_{0}\right)P\left(\bar{r},t|\bar{r}_{0},0\right)P\left(\bar{r}_{0}\right)d^{3}rd^{3}r_{0},
\eeq
where $f^{(m)}(\bar{r})=\frac{1}{r^3}Y_2^{(m)}\left(\Omega \right)$ are the spatial dependencies of the magnetic field, $P(\bar{r}_0)$ is the initial spatial distribution of the nuclear spin ensemble, assumed throughout this work to be uniform, and $P\left(\bar{r},t|\bar{r}_{0},0\right)$ is the propagator from the point $\bar{r_0}$ to $\bar{r}$ given a time difference $t$. The correlation functions $G^{(m_1,m_2)}$ are equal to the Fourier transform of the power spectrum up to a multiplicative of $\left(\frac{\gamma_e\gamma_N \hbar}{4\pi}\right)^2$, therefore sometimes we use these terms interchangeably.   
The first step in calculating these correlation functions is to find the appropriate propagator. Assuming, the ensemble is freely diffusing in the half space above the diamond surface, the propagator $P$ will be the solution to the diffusion equation in the half space with the initial condition $P\left(\bar{r},t=0|\bar{r_0},0\right)=\delta\left(\bar{r}-\bar{r}_{0}\right)$.
We recall that the diffusion equation with the diffusion coefficient $D$ is
\beq
\frac{\partial P}{\partial t}=D\nabla^2 P.
\eeq
The solution of the equation in the whole space,
\beq\label{Diffusionprop}
P\left(\bar{r},t|\bar{r}_{0},0\right)=\left(4\pi Dt\right)^{-3/2}\exp\left[-\frac{\left(\bar{r}-\bar{r}_{0}\right)^{2}}{4Dt}\right].
\eeq
We define a coordinate system in which the NV center is found at the origin and the diamond surface coincides with the $z=d$ plane. 
The appropriate boundary condition is the vanishing of the flux through this plane, meaning: $\left.\partial_{z}P\right|_{z=d}=0$. We can enforce it by using a sum of two solutions to the diffusion equation (method of images),
\begin{align}
	P\left(\bar{r},t|\bar{r}_{0},0\right)=\left(4\pi Dt\right)^{-3/2}e^{-\frac{\left(x-x_{0}\right)^{2}}{4Dt}}e^{-\frac{\left(y-y_{0}\right)^{2}}{4Dt}}\left(e^{-\frac{\left(z-z_{0}\right)^{2}}{4Dt}}+e^{-\frac{\left(z+z_{0}-2d\right)^{2}}{4Dt}}\right).
\end{align}
We can write the total correlation function of the magnetic field as a sum of the real spins contribution and that of the images,
\beq\label{corr0}
G^{\left(m\right)}\left(t\right)=G_{original}^{\left(m\right)}\left(t\right)+G_{reflected}^{\left(m\right)}\left(t\right),
\eeq
\begin{align}
	\label{nodrift0}
	G_{original}^{\left(m\right)}\left(t\right)&=n\left(4\pi Dt\right)^{-3/2}\alpha^{\left(m\right)}\int\int Y_{2}^{\left(m\right)*}\left( \Omega \right)Y_{2}^{\left(m\right)}\left( \Omega_0 \right)exp\left[-\frac{\left(\bar{r}-\bar{r}_{0}\right)^{2}}{4Dt}\right]
	\frac{d^{3} r}{r^3} \frac{d^{3}r_{0}}{r_0^3}, \\\label{nodrift1}
	G_{reflected}^{\left(m\right)}\left(t\right)&=n\left(4\pi Dt\right)^{-3/2}\alpha^{\left(m\right)}\int\int Y_{2}^{\left(m\right)*}\left( \Omega \right)Y_{2}^{\left(m\right)}\left( \Omega_0 \right)exp\left[-\frac{\left(\bar{r}+\bar{r}_{0}-2d\hat{z}\right)^{2}}{4Dt}\right]\frac{d^{3} r}{r^3} \frac{d^{3}r_{0}}{r_0^3} ,
\end{align}
where we substituted $f^{(m)}\left(\bar{r}\right)$ with the dipolar fields spatial dependence. 
We denoted by $\alpha^{\left(m\right)}$ the squared dimensionless coefficient required to transform from the spherical harmonics representation of the dipolar interaction to the common angular notation, and by $n$ the nuclear spin density.
For simplicity, we assumed that the NV's magnetization axis is perpendicular to the diamond surface; namely, the $\hat{z}$ direction; Therefore, the only correlation functions that need to be considered are between $f^{(m_1)}$ and $f^{(m_2)}$ with $m_1=m_2$, since correlation functions with $m_1\neq m_2$ are suppressed because of the resonance requirement discussed in Sec. \ref{Measure}.
It should be noted, that for diffusing particles this will also hold due to the polar symmetry alone. We shall relax the assumption about the NV's magnetization axis later on.

{\it Solution in cylindrical coordinates ---} 
In the following we calculate the correlation functions $G^{(m)}(t)$ analytically and without approximations. The main idea behind the derivation is to use the polar symmetry of the problem, therefore the important tools used are the $2D$ Fourier transform and the Anger-Jacobi expansion \cite{Jacobi–Anger_expansion},
\beq\label{A-JE}
e^{i\bar{k}\cdot\bar{r}}=e^{ikr\cos\theta}=\sum_{m=-\infty}^{\infty}i^m e^{im\theta}J_{m}\left(kr\right),
\eeq
where, $J_m\left(z\right)$ is the $m$'th Bessel function of the first kind.
The main result of this section is
\begin{align}
\label{correlation_no_drift}
G^{\left(0\right)}=2n\pi^{3/2}&\left[\frac{1}{\left(Dt\right)^{3/2}}-\frac{3}{2d^{2}\sqrt{Dt}}+\frac{3\sqrt{Dt}}{d^{4}}-\frac{3Dt\sqrt{\pi}}{2d^{5}}\right.\\\nonumber & \left.+\sqrt{\pi}Erfc\left(\frac{d}{\sqrt{Dt}}\right)e^{\frac{d^{2}}{Dt}}\left(-\frac{d}{\left(Dt\right)^{2}}+\frac{1}{d\left(Dt\right)}-\frac{7}{4d^{3}}+\frac{3Dt}{2d^{5}} \right)+\frac{\sqrt{\pi}}{4d^{3}}\right].
\end{align}
with
\beq\label{same_m}
G^{(0)}=\frac{1}{9}G^{(1)}=\frac{1}{9}G^{(2)}.
\eeq
Evidently, the correlation functions do not decay exponentially as usually assumed, but in a more complex fashion. The characteristic time $\tau_D=\frac{d^2}{D}=\left(\omega_D\right)^{-1}$, dictates a change in the behavior of the function, rather then just the amount of decay.

We now continue with the full derivation.
Let us focus on the first part of the correlation function, \eqref{nodrift0},
\begin{align}\label{corr1}
	G_{original}^{\left(m\right)}\left(t\right)=n\left(4\pi Dt\right)^{-3/2}\alpha^{\left(m\right)}\int\int Y_{2}^{(m)*}\left(\Omega\right)Y_{2}^{(m)}\left(\Omega_{0}\right)exp\left[-\frac{\left(\bar{r}-\bar{r}{}_{0}\right)^{2}}{4Dt}\right]r_{0}^{-3}r^{-3}d^{3}rd^{3}r_{0}\\\label{corr2}
	=n\left(4\pi Dt\right)^{-3/2}\alpha^{\left(m\right)}\int\int Y_{2}^{(m)*}\left(\Omega\right)Y_{2}^{(m)}\left(\Omega_{0}\right)exp\left[-\frac{\left(\bar{\rho}-\bar{\rho_{0}}\right)^{2}}{4Dt}\right]exp\left[-\frac{\left(z-z_{0}\right)^{2}}{4Dt}\right]r_{0}^{-3}r^{-3}d^{3}rd^{3}r_{0}.
\end{align}

We solve this by using the Fourier transform with respect to $\left(\bar{\rho}-\bar{\rho_{0}}\right)$ and then the Anger-Jacobi expansion \eqref{A-JE},
\begin{align}\label{timecorr1}
	G_{original}^{\left(m\right)}=\frac{n\left(4\pi Dt\right)^{-1/2}}{2\pi}\alpha^{\left(m\right)}\sum_{n_{1},n_{2}}\int\int\int &Y_{2}^{(m)*}\left(\Omega\right)Y_{2}^{(m)}\left(\Omega_{0}\right)e^{-Dtk^{2}}\left(-1\right)^{n_{2}}\left(i\right)^{n_{1}+n_{2}}J_{n_{1}}\left[k\rho\right]J_{n_{2}}\left[k\rho_{0}\right]\times\\
	\nonumber &e^{i\left(n_{1}\theta'-n_{2}\theta''\right)}e^{-\frac{\left(z-z_{0}\right)^{2}}{4Dt}}r_{0}^{-3}r^{-3}d^{2}kd^{3}rd^{3}r_{0},
\end{align}
where $\theta'/\theta''$ are the angles between $\bar{k}$ and $\bar{\rho}/\bar{\rho}_{0}$. Since we are in the plane we can write
\beq\label{angeldef1}
\theta'=\varphi'-\varphi,\ \theta''=\varphi'-\varphi_{0},
\eeq
where $\varphi'$ is the free angle of the $\bar{k}$ vector. Substituting \eqref{angeldef1} back into \eqref{timecorr1} and integrating over the polar angles results in
\begin{align}
G_{original}^{\left(m\right)}=n\left(4\pi Dt\right)^{-1/2}\frac{1}{2\pi}\alpha^{\left(m\right)}\sum_{n_{1},n_{2}}\int\int\int& Y_{2}^{(m)*}\left(\Omega\right)Y_{2}^{(m)}\left(\Omega_{0}\right)e^{-Dtk^{2}}\left(-i\right)^{n_{2}}\left(i\right)^{n_{1}}\times\\\nonumber
&J_{n_{1}}\left[k\rho\right]J_{n_{2}}\left[k\rho_{0}\right]e^{i\left(n_{1}\left(\varphi'-\varphi\right)-n_{2}\left(\varphi'-\varphi_{0}\right)\right)}e^{-\frac{\left(z-z_{0}\right)^{2}}{4Dt}}r_{0}^{-3}r^{-3}d^{2}kd^{3}rd^{3}r_{0}
\end{align}
\begin{align}
=n\left(4\pi Dt\right)^{-1/2}\frac{1}{2\pi}\alpha^{\left(m\right)}&\sum_{n_{1},n_{2}}\int\int\int Y_{2}^{(m)*}\left(\theta\right)Y_{2}^{(m)}\left(\theta_{0}\right)e^{-Dtk^{2}}\left(-i\right)^{n_{2}}\left(i\right)^{n_{1}}\times\\\nonumber
&J_{n_{1}}\left[k\rho\right]J_{n_{2}}\left[k\rho_{0}\right]e^{i\left(n_{2}+m\right)\varphi_{0}}e^{-i\left(m+n_{1}\right)\varphi}e^{i\left(n_{1}-n_{2}\right)\varphi'}e^{-\frac{\left(z-z_{0}\right)^{2}}{4Dt}}r_{0}^{-3}r^{-3}d^{2}kd^{3}rd^{3}r_{0}
\end{align}
\begin{align}
=&2n\sqrt{\frac{\pi^{3}}{Dt}}\alpha^{\left(m\right)}\int\int\int Y_{2}^{m*}\left(\theta\right)Y_{2}^{m}\left(\theta_{0}\right)e^{-Dtk^{2}}J_{-m}\left[k\rho\right]J_{-m}\left[k\rho_{0}\right]e^{-\frac{\left(z-z_{0}\right)^{2}}{4Dt}}r_{0}^{-3}r^{-3}kdkd^{2}rd^{2}r_{0}\\\label{timecorr2}
=&2n\sqrt{\frac{\pi^{3}}{Dt}}\alpha^{\left(m\right)}\int\int\int Y_{2}^{m*}\left(\theta\right)Y_{2}^{m}\left(\theta_{0}\right)e^{-Dtk^{2}}J_{m}\left[k\rho\right]J_{m}\left[k\rho_{0}\right]e^{-\frac{\left(z-z_{0}\right)^{2}}{4Dt}}r_{0}^{-3}r^{-3}kdkd^{2}rd^{2}r_{0}.
\end{align}
We shall now pursue the integration over $\rho,\rho_{0}$ for different values of $m$. Starting with $m=0$,
\begin{align}\label{timecorr3}
G_{original}^{\left(0\right)}=2n\sqrt{\frac{\pi^{3}}{Dt}}\int\int\int\left(\frac{3z^{2}}{r^{2}}-1\right)\left(\frac{3z_{0}^{2}}{r_{0}^{2}}-1\right)e^{-Dtk^{2}}J_{0}\left[k\rho\right]J_{0}\left[k\rho_{0}\right]e^{-\frac{\left(z-z_{0}\right)^{2}}{4Dt}}r_{0}^{-3}r^{-3}kdkd^{2}rd^{2}r_{0}.
\end{align}
The integral over $\rho$ is
\begin{align}
\int\limits _{0}^{\infty}\left(\frac{3z^{2}\rho}{r^{5}}-\frac{\rho}{r^{3}}\right)J_{0}\left[k\rho\right]d\rho=&\int\limits _{0}^{\infty}\left(\frac{3z^{2}\rho}{\left(\rho^{2}+z^{2}\right)^{5/2}}-\frac{\rho}{\left(\rho^{2}+z^{2}\right)^{3/2}}\right)J_{0}\left[k\rho\right]d\rho\\\nonumber
&=3z^{2}\frac{e^{-k\left|z\right|}\left(1+k\left|z\right|\right)}{3\left|z\right|^{3}}-\frac{e^{-k\left|z\right|}}{\left|z\right|}=ke^{-k\left|z\right|}.
\end{align}
Substituting this result back into \eqref{timecorr3}
\begin{align}\label{timecorrm0}
G_{original}^{\left(0\right)}=2n\sqrt{\frac{\pi^{3}}{Dt}}\int\int\int k^{3}e^{-k\left(\left|z\right|+\left|z_{0}\right|\right)}e^{-Dtk^{2}}e^{-\frac{\left(z-z_{0}\right)^{2}}{4Dt}}dkdzdz_{0}
\end{align}
Returning to \eqref{timecorr2} with $m=1$,
\begin{align}\label{timecorrm12}
G_{original}^{\left(1\right)}=2n\sqrt{\frac{\pi^{3}}{Dt}}\int\int\int\frac{z\rho^{2}}{r^{5}}\frac{z_{0}\rho_{0}^{2}}{r_{0}^{5}}e^{-Dtk^{2}}J_{1}\left[k\rho\right]J_{1}\left[k\rho_{0}\right]e{-\frac{\left(z-z_{0}\right)^{2}}{4Dt}}kdkd\rho dzd\rho_{0}dz_{0}.
\end{align}
The integral over $\rho$ will now be
\begin{align}
\int\limits _{0}^{\infty}\frac{\rho^{2}}{\left(\rho^{2}+z^{2}\right)^{5/2}}J_{1}\left[k\rho\right]d\rho=\frac{ke^{-k\left|z\right|}}{3\left|z\right|}.
\end{align}
Using this result, the integral \eqref{timecorrm12} reduces to
\begin{align}\label{timecorrm1}
G_{original}^{\left(1\right)}=\frac{2n}{9}\sqrt{\frac{\pi^{3}}{Dt}}\int\int\int k^{3}\frac{zz_{0}}{\left|z\right|\left|z_{0}\right|}e^{-k\left(\left|z\right|+\left|z_{0}\right|\right)}e^{-Dtk^{2}}e^{-\frac{\left(z-z_{0}\right)^{2}}{4Dt}}dkdzdz_{0}.
\end{align}
Returning to \eqref{timecorr2} again with $m=2$,
\begin{align}\label{timecorrm21}
G_{original}^{\left(2\right)}=2n\sqrt{\frac{\pi^{3}}{Dt}}\int\int\int\frac{\rho^{3}}{r^{5}}\frac{\rho_{0}^{3}}{r_{0}^{5}}e^{-Dtk^{2}}J_{2}\left[k\rho\right]J_{2}\left[k\rho_{0}\right]e^{-\frac{\left(z-z_{0}\right)^{2}}{4Dt}}kdkd\rho dzd\rho_{0}dz_{0}.
\end{align}
The integral over $\rho$ is
\begin{align}
\int\limits _{0}^{\infty}\frac{\rho^{3}}{\left(\rho^{2}+z^{2}\right)^{5/2}}J_{2}\left[k\rho\right]d\rho=\frac{1}{3}ke^{-k\left|z\right|}.
\end{align}
Substituting back in \eqref{timecorrm21} we arrive at
\begin{align}\label{timecorrm2}
G_{original}^{\left(2\right)}=\frac{2n}{9}\sqrt{\frac{\pi^{3}}{Dt}}\int\int\int k^{3}e^{-k\left(\left|z\right|+\left|z_{0}\right|\right)}e^{-Dtk^{2}}e^{-\frac{\left(z-z_{0}\right)^{2}}{4Dt}}dkdzdz_{0}.
\end{align}
Eqs. \eqref{timecorrm0}, \eqref{timecorrm1} and \eqref{timecorrm2} validate \eqref{same_m}. Therefore, we turn our focus to $m=0$.
We now integrate over $z_{0},z$, noting that the integration limits are $[d,\infty)$, since the nuclei are confined to the upper half space, 
\begin{align}
G_{original}^{\left(0\right)}&=2n\sqrt{\frac{\pi^{3}}{Dt}}\int\int\int k^{3}e^{-k\left(\left|z\right|+\left|z_{0}\right|\right)}e^{-Dtk^{2}}e^{-\frac{\left(z-z_{0}\right)^{2}}{4Dt}}dkdzdz_{0}\\
&=2n\sqrt{\frac{\pi^{3}}{Dt}}\int dkk^{3}e^{-Dtk^{2}}\int\limits _{d}^{\infty}dz_{0}\int\limits _{d}^{\infty}dze^{-k\left(z+z_{0}\right)}e^{-\frac{\left(z-z_{0}\right)^{2}}{4Dt}}\\
&=2n\pi^{2}\int dkk^{3}e^{-Dtk^{2}}\int\limits _{d}^{\infty}dz_{0}\left.e^{k(Dkt-2z_{0})}Erf\left[\frac{2Dkt+z-z_{0}}{2\sqrt{Dt}}\right]\right|_{z=d}^{\infty}\\\label{timecorr4}
&=2n\pi^{2}\int dkk^{3}\int\limits _{d}^{\infty}dze^{-2kz}Erfc\left[\frac{2Dkt-z+d}{2\sqrt{Dt}}\right].
\end{align}
The integral over z is
\begin{align}\label{timecorrz}
\int\limits _{d}^{\infty}e^{-2kz}Erfc\left(\frac{d+2kDt-z}{2\sqrt{Dt}}\right)&=\frac{1}{2k}\left.\left[e^{-2kd}Erf\left(\frac{-d+2Dkt+z}{2\sqrt{Dt}}\right)-e^{-2kz}Erfc\left(\frac{d+2Dkt-z}{2\sqrt{Dt}}\right)\right]\right|_{z=d}^{\infty}\\\nonumber
&=\frac{1}{k}e^{-2kd}Erfc\left(k\sqrt{Dt}\right).
\end{align}
Substituting \eqref{timecorrz} back into \eqref{timecorr4} we can integrate over $k$,
\begin{align}\label{timecorr5}
G_{original}^{\left(0\right)}&=2n\pi^{2}\int dkk^{2}e^{-2kd}Erfc\left(k\sqrt{Dt}\right)
\\\nonumber&=2n\pi^{2}\left[\frac{1}{2\sqrt{\pi}\left(Dt\right)^{3/2}}-e^{\frac{d^{2}}{Dt}}Erfc\left[\frac{d}{\sqrt{Dt}}\right]\left(\frac{d}{2\left(Dt\right)^{2}}-\frac{1}{4dDt}+\frac{1}{4d^{3}}\right)-\frac{1}{2\sqrt{\pi}d^{2}\sqrt{Dt}}+\frac{1}{4d^{3}}\right].
\end{align}
We now turn our attention to the other part of the correlation function, $G_{reflected}^{\left(0\right)}$, given by \eqref{corr2}. The integration follows the same steps up to \eqref{timecorrm0},
\beq\label{timecorr6}
G_{reflected}^{\left(0\right)}=2n\sqrt{\frac{\pi^{3}}{Dt}}\int\int\int k^{3}e^{-k\left(\left|z\right|+\left|z_{0}\right|\right)}exp\left[-Dtk^{2}\right]exp\left[-\frac{\left(z+z_{0}-2d\right)^{2}}{4Dt}\right]dkdzdz_{0}
.\eeq
The integration over $z,z_0$  yields
\begin{align}
&\int\limits _{d}^{\infty}dz_{0}\int\limits _{d}^{\infty}dze^{-k\left(z+z_{0}\right)}e^{-\frac{\left(z+z_{0}-2d\right)^{2}}{4Dt}}
=\sqrt{\pi Dt}e^{k^{2}Dt-2kd}\int\limits _{d}^{\infty}dz_{0}\left.Erf\left(\frac{-2d+2Dkt+z+z_{0}}{2\sqrt{Dt}}\right)\right|_{z=d}^{\infty}\\
&=\sqrt{\pi Dt}e^{k^{2}Dt-2kd}\int\limits _{d}^{\infty}dz_{0}Erfc\left(\frac{2Dkt-d+z_{0}}{2\sqrt{Dt}}\right)\\
&=\sqrt{\pi Dt}e^{k^{2}Dt-2kd}\left.\left[\left(-d+2Dkt+z\right)Erfc\left(\frac{-d+2Dkt+z}{2\sqrt{Dt}}\right)-2\sqrt{\frac{Dt}{\pi}}e^{-\frac{(-d+2Dkt+z)^{2}}{4Dt}}\right]\right|_{z=d}^{\infty}\\
&=\sqrt{\pi Dt}e^{k^{2}Dt-2kd}\left[-\left(2Dkt\right)Erfc\left(k\sqrt{Dt}\right)+2\sqrt{\frac{Dt}{\pi}}e^{-k^{2}Dt}\right]\\\label{timecorr7}
&=2\left[\left(Dt\right)e^{-2kd}-\sqrt{\pi}\left(Dt\right)^{3/2}ke^{k^{2}Dt-2kd}Erfc\left(k\sqrt{Dt}\right)\right].
\end{align}
Substituting \eqref{timecorr7} back into \eqref{timecorr6},
\begin{align}\label{timecorr8}
G_{reflected}^{\left(0\right)}=4n\sqrt{Dt}\pi^{3/2}\int dkk^{3}\left[e^{-2kd-Dtk^{2}}-\sqrt{\pi Dt}ke^{-2kd}Erfc\left(k\sqrt{Dt}\right)\right].
\end{align}
We solve \eqref{timecorr8} by integrating each term individually,
\begin{align}\label{timecorr9}
&\int\limits _{0}^{\infty}k^{3}e^{-2kd-Dtk^{2}}dk=\frac{d^{2}}{2\left(Dt\right)^{3}}+\frac{1}{2\left(Dt\right)^{2}}-\frac{d\sqrt{\pi}}{\sqrt{Dt}}e^{\frac{d^{2}}{Dt}}\left(\frac{d^{2}}{2\left(Dt\right)^{3}}+\frac{3}{4\left(Dt\right)^{2}}\right)Erfc\left(\frac{d}{\sqrt{Dt}}\right)\\\label{timecorr10}
&-\sqrt{\pi Dt}\int
\limits_0^{\infty} dkk^{4}e^{-2kd}Erfc\left(k\sqrt{Dt}\right)=-\frac{d^{2}}{2\left(Dt\right)^{3}}-\frac{1}{4\left(Dt\right)^{2}}-\frac{1}{2d^{2}\left(Dt\right)}+\frac{3}{2d^{4}}-\frac{3\sqrt{\pi Dt}}{4d^{5}}\\\nonumber
&+\sqrt{\pi}Erfc\left(\frac{d}{\sqrt{Dt}}\right)e^{\frac{d^{2}}{Dt}}\left(\frac{d^{3}}{2\left(Dt\right)^{7/2}}+\frac{d}{2\left(Dt\right)^{5/2}}+\frac{3}{8d\left(Dt\right)^{3/2}}-\frac{3}{4d^{3}\sqrt{Dt}}+\frac{3\sqrt{Dt}}{4d^{5}}\right).
\end{align}
Summing \eqref{timecorr9} and \eqref{timecorr10}, we arrive at
\begin{align}\label{timecorr11}
G_{reflected}^{\left(0\right)}=2n\pi^{3/2}&\left[\frac{1}{2\left(Dt\right)^{3/2}}-\frac{1}{d^{2}\sqrt{Dt}}+\frac{3\sqrt{Dt}}{d^{4}}-\frac{3Dt\sqrt{\pi}}{2d^{5}}\right.\\\nonumber
&\left.+\sqrt{\pi}Erfc\left(\frac{d}{\sqrt{Dt}}\right)e^{\frac{d^{2}}{Dt}}\left(-\frac{d}{2\left(Dt\right)^{2}}+\frac{3}{4d\left(Dt\right)}-\frac{3}{2d^{3}}+\frac{3Dt}{2d^{5}}\right)\right]
.\end{align}
Finally, summing the two parts of the correlation function \eqref{timecorr5} and \eqref{timecorr11} leads the final result \eqref{correlation_no_drift}.

{\it Asymptotic behavior ---} We shall now examine the asymptotic behavior of the correlation function at short and long times. The result \eqref{longtimelimit} presented in this section is used in Fig. 2c of the main text. 
For $t\ll\tau_D$ we can expand eq. \eqref{correlation_no_drift} as follows,
\beq
\label{no_drift_small_times}
G^{\left(0\right)}\approx\frac{n\pi^{2}}{2d^{3}}.
\eeq
This result is expected, since when taking the formal limit $t\rightarrow0$, the diffusion propagator approaches a delta function. Consequently, in this limit $\bar{r}\rightarrow\bar{r}_{0}$, and the integral \eqref{nodrift0},
\beq
G^{0}\sim\int\limits _{d}^{\infty}\frac{1}{r_{0}^{6}}r_{0}^{2}dr_{0}\sim\frac{1}{d^{3}}.
\eeq

For $t\gg\tau_D$ we can expand eq.  \eqref{correlation_no_drift} to
\beq\label{longtimelimit}
G^{(0)}\approx\frac{16n}{15}\frac{\pi^{3/2}}{\left(Dt\right)^{3/2}}.
\eeq
This also can be understood in terms of the integral \eqref{nodrift0}. In this limit, the unnormalized propagator is approximately 1, therefore,
\begin{align}
	G^{(0)}&\propto\left(Dt\right)^{-3/2}\int\limits _{0}^{d/r_{0}}\int\limits _{0}^{d/r}\int\limits _{d}^{r_{max}}\int\limits _{d}^{r_{max}}\left(1-3cos^{2}\theta\right)\left(1-3cos^{2}\theta_{0}\right)\frac{1}{r_{0}}\frac{1}{r}drdr_{0}d\left(cos\theta\right)d\left(cos\theta_{0}\right)\\\nonumber
	&=\left(Dt\right)^{-3/2}\left(\int\limits _{d}^{r_{max}}\left(\frac{d}{r_{0}}\right)^{3}\frac{1}{r_{0}}dr_{0}\right)^{2}\propto\left(Dt\right)^{-3/2}.
\end{align}
The short times limit can be interpreted  intuitively - in short times, the magnetic field squared goes as $r^{-6}$. Since the minimal distance from the NV is $d$, the main contribution comes from particles in about this distance from the NV. Their magnetic field decays as $d^{-6}$, and there are about $nd^3$ such particles in the effective interaction region, which is approximately a hemisphere of radius $\sim d$ on top of the diamond surface near the NV. This limit, up to a prefactor, is equal to $B_{RMS}^2 \tau_D$. For long times the dominant length scale is $\sqrt{\tau_D}$ so the correlation decays as $(D t)^{-3/2}$.

\begin{figure}
	\includegraphics[width=0.6\textwidth]{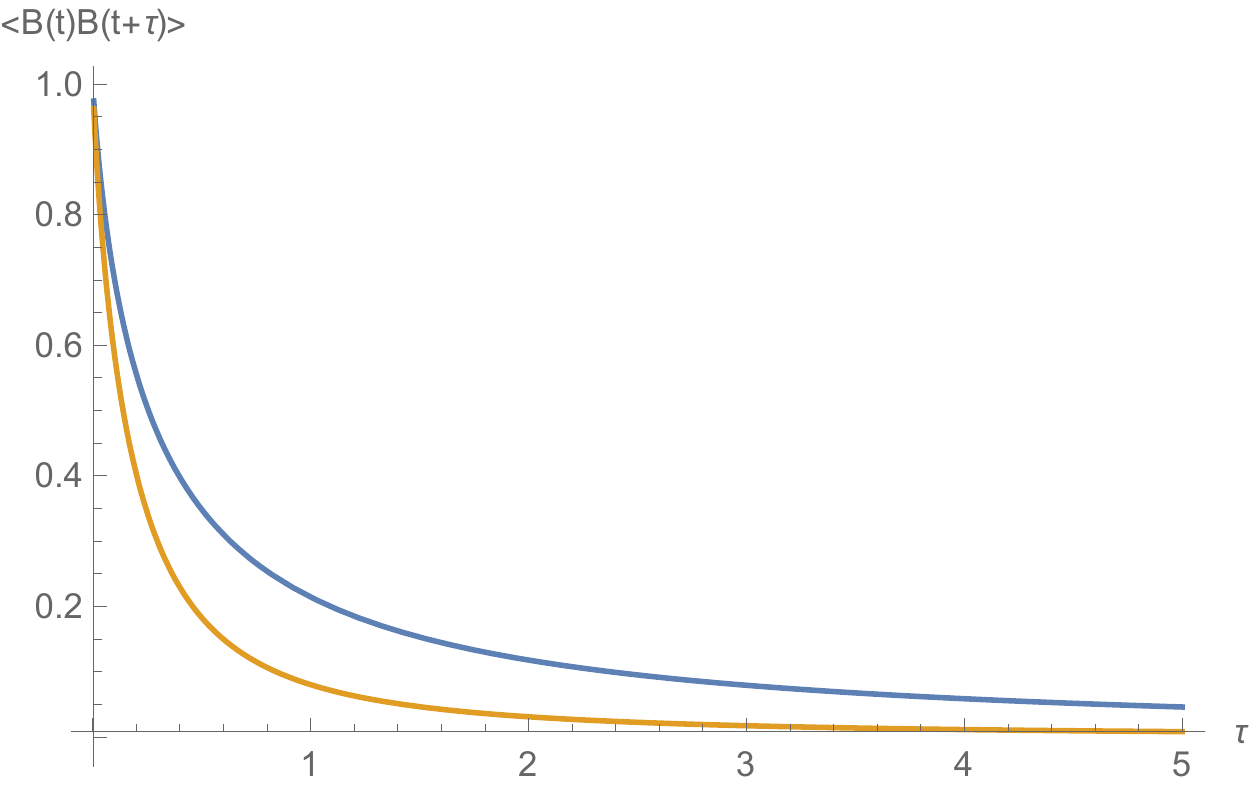} 
	\caption{The temporal correlation function for planar (blue) and spherical (orange) geometries. The correlation function for the planar geometry is plotted according to Eq. \eqref{correlation_no_drift} and the one of the spherical geometry by the numeric Fourier transform of \eqref{exact_B}. Both functions are normalized to their value at $t=0$ and are given for $d=1,\ D=1$.
		\label{Diffusion_different_geometries}}
\end{figure}

A comparison between the correlation function of the spherical geometry, given by the numeric Fourier transform of Eq. \eqref{exact_B} and the correlation function of the planar geometry \eqref{correlation_no_drift} can be found in Fig. \ref{Diffusion_different_geometries}. Though their asymptotic behavior is identical (up to a prefactor), it is evident that the decay of the correlation function is stronger for the spherical geometry.  

\subsection{The correlation function for 2D dynamics}
Some physical systems exhibit diffusion in two spatial dimensions. Therefore, the calculation of the correlation function in 2D is also important. Assuming that the nuclear spins are confined, and are able to diffuse only tangent to the diamond's surface, the calculation until \eqref{timecorrm0} is identical. The correlation should be emended to the following,
\beq
G^{(0)}_{original}=4 n \pi^{2}\int\limits_{0}^{\infty} dk k^3 e^{- 2 k d} e^{- D t k^2},
\eeq  
since $z_0=z=d$ at all times and the factor of $\left(4 \pi D t\right)^{-1/2}$ was discarded as the propagator normalization changed because of the lower dimension of the problem. We note that following the same arguments, we have from \eqref{timecorr6}, $G^{(0)}_{original}=G^{(0)}_{reflected}=\frac{1}{2} G^{(0)}$. These are simple Gaussian integrals, which result in the following,
\beq
G^{0}_{2D}=\frac{4 n \pi^{2}}{\left(D t\right)^2} \left(1+\frac{d^2}{D t}-\sqrt{\pi }e^{\frac{d^2}{D t}}Erfc\left(\frac{d}{\sqrt{D t}}\right)\left(\frac{d^3}{ \left(D t\right)^{3/2}}+\frac{3 d}{4 \left(D t\right)^{1/2}}\right)\right).
\eeq  
For short times we have,
\beq
G^{0}_{2D}\approx\frac{3 n \pi ^{2}}{d^4}-\frac{15 n \pi ^{2} D t}{d^6},
\eeq
while for long times,
\beq
G^{0}_{2D}\approx\frac{4  n \pi ^{2}}{\left(D t\right)^2}.
\eeq

\section{THE TIME CORRELATION OF DRIFTING AND DIFFUSING PARTICLES IN THE WHOLE SPACE}
\label{timevelocity}
In the following section we derive the correlation function of diffusing nuclei subjected to constant drift $\bar{v}=v\hat{x}$. We shall further assume that the nuclei occupy the whole space, while the NV center is found at the origin. Though, this geometry is unrealistic, the asymptotic behavior (scaling) for long times of the correlation function should hold for the planar geometry as well.   

\subsection{The drift-diffusion equation propagator}

The drift-diffusion equation with constant drift is
\beq
\frac{\partial P}{\partial t}=D\nabla^{2}P-\bar{v}\cdot\nabla P.
\eeq
The equation can be rewritten as follows,
\beq
\left[\frac{\partial }{\partial t}- D\nabla^{2}+\bar{v}\cdot\nabla\right]P = 0.
\eeq
In order to solve for the propagator, the equation should be emended to include the source term $\delta\left(\bar{r}-\bar{r_0}\right)\delta\left(t-t_0\right)$ as was shown in \eqref{cont4},
\beq\label{propfine1}
\left[\frac{\partial }{\partial t}- D\nabla^{2}+\bar{v}\cdot\nabla\right]P =\delta\left(\bar{r}-\bar{r_0}\right)\delta\left(t-t_0\right). 
\eeq
We define
\beq\label{propfine2}
\tilde{P}=\exp\left(-\bar{a}\cdot\bar{r}-bt\right)P\left(\bar{r},t|\bar{r}_0,t_0\right)\exp(\bar{a}\cdot\bar{r}_0+bt_0),
\eeq
where $\bar{a}, \ b$ are constants. Using \eqref{propfine2} we can rewrite \eqref{propfine1} as
\beq\label{propfine3}
\left(\frac{\partial}{\partial t}-D\nabla^{2}+\bar{v}\cdot\nabla\right)e^{\bar{a}\cdot\bar{r}+bt}\tilde{P}e^{-\bar{a}\cdot\bar{r}_{0}-bt_{0}}=\delta\left(\bar{r}-\bar{r}_{0}\right)\delta\left(t-t_{0}\right).
\eeq
Multiplying both sides of \eqref{propfine3} by $e^{\bar{a}\cdot\bar{r}_{0}+bt_{0}}$ to the left, and performing the derivatives leads to
\beq\label{propfine4}
e^{\bar{a}\cdot\bar{r}+bt}\left(\frac{\partial}{\partial t}+b-D\left(\nabla^{2}+2\bar{a}\cdot\nabla+a^{2}\right)+\bar{v}\cdot\nabla+\bar{v}\cdot\bar{a}\right)\tilde{P}=e^{\bar{a}\cdot\bar{r}_{0}+bt_{0}}\delta\left(\bar{r}-\bar{r}_{0}\right)\delta\left(t-t_{0}\right).
\eeq
Now multiplying both sides of \eqref{propfine4} by $e^{-\bar{a}\cdot\bar{r}-bt}$ to the left, and using the properties of the Dirac delta function, we arrive at
\beq\label{propfine5}
\left(\frac{\partial}{\partial t}+b-D\left(\nabla^{2}+2\bar{a}\cdot\nabla+a^{2}\right)+\bar{v}\cdot\nabla+\bar{v}\cdot\bar{a}\right)\tilde{P}=\delta\left(\bar{r}-\bar{r}_{0}\right)\delta\left(t-t_{0}\right).
\eeq  
In order to simplify \eqref{propfine5} we choose $\bar{a}=\frac{\bar{v}}{2D}$, which leads to
\beq\label{propfine6}
\left(\frac{\partial}{\partial t}+b-D\nabla^{2}+\frac{v^{2}}{4D}\right)\tilde{P}=\delta\left(\bar{r}-\bar{r}_{0}\right)\delta\left(t-t_{0}\right).
\eeq
We can further simplify \eqref{propfine6} by choosing $b=-\frac{v^2}{4D}$, 
\beq\label{propfine7}
\left(\frac{\partial}{\partial t}-D\nabla^{2}\right)\tilde{P}=\delta\left(\bar{r}-\bar{r}_{0}\right)\delta\left(t-t_{0}\right).
\eeq
This is the standard diffusion equation for $\tilde{P}$ with solution \eqref{Diffusionprop}
\beq\label{propfine8}
\tilde{P}=\frac{1}{\left(4\pi D\Delta t\right)^{3/2}}e^{-\frac{\left(\bar{r}-\bar{r}_{0}\right)^{2}}{4D\Delta t}},
\eeq
where $\Delta{t}=t-t_0$ and we assumed $t>t_0$.
Therefore, the propagator $P$, by \eqref{propfine2} and \eqref{propfine8} is
\beq\label{propfine9}
P\left(\bar{r},t|\bar{r}_0,t_0\right)=\frac{1}{\left(4\pi D\Delta t\right)^{3/2}}e^{\frac{\bar{v}}{2D}\cdot\left(\bar{r}-\bar{r}_{0}\right)}e^{-\frac{v^{2}}{4D}\Delta t}e^{-\frac{\left(\bar{r}-\bar{r}_{0}\right)^{2}}{4D\Delta t}}.
\eeq
The propagator can be rewritten in the more intuitive form
\beq\label{DandDprop}
P\left(\bar{r},t|\bar{r}_0,t_0\right)=\frac{1}{\left(4\pi D\Delta t\right)^{3/2}}e^{-\frac{\left(\bar{r}-\bar{r}_{0}-\bar{v}\Delta{t}\right)^{2}}{4D\Delta t}}.
\eeq

\subsection{Explicit calculation of the correlation function}

In the following we derive explicitly the correlation function $G^{(0)}$. We show that for the specified geometry and $\bar{v}=v\hat{x}$,

\beq\label{DandDtime}
G^{\left(0\right)}\left(t\right)=8\pi^{2}n\left\{ \left(\frac{1}{\left(vt\right)^{3}}-\frac{6Dt}{\left(vt\right)^{5}}\right)Erf\left(\frac{tv}{2\sqrt{Dt}}\right)+\frac{6}{\left(vt\right)^{4}}\sqrt{\frac{Dt}{\pi}}e^{-\frac{\left(vt\right)^{2}}{4\left(Dt\right)}}\right\}.
\eeq  
Note, that this correlation function divergence at the limit of $t \rightarrow 0$. This is expected, since the $B_{RMS}$ should diverge as the distance between the nuclei and the NV could be arbitrarily small.

We start by substituting the propagator \eqref{DandDprop} into \eqref{corrform} together with the dipolar field dependency and taking $t_0=0$,
\beq\label{DandD1}
G^{\left(m\right)}\left(t\right)=\frac{n}{\left(4\pi Dt\right)^{3/2}}\alpha^{\left(m\right)}\int\int d^{3}rd^{3}r_{0}\cdot r^{-3}r_{0}^{-3}\cdot Y_{2}^{\left(m\right)*}\left(\Omega\right)Y_{2}^{\left(m\right)}\left(\Omega_{0}\right)\exp\left(-\frac{\left(x-x_{0}-vt\right)^{2}}{4Dt}\right)\exp\left(-\frac{\left(y-y_{0}\right)^{2}}{4Dt}\right)\exp\left(-\frac{\left(z-z_{0}\right)^{2}}{4Dt}\right).
\eeq
We define the cylindrical coordinate system, $\rho=y^{2}+z^{2},\ y=\rho cos\phi,\ z=\rho sin\phi$, and rewrite eq. \eqref{DandD1} as
\begin{align}
\label{DandD2}
G^{\left(0\right)}\left(t\right)=\frac{n}{\left(4\pi Dt\right)^{3/2}}&\int\int d^{3}rd^{3}r_{0}\cdot\left(\frac{1}{\left(\rho_{0}^{2}+x_{0}^{2}\right)^{3/2}}-\frac{\rho_{0}^{2}sin^{2}\phi_{0}}{\left(\rho_{0}^{2}+x_{0}^{2}\right)^{5/2}}\right)\left(\frac{1}{\left(\rho^{2}+x^{2}\right)^{3/2}}-\frac{\rho^{2}sin^{2}\phi}{\left(\rho^{2}+x^{2}\right)^{5/2}}\right)\\\nonumber
&\times\exp\left(-\frac{\left(x-x_{0}-vt\right)^{2}}{4Dt}\right)\exp\left(-\frac{\left(\bar{\rho}-\bar{\rho}_{0}\right)^{2}}{4Dt}\right).
\end{align}
As in previous section, we use the 2D Fourier transform and the Anger-Jacobi expansion \eqref{A-JE} to arrive at
\begin{align}\label{DandD3}
G^{\left(0\right)}\left(t\right)=\frac{1}{2\pi}\frac{n}{\left(4\pi Dt\right)^{1/2}}\sum_{n_{1},n_{2}}\left(i\right)^{n_{1}}\left(-i\right)^{n_{2}}&\int\int\int d^2k d^{3}rd^{3}r_{0}\cdot\left(\frac{1}{\left(\rho_{0}^{2}+x_{0}^{2}\right)^{3/2}}-\frac{3\rho_{0}^{2}sin^{2}\phi_{0}}{\left(\rho_{0}^{2}+x_{0}^{2}\right)^{5/2}}\right)\left(\frac{1}{\left(\rho^{2}+x^{2}\right)^{3/2}}-\frac{3\rho^{2}sin^{2}\phi}{\left(\rho^{2}+x^{2}\right)^{5/2}}\right)\\\nonumber
&\times\exp\left(-\frac{\left(x-x_{0}-vt\right)^{2}}{4Dt}\right)\exp\left(-k^{2}Dt\right)J_{n_{1}}\left[k\rho\right]J_{n_{2}}\left[k\rho_{0}\right]e^{in_{1}\phi'}e^{-in_{2}\phi_{0}'},
\end{align}
where
\beq\label{DandD4}
\phi'=\phi-\tilde{\phi},\ \phi_{0}'=\phi_{0}-\tilde{\phi}
\eeq
and $\tilde{\phi}$ is the free angle of the $k$ vector. Substituting \eqref{DandD4} into \eqref{DandD3} and integrating over $\tilde{\phi}$ yields
\begin{align}
G_{original}^{\left(0\right)}\left(t\right)=\frac{N}{\left(4\pi Dt\right)^{1/2}}\sum_{n_{1}}\int&\int\int dkd^{3}rd^{3}r_{0}\cdot k\cdot\left(\frac{1}{\left(\rho_{0}^{2}+x_{0}^{2}\right)^{3/2}}-\frac{3\rho_{0}^{2}sin^{2}\phi_{0}}{\left(\rho_{0}^{2}+x_{0}^{2}\right)^{5/2}}\right)\left(\frac{1}{\left(\rho^{2}+x^{2}\right)^{3/2}}-\frac{3\rho^{2}sin^{2}\phi}{\left(\rho^{2}+x^{2}\right)^{5/2}}\right)\\\nonumber
&\times\exp\left(-\frac{\left(x-x_{0}-vt\right)^{2}}{4Dt}\right)\exp\left(-k^{2}Dt\right)J_{n_{1}}\left[k\rho\right]J_{n_{1}}\left[k\rho_{0}\right]e^{in_{1}\phi}e^{-in_{1}\phi_{0}}.
\end{align}
We now continue integrating over the angles and $\rho,\ \rho_0$ term by term. The first integral is
\begin{align}
I_{1}\equiv&\frac{n}{\left(4\pi Dt\right)^{1/2}}\sum_{n_{1}}\int\int\int dkd^{3}rd^{3}r_{0}\cdot k\cdot\frac{1}{\left(\rho_{0}^{2}+x_{0}^{2}\right)^{3/2}}\frac{1}{\left(\rho^{2}+x^{2}\right)^{3/2}}\exp\left(-\frac{\left(x-x_{0}-vt\right)^{2}}{4Dt}\right)\exp\left(-k^{2}Dt\right)J_{n_{1}}\left[k\rho\right]J_{n_{1}}\left[k\rho_{0}\right]e^{in_{1}\phi}e^{-in_{1}\phi_{0}}\\
=&\frac{2\pi^{3/2}n}{\left(Dt\right)^{1/2}}\int\int\int dkd^{2}rd^{2}r_{0}\cdot k\cdot\frac{1}{\left(\rho_{0}^{2}+x_{0}^{2}\right)^{3/2}}\frac{1}{\left(\rho^{2}+x^{2}\right)^{3/2}}exp\left(-\frac{\left(x-x_{0}-vt\right)^{2}}{4Dt}\right)exp\left(-k^{2}Dt\right)J_{0}\left[k\rho\right]J_{0}\left[k\rho_{0}\right]\\\label{DandD5}
=&\frac{2\pi^{3/2}n}{\left(Dt\right)^{1/2}}\int\int\int dkdxdx_{0}\cdot k\cdot\frac{e^{-k\left|x\right|}}{\left|x\right|}\frac{e^{-k\left|x_{0}\right|}}{\left|x_{0}\right|}exp\left(-\frac{\left(x-x_{0}-vt\right)^{2}}{4Dt}\right)exp\left(-k^{2}Dt\right).
\end{align}
The second integral is
\begin{align}
I_{2}&\equiv-\frac{3n}{\left(4\pi Dt\right)^{1/2}}\sum_{n_{1}}\int\int\int dkd^{3}rd^{3}r_{0}\cdot k\cdot\frac{\rho_{0}^{2}sin^{2}\phi_{0}}{\left(\rho_{0}^{2}+x_{0}^{2}\right)^{5/2}}\frac{1}{\left(\rho^{2}+x^{2}\right)^{3/2}}\exp\left(-\frac{\left(x-x_{0}-vt\right)^{2}}{4Dt}\right)\exp\left(-k^{2}Dt\right)J_{n_{1}}\left[k\rho\right]J_{n_{1}}\left[k\rho_{0}\right]e^{in_{1}\phi}e^{-in_{1}\phi_{0}}\\
&=-\frac{3\sqrt{\pi}n}{\left(Dt\right)^{1/2}}\int\int\int dkd^{2}rd^{3}r_{0}\cdot k\cdot\frac{\rho_{0}^{2}sin^{2}\phi_{0}}{\left(\rho_{0}^{2}+x_{0}^{2}\right)^{5/2}}\frac{1}{\left(\rho^{2}+x^{2}\right)^{3/2}}exp\left(-\frac{\left(x-x_{0}-vt\right)^{2}}{4Dt}\right)exp\left(-k^{2}Dt\right)J_{0}\left[k\rho\right]J_{0}\left[k\rho_{0}\right]\\
&=-\frac{3\pi^{3/2}n}{\left(Dt\right)^{1/2}}\int\int\int dkd^{2}rd^{2}r_{0}\cdot k\cdot\frac{\rho_{0}^{2}}{\left(\rho_{0}^{2}+x_{0}^{2}\right)^{5/2}}\frac{1}{\left(\rho^{2}+x^{2}\right)^{3/2}}\exp\left(-\frac{\left(x-x_{0}-vt\right)^{2}}{4Dt}\right)\exp\left(-k^{2}Dt\right)J_{0}\left[k\rho\right]J_{0}\left[k\rho_{0}\right]\\\label{DandD6}
&=-\frac{3\pi^{3/2}n}{\left(Dt\right)^{1/2}}\int\int\int dkdxdx_{0}\cdot k\cdot\frac{x_{0}\left(-k\left|x_{0}\right|+2\right)\left(x_{0}cosh\left(kx_{0}\right)-\left|x_{0}\right|sinh\left(x_{0}\right)\right)}{3\left|x_{0}\right|^{3}}\frac{e^{-k\left|x\right|}}{\left|x\right|}\exp\left(-\frac{\left(x-x_{0}-vt\right)^{2}}{4Dt}\right)\exp\left(-k^{2}Dt\right).
\end{align}
We note that
\beq
xcosh\left(kx\right)-\left|x\right|sinh\left(x\right)=\frac{1}{2}\left[\left(x-\left|x\right|\right)e^{kx}+\left(x+\left|x\right|\right)e^{-kx}\right]=\begin{cases}
	xe^{-kx} & x>0\\
	xe^{kx} & x<0
\end{cases}=xe^{-k\left|x\right|},
\eeq
therefore \eqref{DandD6} can be written as 
\beq\label{DandD7}
I_{2}=-\frac{\pi^{3/2}n}{\left(Dt\right)^{1/2}}\int\int\int dkdxdx_{0}\cdot k\cdot\frac{\left(-k\left|x_{0}\right|+2\right)e^{-k\left|x_{0}\right|}}{\left|x_{0}\right|}\frac{e^{-k\left|x\right|}}{\left|x\right|}exp\left(-\frac{\left(x-x_{0}-v_{x}t\right)^{2}}{4Dt}\right)exp\left(-k^{2}Dt\right).
\eeq

The third integral is the same as the second with the interchange $x_{0}\leftrightarrow x$, therefore,
\beq\label{DandD8}
I_{3}\equiv-\frac{\pi^{3/2}n}{\left(Dt\right)^{1/2}}\int\int\int dkdxdx_{0}\cdot k\cdot\frac{\left(-k\left|x\right|+2\right)e^{-k\left|x\right|}}{\left|x\right|}\frac{e^{-k\left|x_{0}\right|}}{\left|x_{0}\right|}\exp\left(-\frac{\left(x-x_{0}-vt\right)^{2}}{4Dt}\right)\exp\left(-k^{2}Dt\right).
\eeq
The forth integral is
\beq
I_{4}\equiv\frac{9n}{\left(4\pi Dt\right)^{1/2}}\sum_{n_{1}}\int\int\int dkd^{3}rd^{3}r_{0}\cdot k\cdot\frac{\rho_{0}^{2}sin^{2}\phi_{0}}{\left(\rho_{0}^{2}+x_{0}^{2}\right)^{5/2}}\frac{\rho^{2}sin^{2}\phi}{\left(\rho^{2}+x^{2}\right)^{5/2}}\exp\left(-\frac{\left(x-x_{0}-vt\right)^{2}}{4Dt}\right)\exp\left(-k^{2}Dt\right)J_{n_{1}}\left[k\rho\right]J_{n_{1}}\left[k\rho_{0}\right]e^{in_{1}\phi}e^{-in_{1}\phi_{0}}
\eeq
\begin{align}
=-\frac{1}{4}\frac{9n}{\left(4\pi Dt\right)^{1/2}}\sum_{n_{1}}&\int\int\int dkd^{3}rd^{3}r_{0}\cdot k\cdot\frac{\rho_{0}^{2}sin^{2}\phi_{0}}{\left(\rho_{0}^{2}+x_{0}^{2}\right)^{5/2}}\frac{\rho^{2}}{\left(\rho^{2}+x^{2}\right)^{5/2}}\exp\left(-\frac{\left(x-x_{0}-vt\right)^{2}}{4Dt}\right)\exp\left(-k^{2}Dt\right)\\\nonumber
&\times J_{n_{1}}\left[k\rho\right]J_{n_{1}}\left[k\rho_{0}\right]\left(e^{i\phi}-e^{-i\phi}\right)^{2}e^{in_{1}\phi}e^{-in_{1}\phi_{0}}
\end{align}
\begin{align}
=-\frac{1}{4}\frac{9n}{\left(4\pi Dt\right)^{1/2}}\sum_{n_{1}}&\int\int\int dkd^{3}rd^{3}r_{0}\cdot k\cdot\frac{\rho_{0}^{2}sin^{2}\phi_{0}}{\left(\rho_{0}^{2}+x_{0}^{2}\right)^{5/2}}\frac{\rho^{2}}{\left(\rho^{2}+x^{2}\right)^{5/2}}\exp\left(-\frac{\left(x-x_{0}-vt\right)^{2}}{4Dt}\right)\exp\left(-k^{2}Dt\right)\\\nonumber
&\times J_{n_{1}}\left[k\rho\right]J_{n_{1}}\left[k\rho_{0}\right]\left(e^{i\phi\left(n_{1}+2\right)}+e^{i\phi\left(n_{1}-2\right)}-2e^{in_{1}\phi}\right)e^{-in_{1}\phi_{0}}
\end{align}
\begin{align}
=-\frac{1}{2}\frac{9\sqrt{\pi}n}{\left(4Dt\right)^{1/2}}&\int\int\int dkd^{2}rd^{3}r_{0}\cdot k\cdot\frac{\rho_{0}^{2}sin^{2}\phi_{0}}{\left(\rho_{0}^{2}+x_{0}^{2}\right)^{5/2}}\frac{\rho^{2}}{\left(\rho^{2}+x^{2}\right)^{5/2}}\exp\left(-\frac{\left(x-x_{0}-vt\right)^{2}}{4Dt}\right)\exp\left(-k^{2}Dt\right)\\\nonumber
&\times\left[-2J_{0}\left[k\rho\right]J_{0}\left[k\rho_{0}\right]+J_{2}\left[k\rho\right]J_{2}\left[k\rho_{0}\right]\left(e^{i2\phi_{0}}+e^{-i2\phi_{0}}\right)\right]
\end{align}
\begin{align}
&=\frac{9\pi^{3/2}n}{\left(4Dt\right)^{1/2}}\int\int\int dkd^{2}rd^{2}r_{0}\cdot k\cdot\frac{\rho_{0}^{2}}{\left(\rho_{0}^{2}+x_{0}^{2}\right)^{5/2}}\frac{\rho^{2}}{\left(\rho^{2}+x^{2}\right)^{5/2}}\exp\left(-\frac{\left(x-x_{0}-vt\right)^{2}}{4Dt}\right)\exp\left(-k^{2}Dt\right)J_{0}\left[k\rho\right]J_{0}\left[k\rho_{0}\right]\\\label{DandD9}
&=\frac{\pi^{3/2}n}{\left(4Dt\right)^{1/2}}\int\int\int dkdxdx_{0}\cdot k\cdot\frac{\left(-k\left|x_{0}\right|+2\right)e^{-k\left|x_{0}\right|}}{\left|x_{0}\right|}\frac{\left(-k\left|x\right|+2\right)e^{-k\left|x\right|}}{\left|x\right|}\exp\left(-\frac{\left(x-x_{0}-vt\right)^{2}}{4Dt}\right)\exp\left(-k^{2}Dt\right)J_{0}\left[k\rho\right]J_{0}\left[k\rho_{0}\right].
\end{align}

Summing the four results \eqref{DandD5}, \eqref{DandD7}, \eqref{DandD8} and \eqref{DandD9} leads to
\begin{align}\label{DandD10}
G^{\left(0\right)}\left(t\right)=\frac{\pi^{3/2}n}{\left(4Dt\right)^{1/2}}\int\int\int dkdxdx_{0}\cdot k^{3}\cdot \exp\left(-\frac{\left(x-x_{0}-vt\right)^{2}}{4Dt}\right)\exp\left(-k^{2}Dt\right)e^{-k\left(\left|x\right|+\left|x_{0}\right|\right)}.
\end{align}
We now pursue the integration over $x,x_0$. The integral over these coordinate is 
\begin{align}
&\int\int dxdx_{0}\exp\left(-\frac{\left(x-x_{0}-vt\right)^{2}}{4Dt}\right)e^{-k\left(\left|x\right|+\left|x_{0}\right|\right)}\\
&=\int\limits _{0}^{\infty}dx\left[\int\limits _{0}^{\infty}dx_{0}\exp\left(-\frac{\left(x-x_{0}-vt\right)^{2}}{4Dt}\right)e^{-k\left(x+x_{0}\right)}+\int\limits _{-\infty}^{0}dx_{0}\exp\left(-\frac{\left(x-x_{0}-vt\right)^{2}}{4Dt}\right)e^{-k\left(x-x_{0}\right)}\right]\\\nonumber
&+\int\limits _{-\infty}^{0}dx\left[\int\limits _{0}^{\infty}dx_{0}\exp\left(-\frac{\left(x-x_{0}-vt\right)^{2}}{4Dt}\right)e^{-k\left(-x+x_{0}\right)}+\int\limits _{-\infty}^{0}dx_{0}\exp\left(-\frac{\left(x-x_{0}-vt\right)^{2}}{4Dt}\right)e^{k\left(x+x_{0}\right)}\right]\\\label{DandD11}
&=\sqrt{\pi Dt}e^{k^{2}Dt}\left\{ \int\limits _{0}^{\infty}dx\left[e^{kvt}e^{-2kx}Erfc\left[k\sqrt{Dt}+\frac{vt-x}{2\sqrt{Dt}}\right]+e^{-kvt}Erfc\left[k\sqrt{Dt}+\frac{x-vt}{2\sqrt{Dt}}\right]\right]\right.\\\nonumber
&\left.+\int\limits _{-\infty}^{0}dx\left[e^{kvt}Erfc\left[k\sqrt{Dt}+\frac{vt-x}{2\sqrt{Dt}}\right]+e^{-kvt}e^{2kx}Erfc\left[k\sqrt{Dt}+\frac{x-vt}{2\sqrt{Dt}}\right]\right]\right\} .
\end{align}

We solve the integration over $x$ term by term,
\beq\label{DandD12}
e^{kvt}\int\limits _{0}^{\infty}e^{-2kx}Erfc\left[k\sqrt{Dt}+\frac{vt-x}{2\sqrt{Dt}}\right]=\frac{1}{2k}\left[e^{kvt}Erfc\left(\frac{2Dkt+tv}{2\sqrt{Dt}}\right)+e^{-kvt}Erfc\left(\frac{2Dkt-tv}{2\sqrt{Dt}}\right)\right],
\eeq
\beq\label{DandD13}
e^{-kvt}\int\limits _{0}^{\infty}dxErfc\left[k\sqrt{Dt}+\frac{x-vt}{2\sqrt{Dt}}\right]=\left[2\sqrt{\frac{Dt}{\pi}}e^{-k^{2}Dt-\frac{\left(vt\right)^{2}}{4Dt}}-e^{-kvt}\left(2kDt-tv\right)Erfc\left[\frac{2kDt-tv}{2\sqrt{Dt}}\right]\right],
\eeq
\beq\label{DandD14}
e^{kvt}\int\limits _{-\infty}^{0}dxErfc\left[k\sqrt{Dt}+\frac{vt-x}{2\sqrt{Dt}}\right]=\left[2\sqrt{\frac{Dt}{\pi}}e^{-k^{2}Dt-\frac{\left(vt\right)^{2}}{4Dt}}-e^{kvt}\left(2Dkt+tv\right)Erfc\left[\frac{2Dkt+tv}{2\sqrt{Dt}}\right]\right],
\eeq
\beq\label{DandD15}
e^{-kvt}\int\limits _{-\infty}^{0}dxe^{2kx}Erfc\left[k\sqrt{Dt}+\frac{x-vt}{2\sqrt{Dt}}\right]
=\frac{1}{2k}\left(e^{-kvt}Erfc\left(\frac{2Dkt-tv}{2\sqrt{Dt}}\right)+e^{kvt}Erfc\left(\frac{2Dkt+tv}{2\sqrt{Dt}}\right)\right).
\eeq
Substituting the summation of Eqs. \eqref{DandD12}, \eqref{DandD13}, \eqref{DandD14} and \eqref{DandD15} into \eqref{DandD10} together with the factor of $\sqrt{\pi D t}e^{k^2Dt}$ from \eqref{DandD11} yields
\begin{align}\label{DandD16}
G^{\left(0\right)}\left(t\right)=\frac{\pi^{2}n}{2}&\int dk\cdot k^{3}\cdot\left\{ \frac{1}{k}\left[e^{kvt}Erfc\left(\frac{2Dkt+tv}{2\sqrt{Dt}}\right)+e^{-kvt}Erfc\left(\frac{2Dkt-tv}{2\sqrt{Dt}}\right)\right]\right.\\\nonumber
&\left.+4\sqrt{\frac{Dt}{\pi}}e^{-k^{2}Dt-\frac{\left(vt\right)^{2}}{4Dt}}-e^{-kvt}\left(2kDt-tv\right)Erfc\left[\frac{2kDt-tv}{2\sqrt{Dt}}\right]-e^{kvt}\left(2kDt+tv\right)Erfc\left[\frac{2kDt+tv}{2\sqrt{Dt}}\right]\right\}.
\end{align}

Again, we integrate term by term,
\beq\label{DandD17}
\int dk\cdot k^{2}\cdot\left[e^{kvt}Erfc\left(\frac{2Dkt+tv}{2\sqrt{Dt}}\right)+e^{-kvt}Erfc\left(\frac{2Dkt-tv}{2\sqrt{Dt}}\right)\right]
=\frac{4Erf\left(\frac{vt}{2\sqrt{Dt}}\right)}{\left(vt\right)^{3}}-\frac{4}{\sqrt{\pi Dt}}\frac{e^{-\frac{tv^{2}}{4D}}}{\left(vt\right)^{2}},
\eeq
\beq\label{DandD18}
4\sqrt{\frac{Dt}{\pi}}\int dk\cdot k^{3}e^{-k^{2}Dt-\frac{\left(vt\right)^{2}}{4Dt}}=\frac{2}{\sqrt{\pi}\left(Dt\right)^{3/2}}e^{-\frac{\left(vt\right)^{2}}{4Dt}},
\eeq
\begin{align}\label{DandD19}
&\int dk\cdot k^{3}\cdot e^{-kvt}\left(2kDt-tv\right)Erfc\left[\frac{2kDt-tv}{2\sqrt{Dt}}\right]\\\nonumber
=\left(-\frac{48Dt}{\left(vt\right)^{5}}-\frac{6}{\left(vt\right)^{3}}\right)e^{-\frac{\left(vt\right)^{2}}{4Dt}}&+\left(\frac{48Dt}{\left(vt\right)^{5}}-\frac{6}{\left(v_{x}t\right)^{3}}\right)Erfc\left(\frac{-tv}{2\sqrt{Dt}}\right)-\frac{e^{-\frac{\left(vt\right)^{2}}{4Dt}}}{\sqrt{\pi}}\left(\frac{48\sqrt{Dt}}{\left(vt\right)^{4}}+\frac{2}{\sqrt{Dt}\left(vt\right)^{2}}-\frac{1}{\left(Dt\right)^{3/2}}\right),
\end{align}
\begin{align}\label{DandD20}
&\int dk\cdot k^{3}\cdot e^{kvt}\left(2kDt+tv\right)Erfc\left[\frac{2kDt+tv_{x}}{2\sqrt{Dt}}\right]\\\nonumber
=\left(48\frac{Dt}{\left(vt\right)^{5}}+\frac{6}{\left(vt\right)^{3}}\right)e^{-\frac{\left(vt\right)^{2}}{4\left(Dt\right)}}&-\left(\frac{48Dt}{\left(vt\right)^{5}}-\frac{6}{\left(vt\right)^{3}}\right)Erfc\left(\frac{tv}{2\sqrt{Dt}}\right)-\frac{e^{-\frac{\left(vt\right)^{2}}{4\left(Dt\right)}}}{\sqrt{\pi}}\left(\frac{48\sqrt{Dt}}{\left(vt\right)^{4}}+\frac{2}{\sqrt{Dt}\left(vt\right)^{2}}-\frac{1}{\left(Dt\right)^{3/2}}\right).
\end{align}

Substituting Eqs. \eqref{DandD17}, \eqref{DandD18}, \eqref{DandD19} and \eqref{DandD20} into \eqref{DandD16} leads to the final result \eqref{DandDtime}.

\subsection{Asymptotic behavior of the correlation function}

In the following we examine the asymptotic behavior of \eqref{DandDtime}.
First, we would like to check that the formal limit $v\rightarrow0$ recovers the result without drift \eqref{longtimelimit},
\beq\label{nodriftlimit}
\lim_{v\rightarrow0}G^{\left(0\right)}\left(t\right)
=\frac{8\pi^{3/2}n}{15\left(Dt\right)^{3/2}}.
\eeq
Indeed, eq. \eqref{nodriftlimit} recovers \eqref{longtimelimit} up to a prefactor, which is expected due to the difference in geometry. We again note the divergence of the correlation for $t\rightarrow0$, which stems from the non-physical model that allows the nuclear spins to be arbitrarily close to the NV center.
 
The asymptotic behavior for $\sqrt{Dt}\gg vt$, or equivalently $t\ll\frac{v^2}{D}$, is the same.
For $t\gg\frac{v^2}{D}$ we have
\beq\label{DandDlongtime}
G^{\left(0\right)}\left(t\right)\approx\frac{8\pi^{2}n}{\left(vt\right)^{3}},
\eeq
which is expected from dimensional analysis.
These equations are used in Fig. 3c in the main text. 

\section{The power spectrum of diffusing particles}
\label{Powernodrift}

In the following we give an analytic expression for the power spectrum of diffusing particle, and we derive the asymptotic behavior of the power spectrum at low frequencies. 

First we calculate the power spectrum as the Fourier transform of eq. \eqref{correlation_no_drift},

\begin{align}
	&S\left(\omega\right)\propto n\sqrt{2}\pi^{2}\left\{\frac{8}{5}\frac{\left|\omega\right|^{1/2}}{D^{3/2}}\textrm{Re}\left(\,_{1}F_{4}\left(1;\frac{3}{4},\frac{5}{4},\frac{7}{4},\frac{9}{4};-\frac{d^{4}\omega^{2}}{16D^{2}}\right)\right)-\frac{56}{225}\frac{\left|\omega\right|^{3/2}d^{2}}{D^{5/2}}\textrm{Re}\left(\,_{1}F_{4}\left(1;\frac{7}{4},\frac{7}{4},\frac{9}{4},\frac{9}{4};-\frac{d^{4}\omega^{2}}{16D^{2}}\right)\right)\right.\\\nonumber
	&-\frac{224}{225}\frac{\left|\omega\right|^{3/2}d^{2}}{D^{5/2}}\textrm{Re}\left(\,_{1}F_{4}\left(2;\frac{7}{4},\frac{7}{4},\frac{9}{4},\frac{9}{4};-\frac{d^{4}\omega^{2}}{16D^{2}}\right)\right)+\frac{32}{105}\frac{\left|\omega\right|^{3/2}d^{2}}{D^{5/2}}\textrm{Re}\left(\,_{1}F_{4}\left(1;\frac{5}{4},\frac{7}{4},\frac{9}{4},\frac{11}{4};-\frac{d^{4}\omega^{2}}{16D^{2}}\right)\right)\\\nonumber
	&-\frac{64}{225}\frac{\left|\omega\right|^{5/2}d^{4}}{D^{7/2}}\textrm{Re}\left(\,_{1}F_{4}\left(1;\frac{7}{4},\frac{7}{4},\frac{9}{4},\frac{9}{4};-\frac{d^{4}\omega^{2}}{16D^{2}}\right)\right)+\frac{32}{225}\frac{\left|\omega\right|^{5/2}d^{4}}{D^{7/2}}\textrm{Re}\left(\,_{1}F_{4}\left(1;\frac{7}{4},\frac{9}{4},\frac{9}{4},\frac{11}{4};-\frac{d^{4}\omega^{2}}{16D^{2}}\right)\right)\\\nonumber
	&+\frac{128}{1575}\frac{\left|\omega\right|^{7/2}d^{6}}{D^{9/2}}\textrm{Re}\left(\,_{1}F_{4}\left(1;\frac{7}{4},\frac{9}{4},\frac{9}{4},\frac{11}{4};-\frac{d^{4}\omega^{2}}{16D^{2}}\right)\right)
	-\frac{256}{11025}\frac{\left|\omega\right|^{7/2}d^{6}}{D^{9/2}}\textrm{Re}\left(\,_{1}F_{4}\left(1;\frac{9}{4},\frac{9}{4},\frac{11}{4},\frac{11}{4};-\frac{d^{4}\omega^{2}}{16D^{2}}\right)\right)\\\nonumber
	&-\frac{512}{99225}\frac{\left|\omega\right|^{9/2}d^8}{D^{11/2}}\textrm{Re}\left(\,_{1}F_{4}\left(1;\frac{9}{4},\frac{11}{4},\frac{11}{4},\frac{13}{4};-\frac{d^{4}\omega^{2}}{16D^{2}}\right)\right)+\frac{3D}{\sqrt{2}\omega^{2}d^{5}}-\frac{8}{3}\frac{\left|\omega\right|^{1/2}}{D^{3/2}}-\frac{8}{9}\frac{\left|\omega\right|^{3/2}d^{2}}{D^{5/2}}+\frac{32}{45}\frac{\left|\omega\right|^{5/2}d^{4}}{D^{7/2}}\\\nonumber
	&-\frac{3\sqrt{2}}{\left|\omega\right|d^{3}}\left[\frac{\pi}{2}\textrm{ber}_{2}\left(2\sqrt{\frac{\left|\omega\right|}{D}}d\right)+\textrm{kei}_{2}\left(2\sqrt{\frac{\left|\omega\right|}{D}}d\right)\right]+\frac{2\sqrt{2}}{Dd}\left[-\frac{\pi}{2}\textrm{bei}\left(2\sqrt{\frac{\left|\omega\right|}{D}}d\right)+\textrm{ker}\left(2\sqrt{\frac{\left|\omega\right|}{D}}d\right)\right]\\\nonumber
	&+\frac{7}{2}\frac{1}{\left|\omega\right|^{1/2}D^{1/2}d^2}\left[\textrm{ker}_{1}\left(2\sqrt{\frac{\left|\omega\right|}{D}}d\right)-\textrm{kei}_{1}\left(2\sqrt{\frac{\left|\omega\right|}{D}}d\right)-\frac{\pi}{2}\textrm{bei}_{1}\left(2\sqrt{\frac{\left|\omega\right|}{D}}d\right)-\frac{\pi}{2}\textrm{ber}_{1}\left(2\sqrt{\frac{\left|\omega\right|}{D}}d\right)\right]\\\nonumber
	&\left.-2\frac{\left|\omega\right|^{1/2}}{D^{3/2}}\left[\frac{\pi}{2}\textrm{ber}_{1}\left(2\sqrt{\frac{\left|\omega\right|}{D}}d\right)-\frac{\pi}{2}\text{bei}_{1}\left(2\sqrt{\frac{\left|\omega\right|}{D}}d\right)+\textrm{kei}_{1}\left(2\sqrt{\frac{\left|\omega\right|}{D}}d\right)+\textrm{ker}_{1}\left(2\sqrt{\frac{\left|\omega\right|}{D}}d\right)\right]\right\}
\end{align}

For low positive frequencies,
\beq
S\left(\omega\right)\sim \pi ^2 n \left(\frac{3}{4 d D}-\frac{16 \sqrt{2} \sqrt{\omega }}{15 D^{3/2}}+\frac{5 \pi d \omega }{8 D^2}\right),
\eeq
and since the power spectrum is symmetric around $\omega=0$ we can extend the expression to negative frequencies as follows,
\beq
S\left(\omega\right)\sim \pi ^2 n \left(\frac{3}{4 d D}-\frac{16 \sqrt{2} \sqrt{\left|\omega\right| }}{15 D^{3/2}}+\frac{5 \pi d \left|\omega\right| }{8 D^2}\right).
\eeq 
Adding the appropriate proportion constants we arrive at
\beq\label{nodriftlowfreq1}
S\left(\omega\right)\approx\left(\frac{\mu_0 \hbar \gamma_N \gamma_e}{4\pi}\right)^2 n \pi^2\left(\frac{3}{4 d D}-\frac{16 \sqrt{2} \sqrt{\left|\omega\right| }}{15 D^{3/2}}+\frac{5 \pi d \left|\omega\right| }{8 D^2}\right).
\eeq
We can also rewrite \eqref{nodriftlowfreq1} as follows,
\begin{align}\label{nodriftlowfreq2}
S\left(\omega\right)\approx&\left(\frac{\mu_0 \hbar \gamma_N \gamma_e}{4\pi}\right)^2 n\frac{1}{d D} \pi^2\left(\frac{3 }{4}-\frac{16 \sqrt{2} d \sqrt{\left|\omega\right| }}{15 D^{1/2}}+\frac{5 \pi d^2 \left|\omega\right| }{8 D}\right)\\\nonumber
&=\gamma_e^2 B_{RMS}^2 \tau_D \pi^2 \left(\frac{3 }{4}-\frac{16 \sqrt{2}}{15}\sqrt{\frac{\left|\omega\right|}{\omega_D}}+\frac{5 \pi }{8} \frac{\left|\omega\right|}{\omega_D}\right),
\end{align}
where $\tau_D=\omega_D^{-1}=\frac{d^2}{D}$. 
Indeed, we can see that the power spectrum \eqref{nodriftlowfreq1} agrees with the universal estimation \eqref{stat5}, and that the power spectrum at $\omega=0$ is finite and proportional to $\gamma_e^2 B_{RMS}^2 \tau_D$.

\section{THE DRIFT-DIFFUSION PROPAGATOR IN THE FREQUENCY DOMAIN}
\label{propagator}
As explained in the main text, when $T_2>\tau_D,\tau_v$, it is beneficial to estimate the drift velocity from the power spectrum. Since the calculation of the Fourier transform of the temporal correlation function is difficult, we chose a different approach - we calculate the correlation function in the frequency domain directly. 
 
In the following section we take the first step towards this goal, when we derive the drift-diffusion propagator for a constant drift velocity $\bar{v}$,
\beq\label{fourierprop}
P(\bar{r}|\bar{r}_0,\omega)=\frac{1}{4\pi D\left|\bar{r}-\bar{r}_0\right|}e^{ik_{1}\cdot\left|\bar{r}-\bar{r}_0\right|}e^{\frac{\bar{v}\cdot\left(\bar{r}-\bar{r}_0\right)}{2D}}
\eeq
with
\beq\label{polar}
\phi=\pi-tan^{-1}\left(\frac{4D\omega}{v^{2}}\right),R=\sqrt{\left(\frac{v}{2D}\right)^{4}+\left(\frac{\omega}{D}\right)^{2}},\ k_{1}=\sqrt{R}e^{i\phi/2} .
\eeq

We start our derivation with the drift-diffusion equation for the propagator in the time domain (as \eqref{propfine1}),
\beq
\left[\frac{\partial }{\partial t}- D\nabla^{2}+\bar{v}\cdot\nabla\right]P =\delta\left(\bar{r}-\bar{r_0}\right)\delta\left(t\right)  
\eeq
Fourier transform with respect to time leads to
\beq\label{prop0}
\left[-i\omega- D\nabla^{2}+\bar{v}\cdot\nabla\right]P =\delta\left(\bar{r}-\bar{r_0}\right)  
\eeq
As in the previous section, we use the transformation \eqref{propfine2} with $\bar{a}=\frac{\bar{v}}{2D},\ b=0$,
\begin{align}
	\tilde{P}=e^{-\bar{a}\cdot\bar{r}} Pe^{\bar{a}\cdot r_0},
\end{align}
in order to rewrite eq. \eqref{prop0} as
\beq\label{prop2}
\left(-D\nabla^{2}-i\omega+\frac{v^{2}}{4D}\right)\tilde{P}=\delta\left(\bar{r}-\bar{r_0}\right).
\eeq
The solution to this equation is given by
\begin{align}
	\tilde{P}&=\int\frac{d^{3}k}{\left(2\pi\right)^{3}}\frac{e^{i\bar{k}\cdot\left(\bar{r}-\bar{r}_0\right)}}{Dk^{2}-i\omega+\frac{v^{2}}{4D}}=\int\limits _{0}^{\infty}dk\int\limits _{-1}^{1}d\left(cos\theta\right)\frac{k^{2}}{\left(2\pi\right)^{2}}\frac{1}{\left(Dk^{2}-i\omega+\frac{v^{2}}{4D}\right)}e^{ik\cdot\left|\bar{r}-\bar{r}_0\right|cos\theta}\\
	&=\frac{1}{i\left|\bar{r}-\bar{r}_0\right|}\int\limits _{0}^{\infty}\frac{kdk}{\left(2\pi\right)^{2}}\frac{1}{\left(Dk^{2}-i\omega+\frac{v^{2}}{4D}\right)}\left(e^{ik\cdot\left|\bar{r}-\bar{r'}\right|}-e^{-ik\cdot\left|\bar{r}-\bar{r}_0\right|}\right)\\\label{prop4}
	&=\frac{1}{2i\left|\bar{r}-\bar{r}_0\right|}\int\limits _{-\infty}^{\infty}\frac{kdk}{\left(2\pi\right)^{2}}\frac{1}{\left(Dk^{2}-i\omega+\frac{v^{2}}{4D}\right)}\left(e^{ik\cdot\left|\bar{r}-\bar{r}_0\right|}-e^{-ik\cdot\left|\bar{r}-\bar{r}_0\right|}\right).
\end{align}
In order to solve the integral \eqref{prop4}, we use the residue theorem. The poles of the integrand follow the equation
\beq
k^{2}=i\frac{\omega}{D}-\frac{v^{2}}{4D^{2}}\equiv Re^{i\phi},
\eeq
where $R$ and $\phi$ are defined by \eqref{polar}.
The poles are, therefore,
\begin{align}
	k_{1}=\sqrt{R}e^{i\phi/2},k_{2}=\sqrt{R}e^{i\left(\phi/2+\pi\right)}
\end{align}
or alternatively,
\beq\label{kvalue}
k_{1}=\sqrt{R}e^{i\phi/2},\ k_{2}=-k_{1}.
\eeq
If $\omega>0$ the phase $\frac{\pi}{2}\leq\phi\leq\pi$ so $k_{1}$
is in the first quarter and $k_{2}$ is in the third quarter. If $\omega<0$ the phase $\pi\leq\phi\leq\frac{3\pi}{2}$ so $k_{1}$
is in the second quarter and $k_{2}$ is in the fourth quarter. If $\omega=0$ the solutions are on the imaginary axis.
The solution in all of these cases is the same,
\beq
\tilde{P}=\frac{1}{2i\left|\bar{r}-\bar{r}_0\right|}\left[\int\limits _{-\infty}^{\infty}\frac{kdk}{\left(2\pi\right)^{2}}\frac{1}{D\left(k-k_{1}\right)\left(k+k_{1}\right)}e^{ik\cdot\left|\bar{r}-\bar{r}_0\right|}-\int\limits _{-\infty}^{\infty}\frac{kdk}{\left(2\pi\right)^{2}}\frac{1}{D\left(k-k_{1}\right)\left(k+k_{1}\right)}e^{-ik\cdot\left|\bar{r}-\bar{r}_0\right|}\right].
\eeq
We now need to define the required integration paths for the residue theorem. For the first (left) integral to converge, we choose half a circle that goes through the positive imaginary axis, so the relevant singularity is $k_{1}$. For the second integral, we choose half a circle that goes through the lower side of the complex plane, thus, the relevant singularity is $k_{2}$.
These are simple poles, so the integration is trivial,
\beq
\tilde{P}=\frac{1}{4\pi D\left|\bar{r}-\bar{r}_0\right|}\frac{k_{1}}{2k_{1}}e^{ik_{1}\cdot\left|\bar{r}-\bar{r}_0\right|}+\frac{1}{4\pi D\left|\bar{r}-\bar{r'}\right|}\frac{k_{1}}{2k_{1}}e^{ik_{1}\cdot\left|\bar{r}-\bar{r}_0\right|}
=\frac{1}{4\pi D\left|\bar{r}-\bar{r}_0\right|}e^{ik_{1}\cdot\left|\bar{r}-\bar{r}_0\right|}
\eeq
Using the definition of $\tilde{P}$ we arrive at \eqref{fourierprop}.

\section{ASYMPTOTIC BEHAVIOR FOR LOW VELOCITY}
\label{Powerlowdrift}
In the following section we want to extend our previous analysis to include finite drift velocity. We will analyze the correlation functions in the regime
\beq\label{LDL}
\sqrt{\frac{v^2}{D\omega}}\ll1,
\eeq
which we name the "low drift limit". It signifies times where the motion of nuclei is "diffusion dominated", meaning, that the diffusion length scale is much larger than that of the drift - $ vt \ll \sqrt{Dt}$, which translates into \eqref{LDL} in the frequency domain. 
We first continue with the assumption that the NV's magnetization axis is perpendicular to the diamond surface for simplification and we show that it yields a quadratic scaling in the velocity. Later on we show how this assumption can be relaxed yielding a linear scaling in the velocity.

\subsection{NV's magnetization axis coincides with boundary condition}
\label{Appendix: Low_drift1}
In the following we show that in the low drift limit the correlation scales quadratically in the drift velocity; namely  $G^{(m)}_v(\omega)\sim G_{v=0}^{(m)}(\omega)+O(v^2)$. This corresponds to \eqref{Doppler1}, where the linear dependency in the drift velocity is eliminated because of the symmetry.

{\it Approximating the propagator ---}
Since the velocity dependency enters the correlation function only through the propagator, the first step in taking the limit \eqref{LDL} is to approximate the propagator. We recall \eqref{fourierprop},
\beq\label{LDL1}
P=\frac{1}{4\pi D\left|\bar{r}-\bar{r}_0\right|}e^{ik_{1}\cdot\left|\bar{r}-\bar{r}_0\right|}e^{\frac{\bar{v}\cdot\left(\bar{r}-\bar{r}_0\right)}{2D}},
\eeq
where
\beq\label{LDL2}
\phi=\pi+tan^{-1}\left(-\frac{4D\omega}{v^{2}}\right),R=\sqrt{\left(\frac{v}{2D}\right)^{4}+\left(\frac{\omega}{D}\right)^{2}}, \ k_{1}=\sqrt{R}e^{i\phi/2}.
\eeq
In the low drift limit \eqref{LDL} we can approximate \eqref{LDL2} by
\beq
\phi=\pi+tan^{-1}\left(-\frac{4D\omega}{v^{2}}\right)=\pi-tan^{-1}\left(\frac{4D\omega}{v^{2}}\right)\approx\pi-\left(-\left(\frac{4D\omega}{v^{2}}\right)^{-1}+\frac{\pi}{2}\right)=\frac{\pi}{2}+\frac{v^{2}}{4D\omega},
\eeq
\beq
\sqrt{R}=\left(\left(\frac{v}{2D}\right)^{4}+\left(\frac{\omega}{D}\right)^{2}\right)^{1/4}=\sqrt{\frac{\omega}{D}}\left(1+\left(\frac{v^{2}}{4D\omega}\right)^{2}\right)^{1/4}\approx\sqrt{\frac{\omega}{D}}\left(1+\frac{1}{4}\left(\frac{v^{2}}{4D\omega}\right)^{2}\right),
\eeq
\begin{align}
	ik_{1}&\approx e^{i\frac{\pi}{2}}\sqrt{\frac{\omega}{D}}\left(1+\frac{1}{4}\left(\frac{v^{2}}{4D\omega}\right)^{2}\right)e^{i\left(\frac{\pi}{2}+\frac{v^{2}}{4D\omega}\right)/2}=\sqrt{\frac{\omega}{D}}e^{i\frac{3\pi}{4}}e^{i\frac{v^{2}}{8D\omega}}\left(1+\frac{1}{4}\left(\frac{v^{2}}{4D\omega}\right)^{2}\right)\\
	&=ik_{1}\left(v=0\right)e^{i\frac{v^{2}}{8D\omega}}\left(1+\frac{1}{4}\left(\frac{v^{2}}{4D\omega}\right)^{2}\right)\approx ik_{1}\left(v=0\right)\left(1+i\frac{v^{2}}{8D\omega}+\frac{1}{8}\left(\frac{v^{2}}{4D\omega}\right)^{2}\right).
\end{align}
Substituting the last equation into the propagator \eqref{LDL1}, and expanding further yields
\begin{align}
	P&=\frac{1}{4\pi D\left|\bar{r}-\bar{r}_0\right|}e^{ik_{1}\left(v=0\right)e^{i\frac{v^{2}}{8D\omega}}\left(1+\frac{1}{4}\left(\frac{v^{2}}{4D\omega}\right)^{2}\right)\cdot\left|\bar{r}-\bar{r}_0\right|}e^{\frac{\bar{v}\cdot\left(\bar{r}-\bar{r}_0\right)}{2D}}
	\approx\frac{1}{4\pi D\left|\bar{r}-\bar{r}_0\right|}e^{ik_{1}\left(v=0\right)\left(1+i\frac{v^{2}}{8D\omega}\right)\cdot\left|\bar{r}-\bar{r}_0\right|}e^{\frac{\bar{v}\cdot\left(\bar{r}-\bar{r}_0\right)}{2D}}
	\\\label{LDL3}&=\frac{1}{4\pi D\left|\bar{r}-\bar{r}_0\right|}e^{\sqrt{\frac{\omega}{2D}}\left(i-1\right)\left(1+i\frac{v^{2}}{8D\omega}\right)\cdot\left|\bar{r}-\bar{r}_0\right|}e^{\frac{\bar{v}\cdot\left(\bar{r}-\bar{r}_0\right)}{2D}}.
\end{align}

We note that under the assumption that the magnetization axis is perpendicular to surface the
first order will surely be eliminated due to azimuthal symmetry. Therefore, the leading term in the expansion will be of second order. We continue with the spherical Bessel expansion \cite[Eq.~10.60.7]{DLMF} of \eqref{LDL3},

\begin{align}
	P&\approx\frac{1}{4\pi D\left|\bar{r}-\bar{r}_0\right|}e^{e^{i\frac{3\pi}{4}}\sqrt{\frac{\omega}{D}}\left(1+i\frac{v^{2}}{8D\omega}\right)\cdot\left|\bar{r}-\bar{r}_0\right|}e^{\frac{\bar{v}\cdot\left(\bar{r}-\bar{r}_0\right)}{2D}}\\
	&=\frac{1}{4\pi D\left|\bar{r}-\bar{r}_0\right|}e^{e^{i\frac{3\pi}{4}}\sqrt{\frac{\omega}{D}}\left(1+i\frac{v^{2}}{8D\omega}\right)\cdot\left|\bar{r}-\bar{r}_0\right|}\sum_{l=0}^{\infty}\left(2l+1\right)i^{l}j_{l}\left(-i\frac{v}{2D}\left|\bar{r}-\bar{r}_0\right|\right)P_{l}\left(\cos\Theta\right).
\end{align}

Only even values of $l$ will survive the integration due to the symmetry,
\beq\label{LDL4}
P=\frac{1}{4\pi D\left|\bar{r}-\bar{r}_0\right|}e^{e^{i\frac{3\pi}{4}}\sqrt{\frac{\omega}{D}}\left(1+i\frac{v^{2}}{8D\omega}\right)\cdot\left|\bar{r}-\bar{r}_0\right|}\sum_{l=0}^{\infty}\left(-1\right)^{l}\left(4l+1\right)j_{2l}\left(-i\frac{v}{2D}\left|\bar{r}-\bar{r}_0\right|\right)P_{2l}\left(\cos\Theta\right).
\eeq

The direct correlation with the approximated propagator \eqref{LDL4} is given by

\beq\label{LDL5}
G_{O}^{\left(m\right)}=\int\int d^{3}rd^{3}r_{0}Y_{2}^{(m)}\left(\Omega_{0}\right)Y_{2}^{(m)*}\left(\Omega\right)\frac{1}{r^{3}}\frac{1}{r_{0}^{3}}P\left(r,r_{0},\omega\right)
\eeq
\begin{align}\label{LDL6}
	=\frac{1}{4\pi D}\sum_{l=0}^{\infty}\left(-1\right)^{l}\left(4l+1\right)\int\int d^{3}rd^{3}r_{0}Y_{2}^{(m)}\left(\Omega_{0}\right)Y_{2}^{(m)*}\left(\Omega\right)&\frac{1}{r^{3}}\frac{1}{r_{0}^{3}}\frac{1}{\left|\bar{r}-\bar{r}_{0}\right|}e^{e^{i\frac{3\pi}{4}}\sqrt{\frac{\omega}{D}}\left(1+i\frac{v^{2}}{8D\omega}\right)\cdot\left|\bar{r}-\bar{r}_{0}\right|}\times\\\nonumber &j_{2l}\left(-i\frac{v}{2D}\left|\bar{r}-\bar{r}_{0}\right|\right)P_{2l}\left(\cos\Theta\right),
\end{align}
where we, again, omitted the multiplicative factor of the nuclear spin density $n$ for brevity.
We define the dimensionless integration variables
\beq\label{LDL7}
\tilde{r}=\sqrt{\frac{\omega}{D}}r,\ \tilde{r_0}=\sqrt{\frac{\omega}{D}}r_0.
\eeq 
We then write the integral \eqref{LDL6} in terms of the variables \eqref{LDL7}
\begin{align}\label{LDL8}
	G_{O}^{\left(m\right)}=\frac{1}{4\pi D}\sqrt{\frac{\omega}{D}}\sum_{l=0}^{\infty}\left(-1\right)^{l}\left(4l+1\right)\int\int d^{3}\tilde{r}d^{3}\tilde{r}_{0}Y_{2}^{(m)}\left(\Omega_{0}\right)Y_{2}^{(m)*}\left(\Omega\right)&\frac{1}{\tilde{r}^{3}}\frac{1}{\tilde{r}_{0}^{3}}\frac{1}{\left|\bar{\tilde{r}}-\bar{\tilde{r}}_{0}\right|}e^{e^{i\frac{3\pi}{4}}\left(1+i\frac{v^{2}}{8D\omega}\right)\cdot\left|\bar{\tilde{r}}-\bar{\tilde{r}}_{0}\right|}\times\\\nonumber
	&j_{2l}\left(-i\sqrt{\frac{v^{2}}{4D\omega}}\left|\bar{\tilde{r}}-\bar{\tilde{r}}_{0}\right|\right)P_{2l}\left(\cos\Theta\right).
\end{align}

The decaying and oscillating exponents keep the distance $\left|\bar{\tilde{r}}-\bar{\tilde{r}}_0\right|$ of order $1$ allowing us to approximate \eqref{LDL8} by 
\begin{align}\label{LDL9}
	&G_{O}^{\left(m\right)}\approx\frac{1}{4\pi D}\sqrt{\frac{\omega}{D}}\sum_{l=0}^{\infty}\left(4l+1\right)\int\int d^{3}\tilde{r}d^{3}\tilde{r}_{0}Y_{2}^{(m)}\left(\Omega_{0}\right)Y_{2}^{(m)*}\left(\Omega\right)\frac{1}{\tilde{r}^{3}}\frac{1}{\tilde{r}_{0}^{3}}\frac{1}{\left|\bar{\tilde{r}}-\bar{\tilde{r}}_{0}\right|}e^{e^{i\frac{3\pi}{4}}\cdot\left|\bar{\tilde{r}}-\bar{\tilde{r}}_{0}\right|}\times\\\nonumber
	&\left(1+e^{i\frac{5\pi}{4}}\frac{v^{2}}{8D\omega}\cdot\left|\bar{\tilde{r}}-\bar{\tilde{r}}_{0}\right|\right)\left(\sqrt{\frac{v^{2}}{4D\omega}}\left|\bar{\tilde{r}}-\bar{\tilde{r}}_{0}\right|\right)^{2l}\left(\frac{\sqrt{\pi}2^{-2l-1}}{\Gamma\left(2l+\frac{3}{2}\right)}-\frac{\sqrt{\pi}2^{-2l-2}\left(-i\sqrt{\frac{v^{2}}{4D\omega}}\left|\bar{\tilde{r}}-\bar{\tilde{r}}_{0}\right|\right)^{2}}{(4l+3)\Gamma\left(2l+\frac{3}{2}\right)}\right)P_{2l}\left(\cos\Theta\right).
\end{align}
The zeroth order in our small parameter, defined by \eqref{LDL}, reproduces the correlation function without drift. The leading contribution of \eqref{LDL9} will be given by $l=0,1$,
\begin{align}\label{LDL10}
	G_{O}^{\left(m\right)}\approx G_{O}^{\left(m\right)}\left(v=0\right)+\frac{1}{4\pi D}\sqrt{\frac{\omega}{D}}\int\int d^{3}\tilde{r}d^{3}\tilde{r}_{0}Y_{2}^{(m)}\left(\Omega_{0}\right)Y_{2}^{(m)*}\left(\Omega\right)\frac{1}{\tilde{r}^{3}}\frac{1}{\tilde{r}_{0}^{3}}\frac{1}{\left|\bar{\tilde{r}}-\bar{\tilde{r}}_{0}\right|}e^{e^{i\frac{3\pi}{4}}\cdot\left|\bar{\tilde{r}}-\bar{\tilde{r}}_{0}\right|}\times\\\nonumber
	\left[e^{i\frac{5\pi}{4}}\frac{v^{2}}{8D\omega}\cdot\left|\bar{\tilde{r}}-\bar{\tilde{r}}_{0}\right|+\frac{\frac{v^{2}}{4D\omega}\left|\bar{\tilde{r}}-\bar{\tilde{r}}_{0}\right|^{2}}{6}+\frac{1}{30}\frac{v^{2}}{4D\omega}\left|\bar{\tilde{r}}-\bar{\tilde{r}}_{0}\right|^{2}\left(3\cos^{2}\Theta-1\right)\right]
\end{align}
\beq\label{LDL11}
=G_{O}^{\left(m\right)}\left(v=0\right)+\frac{1}{4\pi D}\frac{v^{2}}{8D\omega}\sqrt{\frac{\omega}{D}}\int\int d^{3}\tilde{r}d^{3}\tilde{r}_{0}Y_{2}^{(m)}\left(\Omega_{0}\right)Y_{2}^{(m)*}\left(\Omega\right)\frac{1}{\tilde{r}^{3}}\frac{1}{\tilde{r}_{0}^{3}}e^{e^{i\frac{3\pi}{4}}\cdot\left|\bar{\tilde{r}}-\bar{\tilde{r}}_{0}\right|}\left[e^{i\frac{5\pi}{4}}\cdot+\frac{\left|\bar{\tilde{r}}-\bar{\tilde{r}}_{0}\right|}{15}\left(4+3\cos^{2}\Theta\right)\right].
\eeq

Lets examine the integration over the magnitude of the radial coordinate in \eqref{LDL11}. It consists of two parts, the first is
\beq\label{LDL12}
I_{1}\equiv e^{i\frac{5\pi}{4}}\frac{v^{2}}{32\pi D^{2}\sqrt{\omega D}}\int\limits _{\sqrt{\frac{\omega}{D}}d/\cos\theta_{0}}^{\infty}\frac{d\tilde{r}_{0}}{\tilde{r}_{0}}\int\limits _{\sqrt{\frac{\omega}{D}}d/\cos\theta}^{\infty}\frac{d\tilde{r}}{\tilde{r}}e^{e^{i\frac{3\pi}{4}}\left|\bar{\tilde{r}}-\bar{\tilde{r}}_{0}\right|}
\eeq
We recall that by \eqref{nodriftlowfreq1},
\begin{align}
	G_{O}^{\left(m\right)}\left(v=0\right)\propto\frac{1}{4\pi D}\int\limits _{d/\cos\theta_{0}}^{\infty}\frac{dr_{0}}{r_{0}}\int\limits _{d/\cos\theta}^{\infty}\frac{dr}{r}\frac{1}{\left|\bar{r}-\bar{r_{0}}\right|}e^{e^{i\frac{3\pi}{4}}\sqrt{\frac{\omega}{D}}\left|\bar{r}-\bar{r_{0}}\right|}.
\end{align}
Therefore,
\begin{align}
	&\frac{d}{d\sqrt{\omega}}G_{O}^{\left(m\right)}\left(v=0\right)=e^{i\frac{3\pi}{4}}\sqrt{\frac{1}{D}}\frac{1}{4\pi D}\int\limits _{d/\cos\theta_{0}}^{\infty}\frac{dr_{0}}{r_{0}}\int\limits _{d/\cos\theta}^{\infty}\frac{dr}{r}e^{e^{i\frac{3\pi}{4}}\sqrt{\frac{\omega}{D}}\left|\bar{r}-\bar{r_{0}}\right|}
	\\
	&=e^{i\frac{3\pi}{4}}\sqrt{\frac{1}{D}}\frac{1}{4\pi D}\int\limits _{\sqrt{\frac{\omega}{D}}d/\cos\theta_{0}}^{\infty}\frac{d\tilde{r}_{0}}{\tilde{r}_{0}}\int\limits _{\sqrt{\frac{\omega}{D}}d/\cos\theta}^{\infty}\frac{d\tilde{r}}{\tilde{r}}e^{e^{i\frac{3\pi}{4}}\left|\bar{\tilde{r}}-\bar{\tilde{r}}_{0}\right|},
\end{align}
and it follows that
\beq
I_1=i\frac{v^{2}}{8D\sqrt{\omega}}\frac{d}{d\sqrt{\omega}}G_{O}^{\left(m\right)}\left(v=0\right).
\eeq
We already studied the behavior of $G_O^{(m)}$ at Sec. \ref{Powernodrift}. The derivative by $\sqrt{\omega}$ for $\frac{\omega d^{2}}{D}\ll1$
will be of the form $C1+C2\sqrt{\omega}$, where $C1$ and $C2$ are constants, while for $\frac{\omega d^{2}}{D}\gg1$
it will decay as $\frac{1}{\omega^{5/2}d^{3}}$. We note that the angular integration in \eqref{LDL11} is the same as in \eqref{nodriftlowfreq1}, so the prefactors are also identical. The second part of the integral \eqref{LDL11} will have a similar relation, since
\begin{align}
	\frac{d^{2}}{d\sqrt{\omega}^{2}}G_{O}^{\left(m\right)}\left(v=0\right)=e^{i\frac{6\pi}{4}}\frac{1}{4\pi D \sqrt{D\omega}}\int\limits _{\sqrt{\frac{\omega}{D}}d}^{\infty}\frac{d\tilde{r}_{0}}{\tilde{r}_{0}}\int\limits _{\sqrt{\frac{\omega}{D}}d}^{\infty}\frac{d\tilde{r}}{\tilde{r}}\left|\bar{\tilde{r}}-\bar{\tilde{r}}_{0}\right|e^{e^{i\frac{3\pi}{4}}\left|\bar{\tilde{r}}-\bar{\tilde{r}}_{0}\right|},
\end{align}
it follows that,
\begin{align}
	 I_{2}=e^{i\frac{5\pi}{4}}\frac{v^{2}}{32\pi D^{2}\sqrt{\omega D}}\int\limits _{\sqrt{\frac{\omega}{D}}d}^{\infty}\frac{d\tilde{r}_{0}}{\tilde{r}}_{0}\int\limits _{\sqrt{\frac{\omega}{D}}d}^{\infty}\frac{d\tilde{r}}{\tilde{r}}\left|\bar{\tilde{r}}-\bar{\tilde{r}}_{0}\right|e^{e^{i\frac{3\pi}{4}}\left|\bar{\tilde{r}}-\bar{\tilde{r}}_{0}\right|}=e^{-i\frac{\pi}{4}}\frac{v^{2}}{8D}\frac{d^{2}}{d\sqrt{\omega}^{2}}G_{O}^{\left(m\right)}\left(v=0\right).
\end{align}
Therefore the correlation function will be
\begin{align}
	G_{O}^{\left(m\right)}\left(\omega\right)\approx G_{O}^{\left(m\right)}\left(v=0\right)+\frac{v^{2}}{8D}&\left[\frac{i}{\sqrt{\omega}}\frac{d}{d\sqrt{\omega}}G_{O}^{\left(m\right)}\left(v=0\right)\right.\\\nonumber
	&\left.+e^{-i\frac{\pi}{4}}\int d\Omega_{0}\int d\Omega\frac{\left(4+3\cos^{2}\Theta\right)}{15}\left[\frac{d^{2}}{d\sqrt{\omega}^{2}}G_{O}^{\left(m\right)}\left(v=0\right)\right]_{radial\ part}\right].
\end{align}
A similar expansion can be performed for the images which will result in a prefactor. 
\newpage
\subsection{General Magnetization axis:}
\label{Appendix: Low_drift2}

In the following, we show how the previous calculations can be generalized for an arbitrary magnetization axis of the NV. We then explicitly show that the linear contribution in the velocity is not eliminated in the general case, thus confirming Eqs. \eqref{Doppler7} and \eqref{Doppler31} obtained by the dimensional analysis.  

We would now like to start with the generalization of our previous results for an arbitrary direction of the NV's magnetization axis. We recall that in the previous section, due to the polar symmetry, the linear term in the low drift expansion nullified. 
Since for a general magnetization axis we loose the polar symmetry, there is a chance the linear term will have a finite contribution. This is of main interest, because it could result in an enhanced sensitivity.
The main difference in the calculation is that the magnetization axis defines the spherical harmonics direction. If we have a spherical harmonic at a general orientation, we can write it as a linear combination
of spherical harmonics that are oriented along the $\hat{z}$ axis (the normal to diamond surface) \cite{Rotation}: 
\beq
Y_{2}^{\left(m\right)}\left(\bar{r}\right)=\sum_{m'=-2}^{2}\left[D_{mm'}^{\left(2\right)}\left(\mathcal{R}\right)\right]^{*}Y_{2}^{\left(m'\right)}\left(\bar{r}'\right),\ \bar{r}'=\mathcal{R}\bar{r}
\eeq
where $D_{mm'}^{\left(l\right)}$ is the Wigner D-matrix, and $\mathcal{R}$ is the rotation operator (commonly represented by the Euler angles). Thus, if we look at a general magnetization axis the correlation will be:
\beq
G_{O}^{\left(m\right)}=\int\frac{d^{3}r}{r^{3}}\int\frac{d^{3}r_{0}}{r_{0}^{3}}Y_{2}^{\left(m\right)}\left(\bar{r}_{0}\right)Y_{2}^{\left(m\right)*}\left(\bar{r}\right)\psi\left(\bar{r},\bar{r}_{0}\right)
\label{rotated}
\eeq
\beq
=\sum_{m_{1}=-2}^{2}\sum_{m_{2}=-2}^{2}\left[D_{mm_{1}}^{\left(2\right)}\left(\mathcal{R}\right)\right]^{*}\left[D_{mm_{2}}^{\left(2\right)}\left(\mathcal{R}\right)\right]\int\frac{d^{3}r}{r^{3}}\int\frac{d^{3}r_{0}}{r_{0}^{3}}Y_{2}^{\left(m_{1}\right)}\left(\bar{r}_{0}\right)Y_{2}^{\left(m_{2}\right)*}\left(\bar{r}\right)\psi\left(\bar{r},\bar{r}_{0}\right)
\label{rotated2}
\eeq
where in the last equality we changed our coordinate system without ascribing it a new notation. 
We expand \eqref{rotated2} in the limit \eqref{LDL}, as in the previous section. The linear contribution will be:  
\beq\label{rotated3}
O\left(v\right)=\frac{v}{8\pi D^{2}}\sum_{m_{1}=-2}^{2}\sum_{m_{2}=-2}^{2}\left[D_{mm_{1}}^{\left(2\right)}\left(\mathcal{R}\right)\right]^{*}\left[D_{mm_{2}}^{\left(2\right)}\left(\mathcal{R}\right)\right]\int\frac{d^{3}r}{r^{3}}\int\frac{d^{3}r{}_{0}}{r_{0}^{3}}Y_{2}^{\left(m_{1}\right)}\left(\bar{r}{}_{0}\right)Y_{2}^{\left(m_{2}\right)*}\left(\bar{r}\right)\frac{x-x_{0}}{\left|\bar{r}-\bar{r}_{0}\right|}e^{e^{i\frac{3\pi}{4}}\left|\bar{r}-\bar{r'}\right|}
\eeq
We substitute the expansion \cite[Eq.~10.60.3]{DLMF},
\beq\label{Hankelexpansion}
\frac{e^{ik\cdot\left|\bar{r}-\bar{r}_{0}\right|}}{\left|\bar{r}-\bar{r}_{0}\right|}=ik\sum_{l=0}^{\infty}\left(2l+1\right)j_{l}\left(kr_{<}\right)h_{l}^{\left(1\right)}\left(kr_{>}\right)P_{l}\left(\cos\Theta\right),
\eeq
into \eqref{rotated3}:
\begin{align}
	O\left(v\right)=\frac{v}{2D^{2}}e^{i\frac{3\pi}{4}}\sum_{l=0}^{\infty}\sum_{m_{3}=-l}^{l}\sum_{m_{1}=-2}^{2}\sum_{m_{2}=-2}^{2}&\left[D_{mm_{1}}^{\left(2\right)}\left(\mathcal{R}\right)\right]^{*}\left[D_{mm_{2}}^{\left(2\right)}\left(\mathcal{R}\right)\right]\int\frac{d^{3}r}{r^{3}}\int\frac{d^{3}r{}_{0}}{r_{0}^{3}}Y_{2}^{\left(m_{1}\right)}\left(\bar{r}{}_{0}\right)Y_{2}^{\left(m_{2}\right)*}\left(\bar{r}\right)\times\\\nonumber
	&\left(r\sin\theta\cos\phi-r_{0}\sin\theta_{0}\cos\phi_{0}\right)j_{l}\left(e^{i\frac{\pi}{4}}r_{<}\right)h_{l}^{\left(1\right)}\left(e^{i\frac{\pi}{4}}r_{>}\right)Y_{l}^{m_{3}*}\left(\Omega_{<}\right)Y_{l}^{m_{3}}\left(\Omega_{>}\right)
\end{align}
\begin{align}
	=\frac{v}{2D^{2}}e^{i\frac{3\pi}{4}}&\sum_{l=0}^{\infty}\sum_{m_{3}=-l}^{l}\sum_{m_{1}=-2}^{2}\sum_{m_{2}=-2}^{2}\left[D_{mm_{1}}^{\left(2\right)}\left(\mathcal{R}\right)\right]^{*}\left[D_{mm_{2}}^{\left(2\right)}\left(\mathcal{R}\right)\right]\int d\Omega\int d\Omega_{0}\int\limits _{\frac{d\sqrt{\frac{\omega}{D}}}{\cos\theta}}^{\infty}\frac{dr}{r}\int\limits _{\frac{d\sqrt{\frac{\omega}{D}}}{\cos\theta_{0}}}^{\infty}\frac{dr_{0}}{r_{0}}\times\\\nonumber
	&Y_{2}^{\left(m_{1}\right)}\left(\bar{r}{}_{0}\right)Y_{2}^{\left(m_{2}\right)*}\left(\bar{r}\right)\left(r\sin\theta\cos\phi-r_{0}\sin\theta_{0}\cos\phi_{0}\right)j_{l}\left(e^{i\frac{\pi}{4}}r_{<}\right)h_{l}^{\left(1\right)}\left(e^{i\frac{\pi}{4}}r_{>}\right)Y_{l}^{m_{3}*}\left(\Omega_{<}\right)Y_{l}^{m_{3}}\left(\Omega_{>}\right)
\end{align}
we denote
\beq
\label{rotated4} 
A_{m,m_{1},m_{2}}=\left[D_{mm_{1}}^{\left(2\right)}\left(\mathcal{R}\right)\right]^{*}\left[D_{mm_{2}}^{\left(2\right)}\left(\mathcal{R}\right)\right]
\eeq
and substitute \eqref{rotated4} in \eqref{rotated4}:
\begin{align}
	O\left(v\right)=&\frac{v}{2D^{2}}e^{i\frac{3\pi}{4}}\sum_{l=0}^{\infty}\sum_{m_{3}=-l}^{l}\sum_{m_{1}=-2}^{2}\sum_{m_{2}=-2}^{2}A_{m,m_{1},m_{2}}\int d\Omega\int d\Omega_{0}\int\limits _{\frac{d\sqrt{\frac{\omega}{D}}}{\cos\theta_{0}}}^{\infty}\frac{dr_{0}}{r_{0}}\int\limits _{\frac{d\sqrt{\frac{\omega}{D}}}{\cos\theta}}^{\infty}\frac{dr}{r}\times\\\nonumber
	&Y_{2}^{\left(m_{1}\right)}\left(\bar{r}{}_{0}\right)Y_{2}^{\left(m_{2}\right)*}\left(\bar{r}\right)Y_{l}^{m_{3}*}\left(\Omega_{<}\right)Y_{l}^{m_{3}}\left(\Omega_{>}\right)\left(r\sin\theta\cos\phi-r_{0}\sin\theta_{0}\cos\phi_{0}\right)j_{l}\left(e^{i\frac{\pi}{4}}r_{<}\right)h_{l}^{\left(1\right)}\left(e^{i\frac{\pi}{4}}r_{>}\right)
\end{align}

\begin{align}
	=&\frac{v}{2D^{2}}\sum_{l=0}^{\infty}\sum_{m_{3}=-l}^{l}\sum_{m_{1}=-2}^{2}\sum_{m_{2}=-2}^{2}A_{m,m_{1},m_{2}}\int\limits _{0}^{1}d\cos\theta_{0}\int\limits _{0}^{\cos\theta_{0}}d\cos\theta\int\limits _{\frac{d\sqrt{\frac{\omega}{D}}}{\cos\theta}}^{\infty}dr\times\\\nonumber
	&\left[\int\limits _{\frac{d\sqrt{\frac{\omega}{D}}}{\cos\theta_{0}}}^{r}dr_{0}\int d\phi\int d\phi_{0}Y_{2}^{\left(m_{1}\right)}\left(\bar{r}{}_{0}\right)Y_{2}^{\left(m_{2}\right)*}\left(\bar{r}\right)Y_{l}^{m_{3}*}\left(\Omega_{0}\right)Y_{l}^{m_{3}}\left(\Omega\right)\left(\frac{j_{l}\left(e^{i\frac{\pi}{4}}r_{0}\right)}{r_{0}}h_{l}^{\left(1\right)}\left(e^{i\frac{\pi}{4}}r\right)\sin\theta\cos\phi\right.\right.\\\nonumber
	&\left.-j_{l}\left(e^{i\frac{\pi}{4}}r_{0}\right)\frac{h_{l}^{\left(1\right)}\left(e^{i\frac{\pi}{4}}r\right)}{r}\sin\theta_{0}\cos\phi_{0}\right)+\int\limits _{r}^{\infty}dr_{0}\int d\phi\int d\phi_{0}Y_{2}^{\left(m_{1}\right)}\left(\bar{r}{}_{0}\right)Y_{2}^{\left(m_{2}\right)*}\left(\bar{r}\right)Y_{l}^{m_{3}*}\left(\Omega\right)Y_{l}^{m_{3}}\left(\Omega_{0}\right)\times\\\nonumber
	&\left.\left(j_{l}\left(e^{i\frac{\pi}{4}}r\right)\frac{h_{l}^{\left(1\right)}\left(e^{i\frac{\pi}{4}}r_{0}\right)}{r_{0}}\sin\theta\cos\phi-\frac{j_{l}\left(e^{i\frac{\pi}{4}}r\right)}{r}h_{l}^{\left(1\right)}\left(e^{i\frac{\pi}{4}}r_{0}\right)\sin\theta_{0}\cos\phi_{0}\right)\right]
\end{align}
\begin{align}\label{rotated5}
	+&\frac{v}{2D^{2}}\sum_{l=0}^{\infty}\sum_{m_{3}=-l}^{l}\sum_{m_{1}=-2}^{2}\sum_{m_{2}=-2}^{2}A_{m,m_{1},m_{2}}\int\limits _{0}^{1}d\cos\theta_{0}\int\limits _{\cos\theta_{0}}^{1}d\cos\theta\int\limits _{\frac{d\sqrt{\frac{\omega}{D}}}{\cos\theta_{0}}}^{\infty}dr_{0}\times\\\nonumber
	&\left[\int\limits _{\frac{d\sqrt{\frac{\omega}{D}}}{\cos\theta}}^{r_{0}}dr\int d\phi\int d\phi_{0}Y_{2}^{\left(m_{1}\right)}\left(\bar{r}{}_{0}\right)Y_{2}^{\left(m_{2}\right)*}\left(\bar{r}\right)Y_{l}^{m_{3}*}\left(\Omega_{0}\right)Y_{l}^{m_{3}}\left(\Omega\right)\left(j_{l}\left(e^{i\frac{\pi}{4}}r\right)\frac{h_{l}^{\left(1\right)}\left(e^{i\frac{\pi}{4}}r_{0}\right)}{r_{0}}\sin\theta\cos\phi\right.\right.\\\nonumber
	&\left.-\frac{j_{l}\left(e^{i\frac{\pi}{4}}r\right)}{r}h_{l}^{\left(1\right)}\left(e^{i\frac{\pi}{4}}r_{0}\right)\sin\theta_{0}\cos\phi_{0}\right)+\int\limits _{r_{0}}^{\infty}dr\int d\phi\int d\phi_{0}Y_{2}^{\left(m_{1}\right)}\left(\bar{r}{}_{0}\right)Y_{2}^{\left(m_{2}\right)*}\left(\bar{r}\right)Y_{l}^{m_{3}*}\left(\Omega_{0}\right)Y_{l}^{m_{3}}\left(\Omega\right)\times\\\nonumber
	&\left.\left(\frac{j_{l}\left(e^{i\frac{\pi}{4}}r_{0}\right)}{r_{0}}h_{l}^{\left(1\right)}\left(e^{i\frac{\pi}{4}}r\right)\sin\theta\cos\phi-j_{l}\left(e^{i\frac{\pi}{4}}r_{0}\right)\frac{h_{l}^{\left(1\right)}\left(e^{i\frac{\pi}{4}}r\right)}{r}\sin\theta_{0}\cos\phi_{0}\right)\right]
\end{align}
We perform the integration over the polar coordinates in \eqref{rotated5}
\begin{align}\label{rotated6}
	O\left(v\right)=&\frac{\pi^{2}v}{D^{2}}\sum_{m_{1}=-2}^{2}\sum_{l=\min\left\{ m\right\} }^{\infty}A_{m,m_{1},m_{1}\pm1}\int\limits _{0}^{1}d\cos\theta_{0}\int\limits _{0}^{\cos\theta_{0}}d\cos\theta\int\limits _{\frac{d\sqrt{\frac{\omega}{D}}}{\cos\theta}}^{\infty}dr\times\\\nonumber
	&\left[\int\limits _{\frac{d\sqrt{\frac{\omega}{D}}}{\cos\theta_{0}}}^{r}dr_{0}\left(\frac{j_{l}\left(e^{i\frac{\pi}{4}}r_{0}\right)}{r_{0}}h_{l}^{\left(1\right)}\left(e^{i\frac{\pi}{4}}r\right)\sin\theta Y_{2}^{m_{1}}\left(\cos\theta_{0}\right)Y_{l}^{m_{1}}\left(\cos\theta_{0}\right)Y_{2}^{m_{1}\pm1}\left(\cos\theta\right)Y_{l}^{m_{1}}\left(\cos\theta\right)\right.\right.\\\nonumber
	&\left.-j_{l}\left(e^{i\frac{\pi}{4}}r_{0}\right)\frac{h_{l}^{\left(1\right)}\left(e^{i\frac{\pi}{4}}r\right)}{r}\sin\theta_{0}Y_{2}^{m_{1}\pm1}\left(\cos\theta_{0}\right)Y_{l}^{m_{1}}\left(\cos\theta_{0}\right)Y_{2}^{m_{1}}\left(\cos\theta\right)Y_{l}^{m_{1}}\left(\cos\theta\right)\right)\\\nonumber
	&+\int\limits _{r}^{\infty}dr_{0}\left(j_{l}\left(e^{i\frac{\pi}{4}}r\right)\frac{h_{l}^{\left(1\right)}\left(e^{i\frac{\pi}{4}}r_{0}\right)}{r_{0}}Y_{2}^{m_{1}}\left(\cos\theta_{0}\right)Y_{l}^{m_{1}}\left(\cos\theta_{0}\right)Y_{2}^{m_{1}\pm1}\left(\cos\theta\right)Y_{l}^{m_{1}}\left(\cos\theta\right)\sin\theta\right.\\\nonumber
	&\left.\left.-\frac{j_{l}\left(e^{i\frac{\pi}{4}}r\right)}{r}h_{l}^{\left(1\right)}\left(e^{i\frac{\pi}{4}}r_{0}\right)\sin\theta_{0}Y_{2}^{m_{1}\pm1}\left(\cos\theta_{0}\right)Y_{l}^{m_{1}}\left(\cos\theta_{0}\right)Y_{2}^{m_{1}}\left(\cos\theta\right)Y_{l}^{m_{1}}\left(\cos\theta\right)\right)\right]
\end{align}
\begin{align}\label{rotated7}
	&+\frac{\pi^{2}v}{D^{2}}\sum_{m_{1}=-2}^{2}\sum_{l=\min\left\{ m\right\} }^{\infty}A_{m,m_{1},m_{1}\pm1}\int\limits _{0}^{1}d\cos\theta_{0}\int\limits _{\cos\theta_{0}}^{1}d\cos\theta\int\limits _{\frac{d\sqrt{\frac{\omega}{D}}}{\cos\theta_{0}}}^{\infty}dr_{0}\times\\\nonumber
	&\left[\int\limits _{\frac{d\sqrt{\frac{\omega}{D}}}{\cos\theta}}^{r_{0}}dr\left(j_{l}\left(e^{i\frac{\pi}{4}}r\right)\frac{h_{l}^{\left(1\right)}\left(e^{i\frac{\pi}{4}}r_{0}\right)}{r_{0}}\sin\theta Y_{2}^{m_{1}}\left(\cos\theta_{0}\right)Y_{l}^{m_{1}}\left(\cos\theta_{0}\right)Y_{2}^{m_{1}\pm1}\left(\cos\theta\right)Y_{l}^{m_{1}}\left(\cos\theta\right)\right.\right.\\\nonumber
	&\left.-\frac{j_{l}\left(e^{i\frac{\pi}{4}}r\right)}{r}h_{l}^{\left(1\right)}\left(e^{i\frac{\pi}{4}}r_{0}\right)\sin\theta_{0}Y_{2}^{m_{1}\pm1}\left(\cos\theta_{0}\right)Y_{l}^{m_{1}}\left(\cos\theta_{0}\right)Y_{2}^{m_{1}}\left(\cos\theta\right)Y_{l}^{m_{1}}\left(\cos\theta\right)\right)\\\nonumber
	&+\int\limits _{r_{0}}^{\infty}dr\left(\frac{j_{l}\left(e^{i\frac{\pi}{4}}r_{0}\right)}{r_{0}}h_{l}^{\left(1\right)}\left(e^{i\frac{\pi}{4}}r\right)\sin\theta Y_{2}^{m_{1}}\left(\cos\theta_{0}\right)Y_{l}^{m_{1}}\left(\cos\theta_{0}\right)Y_{2}^{m_{1}\pm1}\left(\cos\theta\right)Y_{l}^{m_{1}}\left(\cos\theta\right)\right.\\\nonumber
	&\left.\left.-j_{l}\left(e^{i\frac{\pi}{4}}r_{0}\right)\frac{h_{l}^{\left(1\right)}\left(e^{i\frac{\pi}{4}}r\right)}{r}\sin\theta_{0}Y_{2}^{m_{1}\pm1}\left(\cos\theta_{0}\right)Y_{l}^{m_{1}}\left(\cos\theta_{0}\right)Y_{2}^{m_{1}}\left(\cos\theta\right)Y_{l}^{m_{1}}\left(\cos\theta\right)\right)\right]
\end{align}

Let us define the integrals:

\beq\label{rotated8}
I_{l}^{1}\left(\theta,\theta_{0}\right)=\int\limits _{d\sqrt{\frac{\omega}{D}}/\cos\theta}^{\infty}dr\int\limits _{d\sqrt{\frac{\omega}{D}}/\cos\theta_{0}}^{r}dr_{0}\left(\frac{j_{l}\left(e^{i\frac{\pi}{4}}r_{0}\right)}{r_{0}}h_{l}^{\left(1\right)}\left(e^{i\frac{\pi}{4}}r\right)f_{l}^{m_{1}}\left(\theta,\theta_{0}\right)-j_{l}\left(e^{i\frac{\pi}{4}}r_{0}\right)\frac{h_{l}^{\left(1\right)}\left(e^{i\frac{\pi}{4}}r\right)}{r}f_{l}^{m_{1}}\left(\theta_{0},\theta\right)\right)
\eeq
\beq\label{rotated9}
I_{l}^{2}\left(\theta,\theta_{0}\right)=\int\limits _{d\sqrt{\frac{\omega}{D}}/\cos\theta}^{\infty}dr\int\limits _{r}^{\infty}dr_{0}\left(j_{l}\left(e^{i\frac{\pi}{4}}r\right)\frac{h_{l}^{\left(1\right)}\left(e^{i\frac{\pi}{4}}r_{0}\right)}{r_{0}}f_{l}^{m_{1}}\left(\theta,\theta_{0}\right)-\frac{j_{l}\left(e^{i\frac{\pi}{4}}r\right)}{r}h_{l}^{\left(1\right)}\left(e^{i\frac{\pi}{4}}r_{0}\right)f_{l}^{m_{1}}\left(\theta_{0},\theta\right)\right)
\eeq
where:

\beq\label{rotated10}
f_{l}^{m_{1}}\left(\theta,\theta_{0}\right)\equiv\sin\theta Y_{2}^{m_{1}\pm1}\left(\cos\theta\right)Y_{l}^{m_{1}}\left(\cos\theta\right)Y_{2}^{m_{1}}\left(\cos\theta_{0}\right)Y_{l}^{m_{1}}\left(\cos\theta_{0}\right)
\eeq
with the definitions \eqref{rotated8},\eqref{rotated9},\eqref{rotated10} the linear term \eqref{rotated7} can be written compactly:

\begin{align}\label{rotated11}
	O\left(v\right)=\frac{\pi^{2}v}{D^{2}}\sum_{m_{1}=-2}^{2}\sum_{l=\min\left\{ m\right\} }^{\infty}A_{m,m_{1},m_{1}\pm1}\int\limits _{0}^{1}d\cos\theta_{0}&\left[\int\limits _{0}^{\cos\theta_{0}}d\cos\theta\left(I_{l}^{1}\left(\theta,\theta_{0}\right)+I_{l}^{2}\left(\theta,\theta_{0}\right)\right)\right.\\\nonumber
	&\left.-\int\limits _{\cos\theta_{0}}^{1}d\cos\theta\left(I_{l}^{1}\left(\theta_{0},\theta\right)+I_{l}^{2}\left(\theta_{0},\theta\right)\right)\right]
\end{align}

As in the previous sections, we expect the most significant contribution to come from $l=0$, for which $m_{1}=0$. 
Let us examine \eqref{rotated10} in the case where $m_{1}=0$:

\begin{align}\label{rotated12}
	&f_{l}^{0}\left(\theta,\theta_{0}\right)+f_{l}^{0}\left(\theta_{0},\theta\right)=\sin\theta\left\{ \left(A_{m,0,1}Y_{2}^{1}\left(\cos\theta\right)+A_{m,0,-1}Y_{2}^{-1}\left(\cos\theta\right)\right)Y_{l}^{0}\left(\cos\theta\right)Y_{2}^{0}\left(\cos\theta_{0}\right)Y_{l}^{0}\left(\cos\theta_{0}\right)\right\}\\\nonumber
	&=\left[D_{m0}^{\left(2\right)}\left(\mathcal{R}\right)\right]^{*}\sin\theta\left\{ \left(\left[D_{m1}^{\left(2\right)}\left(\mathcal{R}\right)\right]Y_{2}^{1}\left(\cos\theta\right)+\left[D_{m-1}^{\left(2\right)}\left(\mathcal{R}\right)\right]Y_{2}^{-1}\left(\cos\theta\right)\right)Y_{l}^{0}\left(\cos\theta\right)Y_{2}^{0}\left(\cos\theta_{0}\right)Y_{l}^{0}\left(\cos\theta_{0}\right)\right\}
\end{align}

this term does not exist for $m=0$ since $\left[D_{01}^{\left(2\right)}\left(\mathcal{R}\right)\right]$
is not defined, but for $m=1$ it will not vanish.

\subsubsection{Evaluation for low frequencies $\left(d\sqrt{\frac{\omega}{D}}\ll1\right)$:}

First, we note that combining the low drift limit \eqref{LDL} with the low frequency limit $\left(d\sqrt{\frac{\omega}{D}}\ll1\right)$ gives the restriction:
$\frac{v^{2}}{D}\ll\omega\ll\frac{D}{d^{2}}$. Subsequently, the limit $\omega\rightarrow0$
does not exist in this regime. 
We wish to expand the integrals \eqref{rotated8},\eqref{rotated8} in the low frequency regime. We shall evaluate the first order of \eqref{rotated8} by taking the derivative:

\begin{align}\label{rotated13}
	&\frac{\partial}{\partial\sqrt{\omega}}I_{l}^{1}\left(\theta,\theta_{0}\right)=-\frac{d}{\sqrt{D}}\left\{ \frac{1}{\cos\theta}\left(\left[e^{i\pi\frac{3}{4}}\left(\frac{\sin\left(e^{i\frac{\pi}{4}}\frac{d\sqrt{\frac{\omega}{D}}}{\cos\theta}\right)}{\frac{d\sqrt{\frac{\omega}{D}}}{\cos\theta}}-\frac{\sin\left(e^{i\frac{\pi}{4}}\frac{d\sqrt{\frac{\omega}{D}}}{\cos\theta_{0}}\right)}{\frac{d\sqrt{\frac{\omega}{D}}}{\cos\theta_{0}}}\right)-\text{Ci}\left(e^{i\frac{\pi}{4}}\frac{d\sqrt{\frac{\omega}{D}}}{\cos\theta_{0}}\right)\times\right.\right.\right.\\\nonumber
	&\left.+\text{Ci}\left(e^{i\frac{\pi}{4}}\frac{d\sqrt{\frac{\omega}{D}}}{\cos\theta}\right)\right]h_{l}^{\left(1\right)}\left(e^{i\frac{\pi}{4}}\frac{d\sqrt{\frac{\omega}{D}}}{\cos\theta}\right)f_{l}^{m_{1}}\left(\theta,\theta_{0}\right)-e^{i\pi\frac{3}{4}}\left(\text{Si}\left(e^{i\frac{\pi}{4}}\frac{d\sqrt{\frac{\omega}{D}}}{\cos\theta}\right)-\text{Si}\left(e^{i\frac{\pi}{4}}\frac{d\sqrt{\frac{\omega}{D}}}{\cos\theta_{0}}\right)\right)\times\\\nonumber
	&\left.\frac{h_{l}^{\left(1\right)}\left(e^{i\frac{\pi}{4}}d\sqrt{\frac{\omega}{D}}/\cos\theta\right)}{d\sqrt{\frac{\omega}{D}}/\cos\theta}f_{l}^{m_{1}}\left(\theta_{0},\theta\right)\right)+\frac{1}{\cos\theta_{0}}\left(\frac{j_{l}\left(e^{i\frac{\pi}{4}}d\sqrt{\frac{\omega}{D}}/\cos\theta_{0}\right)}{d\sqrt{\frac{\omega}{D}}/\cos\theta_{0}}\left(e^{i\frac{\pi}{4}}\left(\text{Ei}\left(e^{i\frac{3\pi}{4}}d\sqrt{\frac{\omega}{D}}/\cos\theta\right)-i\pi\right)\right)\times\right.\\\nonumber
	&\left.\left.f_{l}^{m_{1}}\left(\theta,\theta_{0}\right)-j_{l}\left(e^{i\frac{\pi}{4}}d\sqrt{\frac{\omega}{D}}/\cos\theta_{0}\right)\left(-\text{Ei}\left(e^{i\frac{3\pi}{4}}d\sqrt{\frac{\omega}{D}}/\cos\theta\right)-\frac{e^{i\frac{\pi}{4}}e^{e^{i\frac{3\pi}{4}}d\sqrt{\frac{\omega}{D}}/\cos\theta}}{d\sqrt{\frac{\omega}{D}}/\cos\theta}+i\pi\right)f_{l}^{m_{1}}\left(\theta_{0},\theta\right)\right)\right\}
\end{align}
\begin{align}\label{rotated14}
	&=-\frac{d}{\sqrt{D}}\left\{ \frac{ \left(d\sqrt{\frac{\omega}{D}}\right)^{l}}{\cos\theta}\left[\sqrt{\pi}e^{i\frac{\pi}{4}l}2^{-l-2}\Gamma\left(\frac{l}{2}\right)\,_{1}\tilde{F}_{2}\left(\frac{l}{2};\frac{l}{2}+1,l+\frac{3}{2};-\frac{i}{4}\left(\frac{d\sqrt{\frac{\omega}{D}}}{\cos\theta_{0}}\right)^{2}\right)\right.\right.\\\nonumber
	&\left(-\frac{1}{\cos^{l}\theta_{0}}+\frac{1}{\cos^{l}\theta}\right)h_{l}^{\left(1\right)}\left(\frac{e^{i\frac{\pi}{4}}d\sqrt{\frac{\omega}{D}}}{\cos\theta}\right)f_{l}^{m_{1}}\left(\theta,\theta_{0}\right)-\sqrt{\pi}2^{-l-2}e^{i\frac{\pi}{4}l}\Gamma\left(\frac{l+1}{2}\right)\times\\\nonumber
	&\left.\,_{1}\tilde{F}_{2}\left(\frac{l+1}{2};\frac{l+3}{2},l+\frac{3}{2};-\frac{i}{4}\left(\frac{d\sqrt{\frac{\omega}{D}}}{\cos\theta_{0}}\right)^{2}\right)\left(-\frac{\cos\theta}{\cos^{l+1}\theta_{0}}+\frac{1}{\cos^{l}\theta}\right)h_{l}^{\left(1\right)}\left(\frac{e^{i\frac{\pi}{4}}d\sqrt{\frac{\omega}{D}}}{\cos\theta}\right)f_{l}^{m_{1}}\left(\theta_{0},\theta\right)\right]\\\nonumber
	&\left.+\frac{1}{\cos\theta_{0}}\int\limits _{\frac{d\sqrt{\frac{\omega}{D}}}{\cos\theta}}^{\infty}dr\left(\frac{j_{l}\left(\frac{e^{i\frac{\pi}{4}}d\sqrt{\frac{\omega}{D}}}{\cos\theta_{0}}\right)}{\frac{d\sqrt{\frac{\omega}{D}}}{\cos\theta_{0}}}h_{l}^{\left(1\right)}\left(e^{i\frac{\pi}{4}}r\right)f_{l}^{m_{1}}\left(\theta,\theta_{0}\right)-j_{l}\left(\frac{e^{i\frac{\pi}{4}}d\sqrt{\frac{\omega}{D}}}{\cos\theta_{0}}\right)\frac{h_{l}^{\left(1\right)}\left(e^{i\frac{\pi}{4}}r\right)}{r}f_{l}^{m_{1}}\left(\theta_{0},\theta\right)\right)\right\} 
\end{align}
for $l=0$ \eqref{rotated14} simplifies to:
\begin{align}\label{rotated15}
	&\frac{\partial}{\partial\sqrt{\omega}}I_{0}^{1}\left(\theta,\theta_{0}\right)=-\frac{d}{\sqrt{D}}\left\{ \frac{1}{\cos\theta}\left[\left(e^{i\pi\frac{3}{4}}\left(\frac{\sin\left(e^{i\frac{\pi}{4}}\frac{d\sqrt{\frac{\omega}{D}}}{\cos\theta}\right)}{\frac{d\sqrt{\frac{\omega}{D}}}{\cos\theta}}-\frac{\sin\left(e^{i\frac{\pi}{4}}\frac{d\sqrt{\frac{\omega}{D}}}{\cos\theta_{0}}\right)}{\frac{d\sqrt{\frac{\omega}{D}}}{\cos\theta_{0}}}\right)-\text{Ci}\left(\frac{e^{i\frac{\pi}{4}}d\sqrt{\frac{\omega}{D}}}{\cos\theta_{0}}\right)\right.\right.\right.\\\nonumber
	&\left.+\text{Ci}\left(\frac{e^{i\frac{\pi}{4}}d\sqrt{\frac{\omega}{D}}}{\cos\theta}\right)\right)h_{l}^{\left(1\right)}\left(e^{i\frac{\pi}{4}}\frac{d\sqrt{\frac{\omega}{D}}}{\cos\theta}\right)f_{l}^{m_{1}}\left(\theta,\theta_{0}\right)-e^{i\pi\frac{3}{4}}\left(\text{Si}\left(e^{i\frac{\pi}{4}}\frac{d\sqrt{\frac{\omega}{D}}}{\cos\theta}\right)-\text{Si}\left(e^{i\frac{\pi}{4}}\frac{d\sqrt{\frac{\omega}{D}}}{\cos\theta_{0}}\right)\right)\times\\\nonumber
	&\left.\frac{h_{0}^{\left(1\right)}\left(\frac{e^{i\frac{\pi}{4}}d\sqrt{\frac{\omega}{D}}}{\cos\theta}\right)}{\frac{d\sqrt{\frac{\omega}{D}}}{\cos\theta}}f_{l}^{m_{1}}\left(\theta_{0},\theta\right)\right]+\frac{1}{\cos\theta_{0}}\left[\frac{j_{0}\left(\frac{e^{i\frac{\pi}{4}}d\sqrt{\frac{\omega}{D}}}{\cos\theta_{0}}\right)}{\frac{d\sqrt{\frac{\omega}{D}}}{\cos\theta_{0}}}\left(e^{i\frac{\pi}{4}}\left(\text{Ei}\left(\frac{e^{i\frac{3\pi}{4}}d\sqrt{\frac{\omega}{D}}}{\cos\theta}\right)-i\pi\right)\right)f_{0}^{m_{1}}\left(\theta,\theta_{0}\right)\right.\\\nonumber
	&\left.\left.-j_{0}\left(\frac{e^{i\frac{\pi}{4}}d\sqrt{\frac{\omega}{D}}}{\cos\theta_{0}}\right)\left(-\text{Ei}\left(\frac{e^{i\frac{3\pi}{4}}d\sqrt{\frac{\omega}{D}}}{\cos\theta}\right)-\frac{e^{i\frac{\pi}{4}}e^{e^{i\frac{3\pi}{4}}\frac{d\sqrt{\frac{\omega}{D}}}{\cos\theta}}}{\frac{d\sqrt{\frac{\omega}{D}}}{\cos\theta}}+i\pi\right)f_{l}^{m_{1}}\left(\theta_{0},\theta\right)\right]\right\} 
\end{align}
The exponential, sine and cosine integrals in \eqref{rotated15} can be expanded in the low frequency limit:
\begin{align}
	&\frac{\partial}{\partial\sqrt{\omega}}I_{0}^{1}\left(\theta,\theta_{0}\right)\approx-\frac{d}{\sqrt{D}}\left\{ \frac{1}{\cos\theta}\left[\left(\frac{1}{2}e^{i\frac{3\pi}{4}}d\sqrt{\frac{\omega}{D}}\left(\frac{1}{\cos\theta}\log\left(\frac{\cos\theta_{0}}{\cos\theta}\right)+\frac{\cos\theta}{6}\left(\frac{1}{\cos^{2}\theta_{0}}-\frac{1}{\cos^{2}\theta}\right)\right)+\frac{e^{i\frac{\pi}{4}}\cos\theta}{d\sqrt{\frac{\omega}{D}}}\log\left(\frac{\cos\theta_{0}}{\cos\theta}\right)\right.\right.\right.\\\nonumber
	&\left.\left.+\log\left(\frac{\cos\theta_{0}}{\cos\theta}\right)\right)f_{0}^{m_{1}}\left(\theta,\theta_{0}\right)-\left(\frac{1}{\cos\theta_{0}}-\frac{1}{\cos\theta}\right)\left(-\frac{i}{6}\left(\frac{d\sqrt{\frac{\omega}{D}}}{\cos\theta}\right)+\frac{1}{2}e^{i\frac{3\pi}{4}}+e^{i\frac{\pi}{4}}\left(\frac{\cos\theta}{d\sqrt{\frac{\omega}{D}}}\right)^{2}+\frac{\cos\theta}{d\sqrt{\frac{\omega}{D}}}\right)f_{l}^{m_{1}}\left(\theta_{0},\theta\right)\right]\\\nonumber
	&+\frac{1}{\cos\theta_{0}}\left[-\frac{1}{4}e^{i\frac{3\pi}{4}}d\sqrt{\frac{\omega}{D}}\frac{\cos\theta_{0}}{\cos^{2}\theta}-\frac{\cos\theta_{0}}{\cos\theta}+\left(\frac{\cos\theta_{0}}{d\sqrt{\frac{\omega}{D}}}-\frac{1}{6}e^{\frac{\pi i}{2}}\frac{d\sqrt{\frac{\omega}{D}}}{\cos\theta_{0}}\right)\left(e^{i\frac{\pi}{4}}\log\left(\frac{d\sqrt{\frac{\omega}{D}}}{\cos\theta}\right)-\frac{1}{4}e^{i\frac{3\pi}{4}}\pi+e^{i\frac{\pi}{4}}\gamma\right)f_{l}^{m_{1}}\left(\theta,\theta_{0}\right)\right.\\\nonumber
	&\left.\left.-\left(\frac{1}{2}e^{i\frac{3\pi}{4}}d\sqrt{\frac{\omega}{D}}\left(\frac{1}{3}\frac{\cos\theta}{\cos^{2}\theta_{0}}-\frac{1}{\cos\theta}\right)-e^{i\frac{\pi}{4}}\frac{\cos\theta}{d\sqrt{\frac{\omega}{D}}}-\log(\frac{d\sqrt{\frac{\omega}{D}}}{\cos\theta})+1-\gamma+\frac{i\pi}{4}\right)f_{l}^{m_{1}}\left(\theta_{0},\theta\right)\right]\right\} 
\end{align} 
\begin{align}\label{rotated16}
	&\approx-\frac{d}{\sqrt{D}}\left\{ \frac{1}{\cos\theta}\left[\left(\frac{e^{i\frac{\pi}{4}}\cos\theta}{d\sqrt{\frac{\omega}{D}}}\log\left(\frac{\cos\theta_{0}}{\cos\theta}\right)+\log\left(\frac{\cos\theta_{0}}{\cos\theta}\right)\right)f_{0}^{m_{1}}\left(\theta,\theta_{0}\right)-\left(\frac{1}{\cos\theta_{0}}-\frac{1}{\cos\theta}\right)\times\right.\right.\\\nonumber
	&\left.\left(e^{i\frac{\pi}{4}}\left(\frac{\cos\theta}{d\sqrt{\frac{\omega}{D}}}\right)^{2}+\frac{\cos\theta}{d\sqrt{\frac{\omega}{D}}}\right)f_{l}^{m_{1}}\left(\theta_{0},\theta\right)\right]+\frac{1}{\cos\theta_{0}}\left[e^{i\frac{\pi}{4}}\frac{\cos\theta_{0}}{d\sqrt{\frac{\omega}{D}}}\log\left(\frac{d\sqrt{\frac{\omega}{D}}}{\cos\theta}\right)f_{l}^{m_{1}}\left(\theta,\theta_{0}\right)+e^{i\frac{\pi}{4}}\frac{\cos\theta}{d\sqrt{\frac{\omega}{D}}}\right.\\\nonumber
	&\left.\left.+\log(\frac{d\sqrt{\frac{\omega}{D}}}{\cos\theta})f_{l}^{m_{1}}\left(\theta_{0},\theta\right)\right]\right\} 
\end{align}
We recall that for $l=0$ \eqref{rotated10} can be explicitly written as
\begin{align}\label{rotated17}
	&f_{0}^{m_{1}}\left(\theta,\theta_{0}\right)=\sin\theta Y_{2}^{m_{1}\pm1}\left(\cos\theta\right)Y_{0}^{m_{1}}\left(\cos\theta\right)Y_{2}^{m_{1}}\left(\cos\theta_{0}\right)Y_{0}^{m_{1}}\left(\cos\theta_{0}\right)
	\\
	&=\frac{1}{4\pi}\sin\theta Y_{2}^{\pm1}\left(\cos\theta\right)Y_{2}^{0}\left(\cos\theta_{0}\right)=\frac{1}{4\pi}\left(3\cos^{2}\theta_{0}-1\right)\sin^{2}\theta\cos\theta
\end{align}

The most significant term in \eqref{rotated16} is:
\beq\label{rotated18}
\frac{\partial}{\partial\sqrt{\omega}}I_{0}^{1}\left(\theta,\theta_{0}\right)\sim e^{i\frac{\pi}{4}}\frac{\sqrt{D}}{d\omega}\left(\frac{\cos^{2}\theta}{\cos\theta_{0}}-\cos\theta\right)\left(3\cos^{2}\theta_{0}-1\right)\sin^{2}\theta_{0}\cos\theta_{0}
\eeq
For this term to have a finite contribution, We need to make sure that the angular integration is non-zero. Taking the angular dependence of \eqref{rotated18} and substituting into \eqref{rotated11}:
\begin{align}
	&\int\limits _{0}^{1}d\cos\theta_{0}\int\limits _{0}^{\cos\theta_{0}}d\cos\theta\left(\frac{\cos^{2}\theta}{\cos\theta_{0}}-\cos\theta\right)\left(3\cos^{2}\theta_{0}-1\right)\left(1-\cos^{2}\theta_{0}\right)\cos\theta_{0}
	\\
	&=-\frac{1}{6}\int\limits _{0}^{1}d\cos\theta_{0}\left(4\cos^{5}\theta_{0}-3\cos^{7}\theta_{0}-\cos^{3}\theta_{0}\right)=-\frac{1}{6}\left(\frac{2}{3}-\frac{5}{8}\right)\neq0
\end{align}

So the first correction to the correlation function goes as: $-e^{i\frac{\pi}{4}}\frac{v}{D^{2}}\sqrt{\frac{D}{d^{2}\omega}}$.

We can see that if we substitute $\omega=\frac{v^{2}}{D}$
the dependency on $v$ vanishes as expected by our expansion in the limit $\omega\gg\frac{v^2}{D}$.
To insure that this term does not cancel out, we have to calculate the leading contribution of $I_{0}^{2}$ defined in \eqref{rotated9}:
\beq
I_{0}^{2}\left(\theta,\theta_{0}\right)=\int\limits _{d\sqrt{\frac{\omega}{D}}/\cos\theta}^{\infty}dr\int\limits _{r}^{\infty}dr_{0}\left(j_{0}\left(e^{i\frac{\pi}{4}}r\right)\frac{h_{0}^{\left(1\right)}\left(e^{i\frac{\pi}{4}}r_{0}\right)}{r_{0}}f_{0}^{m_{1}}\left(\theta,\theta_{0}\right)-\frac{j_{0}\left(e^{i\frac{\pi}{4}}r\right)}{r}h_{0}^{\left(1\right)}\left(e^{i\frac{\pi}{4}}r_{0}\right)f_{0}^{m_{1}}\left(\theta_{0},\theta\right)\right)
\eeq
\begin{align}\label{rotated19}
	=\int\limits _{\frac{d\sqrt{\frac{\omega}{D}}}{\cos\theta}}^{\infty}dr\left[\left(-\text{Ci}\left(e^{i\frac{\pi}{4}}r\right)-i\text{Si}\left(e^{i\frac{\pi}{4}}r\right)-e^{i\frac{\pi}{4}}\frac{e^{e^{i3\frac{\pi}{4}}r}}{r}+\frac{i\pi}{2}\right)j_{0}\left(e^{i\frac{\pi}{4}}r\right)f_{0}^{m_{1}}\left(\theta,\theta_{0}\right)\right.\\\nonumber
	\left.-e^{i\frac{\pi}{4}}\frac{j_{0}\left(e^{i\frac{\pi}{4}}r\right)}{r}\left(\text{Ci}\left(e^{i\frac{\pi}{4}}r\right)-i\frac{\pi}{2}+i\text{Si}\left(e^{i\frac{\pi}{4}}r\right)\right)f_{0}^{m_{1}}\left(\theta_{0},\theta\right)\right]
\end{align}
We use the same approach as before and take the derivative of \eqref{rotated19}
\begin{align}
	&\frac{\partial}{\partial\sqrt{\omega}}I_{0}^{2}\left(\theta,\theta_{0}\right)=-\frac{d}{\sqrt{D}\cos\theta}\left[\left(-\text{Ci}\left(e^{i\frac{\pi}{4}}\frac{d\sqrt{\frac{\omega}{D}}}{\cos\theta}\right)-i\text{Si}\left(e^{i\frac{\pi}{4}}\frac{d\sqrt{\frac{\omega}{D}}}{\cos\theta}\right)-e^{i\frac{\pi}{4}}\frac{e^{e^{i3\frac{\pi}{4}}\frac{d\sqrt{\frac{\omega}{D}}}{\cos\theta}}}{\frac{d\sqrt{\frac{\omega}{D}}}{\cos\theta}}+\frac{i\pi}{2}\right)\times\right.\\\nonumber
	&\left.j_{0}\left(e^{i\frac{\pi}{4}}\frac{d\sqrt{\frac{\omega}{D}}}{\cos\theta}\right)f_{0}^{m_{1}}\left(\theta,\theta_{0}\right)-e^{i\frac{\pi}{4}}\frac{j_{0}\left(e^{i\frac{\pi}{4}}\frac{d\sqrt{\frac{\omega}{D}}}{\cos\theta}\right)}{\frac{d\sqrt{\frac{\omega}{D}}}{\cos\theta}}\left(\text{Ci}\left(e^{i\frac{\pi}{4}}\frac{d\sqrt{\frac{\omega}{D}}}{\cos\theta}\right)-i\frac{\pi}{2}+i\text{Si}\left(e^{i\frac{\pi}{4}}\frac{d\sqrt{\frac{\omega}{D}}}{\cos\theta}\right)\right)f_{0}^{m_{1}}\left(\theta_{0},\theta\right)\right]
\end{align}
\beq
\approx-\frac{d}{\sqrt{D}\cos\theta}\left(-\frac{e^{i\frac{\pi}{4}}}{\frac{d\sqrt{\frac{\omega}{D}}}{\cos\theta}}f_{0}^{m_{1}}\left(\theta,\theta_{0}\right)-\frac{\left(2\log\left(\frac{d\sqrt{\frac{\omega}{D}}}{\cos\theta}\right)+2\gamma-i\frac{\pi}{2}\right)}{2\frac{d\sqrt{\frac{\omega}{D}}}{\cos\theta}}f_{0}^{m_{1}}\left(\theta_{0},\theta\right)\right)
\eeq

We can see that $I_{0}^{2}$ does not have a contribution to the order $\sim\frac{1}{\sqrt{\omega}}$.

Finally, we have to deal with the image contribution:
\begin{align}\label{rotatedim1}
	O_{R}\left(v\right)=\frac{v}{8\pi D^{2}}&\sum_{m_{1}=-2}^{2}\sum_{m_{2}=-2}^{2}\left[D_{mm_{1}}^{\left(2\right)}\left(\mathcal{R}\right)\right]^{*}\left[D_{mm_{2}}^{\left(2\right)}\left(\mathcal{R}\right)\right]\int\frac{d^{3}r}{r^{3}}\int\frac{d^{3}r{}_{0}}{r_{0}^{3}}Y_{2}^{\left(m_{1}\right)}\left(\bar{r}{}_{0}\right)Y_{2}^{\left(m_{2}\right)*}\left(\bar{r}\right)\times\\\nonumber
	&\frac{x-x_{0}}{\sqrt{\left(x-x_{0}\right)^{2}+\left(y-y_{0}\right)^{2}+\left(z+z_{0}-2\sqrt{\frac{\omega}{D}}d\right)^{2}}}e^{e^{i\frac{3\pi}{4}}\sqrt{\left(x-x_{0}\right)^{2}+\left(y-y_{0}\right)^{2}+\left(z+z_{0}-2d\sqrt{\frac{\omega}{D}}\right)^{2}}}
\end{align}
We rewrite the integral \eqref{rotatedim1} in terms of the new variables $z'=z-d\sqrt{\frac{\omega}{D}},\ z'_0=z_0-d\sqrt{\frac{\omega}{D}}$: 
\begin{align}\label{rotatedim2}
	O_{R}\left(v\right)=&\frac{v}{8\pi D^{2}}\sum_{m_{1}=-2}^{2}\sum_{m_{2}=-2}^{2}\left[D_{mm_{1}}^{\left(2\right)}\left(\mathcal{R}\right)\right]^{*}\left[D_{mm_{2}}^{\left(2\right)}\left(\mathcal{R}\right)\right]\int\frac{d^{3}r}{\left(x^{2}+y^{2}+\left(z+\sqrt{\frac{\omega}{D}}d\right)^{2}\right)^{3/2}}\times\\\nonumber
	&\int\frac{d^{3}r_{0}}{\left(x_{0}^{2}+y_{0}^{2}+\left(z_{0}+\sqrt{\frac{\omega}{D}}d\right)^{2}\right)^{3/2}}Y_{2}^{\left(m_{1}\right)}\left(\bar{r}{}_{0}\right)Y_{2}^{\left(m_{2}\right)*}\left(\bar{r}\right)\frac{x-x_{0}}{\sqrt{\left(x-x_{0}\right)^{2}+\left(y-y_{0}\right)^{2}+\left(z+z_{0}\right)^{2}}}e^{e^{i\frac{3\pi}{4}}\sqrt{\left(x-x_{0}\right)^{2}+\left(y-y_{0}\right)^{2}+\left(z+z_{0}\right)^{2}}}
\end{align}
where we eliminated the new notation for shorts. We now use transformation $\cos\theta_{0}\rightarrow-\cos\theta_{0}$ so the integral \eqref{rotatedim2} changes to:
\begin{align}
	O_{R}\left(v\right)=&\frac{v}{8\pi D^{2}}\sum_{m_{1}=-2}^{2}\left(-1\right)^{m_1}\sum_{m_{2}=-2}^{2}\left[D_{mm_{1}}^{\left(2\right)}\left(\mathcal{R}\right)\right]^{*}\left[D_{mm_{2}}^{\left(2\right)}\left(\mathcal{R}\right)\right]\int\limits _{0}^{2\pi}d\phi\int\limits _{0}^{1}d\cos\theta\int\limits _{0}^{\infty}r^{2}dr\int\limits _{0}^{2\pi}d\phi_{0}\int\limits _{-1}^{0}d\cos\theta_{0}\int\limits _{0}^{\infty}r_{0}^{2}dr_{0}\times\\\nonumber
	&\frac{1}{\left(x^{2}+y^{2}+\left(z+\sqrt{\frac{\omega}{D}}d\right)^{2}\right)^{3/2}}\frac{1}{\left(x_{0}^{2}+y_{0}^{2}+\left(z_{0}-\sqrt{\frac{\omega}{D}}d\right)^{2}\right)^{3/2}}Y_{2}^{\left(m_{1}\right)}\left(\bar{r}{}_{0}\right)Y_{2}^{\left(m_{2}\right)*}\left(\bar{r}\right)\times\\\nonumber
	&\frac{x-x_{0}}{\sqrt{\left(x-x_{0}\right)^{2}+\left(y-y_{0}\right)^{2}+\left(z-z_{0}\right)^{2}}}e^{e^{i\frac{3\pi}{4}}\sqrt{\left(x-x_{0}\right)^{2}+\left(y-y_{0}\right)^{2}+\left(z-z_{0}\right)^{2}}}
\end{align}
\begin{align}\label{rotatedim3}
	&=\frac{v}{8\pi D^{2}}\sum_{m_{1}=-2}^{2}\left(-1\right)^{m_1}\sum_{m_{2}=-2}^{2}\left[D_{mm_{1}}^{\left(2\right)}\left(\mathcal{R}\right)\right]^{*}\left[D_{mm_{2}}^{\left(2\right)}\left(\mathcal{R}\right)\right]\int\limits _{0}^{2\pi}d\phi\int\limits _{0}^{1}d\cos\theta\int\limits _{0}^{\infty}r^{2}dr\int\limits _{0}^{2\pi}d\phi_{0}\int\limits _{-1}^{0}d\cos\theta_{0}\int\limits _{0}^{\infty}r_{0}^{2}dr_{0}\times\\\nonumber
	&\frac{1}{\left(x^{2}+y^{2}+\left(z+\sqrt{\frac{\omega}{D}}d\right)^{2}\right)^{3/2}}\frac{1}{\left(x_{0}^{2}+y_{0}^{2}+\left(z_{0}-\sqrt{\frac{\omega}{D}}d\right)^{2}\right)^{3/2}}Y_{2}^{\left(m_{1}\right)}\left(\bar{r}{}_{0}\right)Y_{2}^{\left(m_{2}\right)*}\left(\bar{r}\right)\frac{x-x_{0}}{\left|\bar{r}-\bar{r}_{0}\right|}e^{e^{i\frac{3\pi}{4}}\left|\bar{r}-\bar{r}_{0}\right|}
\end{align}
We would like to expand \eqref{rotatedim3} in a way which is similar to a multipole expansion. We can write,
\begin{align}\label{nodriftlowfreqim4}
&\frac{1}{\left(r^{2}\mp2r\cos\theta' d+d^{2}\right)^{3/2}}=\begin{cases}
\frac{1}{d^{3}}\frac{1}{\left(\left(r/d\right)^{2}\mp2\left(r/d\right)\cos\theta'+1\right)^{3/2}} & r<d\\
\frac{1}{r^{3}}\frac{1}{\left(\left(d/r\right)^{2}\mp2\left(d/r\right)\cos\theta'+1\right)^{3/2}} & r>d
\end{cases}\\\nonumber
&\underbrace{=}_{x=\cos\theta',\ y=r/d}\begin{cases}
\frac{1}{d^{3}}\frac{1}{\left(y^{2}\mp2yx+1\right)^{3/2}} & y<1\\
\frac{1}{r^{3}}\frac{1}{\left(\left(1/y\right)^{2}\mp2\left(1/y\right)x+1\right)^{3/2}} & y>1
\end{cases}
=\begin{cases}
\pm\frac{1}{d^{3}}\frac{1}{y}\frac{d}{dx}\frac{1}{\left(y^{2}\mp2yx+1\right)^{1/2}} & y<1\\\nonumber
\pm\frac{1}{r^{3}}y\frac{d}{dx}\frac{1}{\left(\left(1/y\right)^{2}\mp2\left(1/y\right)x+1\right)^{1/2}} & y>1
\end{cases}\\\nonumber
&\underbrace{=}_{\left|x\right|\leq1}\begin{cases}
\pm\frac{1}{d^{3}}\frac{1}{y}\frac{d}{dx}\sum_{l=0}^{\infty}\left(\pm y\right)^{l}P_{l}\left(x\right) & y<1\\
\pm\frac{1}{r^{3}}y\frac{d}{dx}\sum_{l=0}^{\infty}\left(\pm\frac{1}{y}\right)^{l}P_{l}\left(x\right) & y>1
\end{cases}
=\begin{cases}
\pm\frac{1}{d^{3}}\frac{1}{y}\sum_{l=0}^{\infty}\left(\pm y\right)^{l}\frac{d}{dx}P_{l}\left(x\right) & y<1\\
\pm\frac{1}{r^{3}}y\sum_{l=0}^{\infty}\left(\pm\frac{1}{y}\right)^{l}\frac{d}{dx}P_{l}\left(x\right) & y>1
\end{cases} .
\end{align}
For $l>0$ \cite[Eq.~14.10.5]{DLMF},
\beq\label{nodriftlowfreqim5}
\frac{d}{dx}P_{l}\left(x\right)=\frac{l}{x^{2}-1}\left(xP_{l}\left(x\right)-P_{l-1}\left(x\right)\right)\ .
\eeq
Therefore, by substituting \eqref{nodriftlowfreqim5} into \eqref{nodriftlowfreqim4},
\beq\label{nodriftlowfreqim6}
\frac{1}{\left(r^{2}\mp2r\cos\theta' d+d^{2}\right)^{3/2}}=\begin{cases}
	\frac{1}{d^{3}}\sum_{l=1}^{\infty}\left(\pm y\right)^{l-1}\frac{l}{x^{2}-1}\left(xP_{l}\left(x\right)-P_{l-1}\left(x\right)\right) & y<1\\
	\frac{1}{r^{3}}\sum_{l=1}^{\infty}\left(\pm\frac{1}{y}\right)^{l-1}\frac{l}{x^{2}-1}\left(xP_{l}\left(x\right)-P_{l-1}\left(x\right)\right) & y>1
\end{cases}\ .
\eeq 

The integral \eqref{rotatedim3} can, therefore, be expanded using \eqref{nodriftlowfreqim6} in terms of the normalized variables $x_i->\sqrt{\frac{\omega}{D}}x_i$:
\begin{align}\label{rotatedim4}
	&O_{R}\left(v\right)\approx\frac{v}{8\pi D^{2}}\sum_{m_{1}=-2}^{2}\sum_{m_{2}=-2}^{2}\sum_{l,k=1}^{\infty}\left(-1\right)^{m_1+l-1}\left(\sqrt{\frac{\omega}{D}d}\right)^{l+k-2}lk\left[D_{mm_{1}}^{\left(2\right)}\left(\mathcal{R}\right)\right]^{*}\left[D_{mm_{2}}^{\left(2\right)}\left(\mathcal{R}\right)\right]\int\limits _{0}^{2\pi}d\phi\int\limits _{0}^{1}d\cos\theta\int\limits _{0}^{2\pi}d\phi_{0}\int\limits _{-1}^{0}d\cos\theta_{0}\times\\\nonumber
	&\frac{1}{\cos^{2}\theta-1}\left(\cos\theta P_{l}\left(\cos\theta\right)-P_{l-1}\left(\cos\theta\right)\right)\frac{1}{\cos^{2}\theta_{0}-1}\left(\cos\theta_{0}P_{k}\left(\cos\theta_{0}\right)-P_{k-1}\left(\cos\theta_{0}\right)\right)Y_{2}^{\left(m_{1}\right)}\left(\bar{r}{}_{0}\right)Y_{2}^{\left(m_{2}\right)*}\left(\bar{r}\right)\times\\\nonumber
	&\int\limits _{\sqrt{\frac{\omega}{D}}d}^{\infty}\int\limits _{\sqrt{\frac{\omega}{D}}d}^{\infty}drdr{}_{0}\left(\frac{1}{r}\right)^{l}\left(\frac{1}{r_{0}}\right)^{k}\frac{x-x_{0}}{\left|\bar{r}-\bar{r}_{0}\right|}e^{e^{i\frac{3\pi}{4}}\left|\bar{r}-\bar{r}_{0}\right|}
\end{align}

As in the calculation without drift, we expect the most significant contribution to come
from $l=k=1$ in \eqref{rotatedim4}:
\begin{align}\label{rotatedim5}
	O_{R}\left(v\right)\approx\frac{v}{8\pi D^{2}}&\sum_{m_{1}=-2}^{2}\left(-1\right)^{m_1}\sum_{m_{2}=-2}^{2}\left[D_{mm_{1}}^{\left(2\right)}\left(\mathcal{R}\right)\right]^{*}\left[D_{mm_{2}}^{\left(2\right)}\left(\mathcal{R}\right)\right]\int\limits _{0}^{2\pi}d\phi\int\limits _{0}^{1}d\cos\theta\int\limits _{0}^{2\pi}d\phi_{0}\int\limits _{-1}^{0}d\cos\theta_{0}\times\\\nonumber
	&Y_{2}^{\left(m_{1}\right)}\left(\bar{r}{}_{0}\right)Y_{2}^{\left(m_{2}\right)*}\left(\bar{r}\right)\int\limits _{\sqrt{\frac{\omega}{D}}d}^{\infty}\int\limits _{\sqrt{\frac{\omega}{D}}d}^{\infty}drdr{}_{0}\frac{1}{r}\frac{1}{r_{0}}\frac{x-x_{0}}{\left|\bar{r}-\bar{r}_{0}\right|}e^{e^{i\frac{3\pi}{4}}\left|\bar{r}-\bar{r}_{0}\right|}
\end{align}

The integral \eqref{rotatedim5} is very similar to \eqref{rotated3}, therefore, we use the same expansion \eqref{Hankelexpansion}: 
\begin{align}
	&O_{R}\left(v\right)\approx\frac{v}{2D^{2}}\sum_{m_{1},m_{2}=-2}^{2} \left(-1\right)^{m_1}\sum_{l=0}^{\infty}\sum_{m_{3}}\left[D_{mm_{1}}^{\left(2\right)}\left(\mathcal{R}\right)\right]^{*}\left[D_{mm_{2}}^{\left(2\right)}\left(\mathcal{R}\right)\right]\int\limits _{0}^{2\pi}d\phi\int\limits _{0}^{1}d\cos\theta\int\limits _{0}^{2\pi}d\phi_{0}\int\limits _{-1}^{0}d\cos\theta_{0}\times\\\nonumber
	&Y_{2}^{\left(m_{1}\right)}\left(\bar{r}{}_{0}\right)Y_{2}^{\left(m_{2}\right)*}\left(\bar{r}\right)\int\limits _{\sqrt{\frac{\omega}{D}}d}^{\infty}\int\limits _{\sqrt{\frac{\omega}{D}}d}^{\infty}drdr{}_{0}\frac{1}{r}\frac{1}{r_{0}}Y_{l}^{m_{3}*}\left(\Omega_{<}\right)Y_{l}^{m_{3}}\left(\Omega_{>}\right)\left(r\sin\theta\cos\phi-r_{0}\sin\theta_{0}\cos\phi_{0}\right)j_{l}\left(e^{i\frac{\pi}{4}}r_{<}\right)h_{l}^{\left(1\right)}\left(e^{i\frac{\pi}{4}}r_{>}\right)
\end{align}
\begin{align}\label{rotatedim6}
	&=\frac{v}{2D^{2}}\sum_{m_{1},m_{2}=-2}^{2}\left(-1\right)^{m_1}\sum_{l=0}^{\infty}\sum_{m_{3}}\left[D_{mm_{1}}^{\left(2\right)}\left(\mathcal{R}\right)\right]^{*}\left[D_{mm_{2}}^{\left(2\right)}\left(\mathcal{R}\right)\right]\int\limits _{0}^{2\pi}d\phi\int\limits _{0}^{1}d\cos\theta\int\limits _{0}^{2\pi}d\phi_{0}\int\limits _{-1}^{0}d\cos\theta_{0}\times\\\nonumber
	&Y_{2}^{\left(m_{1}\right)}\left(\bar{r}{}_{0}\right)Y_{2}^{\left(m_{2}\right)*}\left(\bar{r}\right)\int\limits _{\sqrt{\frac{\omega}{D}}d}^{\infty}dr_{0}\left[\int\limits _{\sqrt{\frac{\omega}{D}}d}^{r_{0}}dr\frac{1}{r}\frac{1}{r_{0}}Y_{l}^{m_{3}*}\left(\Omega\right)Y_{l}^{m_{3}}\left(\Omega_{0}\right)\left(r\sin\theta\cos\phi-r_{0}\sin\theta_{0}\cos\phi_{0}\right)j_{l}\left(e^{i\frac{\pi}{4}}r\right)h_{l}^{\left(1\right)}\left(e^{i\frac{\pi}{4}}r_{0}\right)\right.\\\nonumber
	&\left.+\int\limits _{r_{0}}^{\infty}dr\frac{1}{r}\frac{1}{r_{0}}Y_{l}^{m_{3}*}\left(\Omega_{0}\right)Y_{l}^{m_{3}}\left(\Omega\right)\left(r\sin\theta\cos\phi-r_{0}\sin\theta_{0}\cos\phi_{0}\right)j_{l}\left(e^{i\frac{\pi}{4}}r_{0}\right)h_{l}^{\left(1\right)}\left(e^{i\frac{\pi}{4}}r\right)\right]
\end{align}
Carrying out the polar integration in \eqref{rotatedim6} and using the notation \eqref{rotated10} we arrive at
\begin{align}\label{rotatedim7}
	&O_{R}\left(v\right)=\frac{\pi^{2}v}{D^{2}}\sum_{m_{1}=-2}^{2}\left(-1\right)^{m_1}\sum_{l=\min\ m}^{\infty}A_{m,m_{1},m_{1}\pm1}\int\limits _{0}^{1}d\cos\theta\int\limits _{-1}^{0}d\cos\theta_{0}\int\limits _{\sqrt{\frac{\omega}{D}}d}^{\infty}dr_{0}\times\\\nonumber
	&\left[\int\limits _{\sqrt{\frac{\omega}{D}}d}^{r_{0}}dr\frac{1}{r}\frac{1}{r_{0}}\left(rf_{l}^{m_{1}}\left(\theta,\theta_{0}\right)-r_{0}f_{l}^{m_{1}}\left(\theta_{0},\theta\right)\right)j_{l}\left(e^{i\frac{\pi}{4}}r\right)h_{l}^{\left(1\right)}\left(e^{i\frac{\pi}{4}}r_{0}\right)\right.\\\nonumber
	&\left.+\int\limits _{r_{0}}^{\infty}dr\frac{1}{r}\frac{1}{r_{0}}\left(rf_{l}^{m_{1}}\left(\theta,\theta_{0}\right)-r_{0}f_{l}^{m_{1}}\left(\theta_{0},\theta\right)\right)j_{l}\left(e^{i\frac{\pi}{4}}r_{0}\right)h_{l}^{\left(1\right)}\left(e^{i\frac{\pi}{4}}r\right)\right]
\end{align}

We recall that by \eqref{rotated10}:
\beq
f_{l}^{m_{1}}\left(\theta,\theta_{0}\right)=\sin\theta Y_{2}^{m_{1}\pm1}\left(\cos\theta\right)Y_{l}^{m_{1}}\left(\cos\theta\right)Y_{2}^{m_{1}}\left(\cos\theta_{0}\right)Y_{l}^{m_{1}}\left(\cos\theta_{0}\right)
\eeq
Thus, if we use the transformation $\cos\theta_{0}\rightarrow-\cos\theta_{0}$ in 
\eqref{rotatedim7} we get:
\begin{align}\label{rotatedim8}
	&O_{R}\left(v\right)\approx\frac{\pi^{2}v}{D^{2}}\sum_{m_{1}=-2}^{2}\left(-1\right)^{m_1}\sum_{l=\min\ m}^{\infty}\left(-1\right)^{l}A_{m,m_{1},m_{1}\pm1}\int\limits _{0}^{1}d\cos\theta\int\limits _{0}^{1}d\cos\theta_{0}\int\limits _{\sqrt{\frac{\omega}{D}}d}^{\infty}dr_{0}\times\\\nonumber
	&\left[\int\limits _{\sqrt{\frac{\omega}{D}}d}^{r_{0}}dr\left(\frac{h_{l}^{\left(1\right)}\left(e^{i\frac{\pi}{4}}r_{0}\right)}{r_{0}}j_{l}\left(e^{i\frac{\pi}{4}}r\right)f_{l}^{m_{1}}\left(\theta,\theta_{0}\right)+h_{l}^{\left(1\right)}\left(e^{i\frac{\pi}{4}}r_{0}\right)\frac{j_{l}\left(e^{i\frac{\pi}{4}}r\right)}{r}f_{l}^{m_{1}}\left(\theta_{0},\theta\right)\right)\right.\\\nonumber
	&\left.+\int\limits _{r_{0}}^{\infty}dr\left(\frac{j_{l}\left(e^{i\frac{\pi}{4}}r_{0}\right)}{r_{0}}h_{l}^{\left(1\right)}\left(e^{i\frac{\pi}{4}}r\right)f_{l}^{m_{1}}\left(\theta,\theta_{0}\right)+\frac{h_{l}^{\left(1\right)}\left(e^{i\frac{\pi}{4}}r\right)}{r}j_{l}\left(e^{i\frac{\pi}{4}}r_{0}\right)f_{l}^{m_{1}}\left(\theta_{0},\theta\right)\right)\right]
\end{align}
Taking the derivative of \eqref{rotatedim8} yields:
\begin{align}
	&\frac{\partial}{\partial\sqrt{\omega}}O_{R}\left(v\right)\approx-\frac{d}{\sqrt{D}}\frac{\pi^{2}v}{D^{2}}\sum_{m_{1}=-2}^{2}\sum_{l=\min\ m}^{\infty}\left(-1\right)^{m_1+l}A_{m,m_{1},m_{1}\pm1}\int\limits _{0}^{1}d\cos\theta\int\limits _{0}^{1}d\cos\theta_{0}\times\\\nonumber
	&\left[\int\limits _{\sqrt{\frac{\omega}{D}}d}^{\infty}dr_{0}\left(\frac{h_{l}^{\left(1\right)}\left(e^{i\frac{\pi}{4}}r_{0}\right)}{r_{0}}j_{l}\left(e^{i\frac{\pi}{4}}\sqrt{\frac{\omega}{D}}d\right)f_{l}^{m_{1}}\left(\theta,\theta_{0}\right)+h_{l}^{\left(1\right)}\left(e^{i\frac{\pi}{4}}r_{0}\right)\frac{j_{l}\left(e^{i\frac{\pi}{4}}\sqrt{\frac{\omega}{D}}d\right)}{\sqrt{\frac{\omega}{D}}d}f_{l}^{m_{1}}\left(\theta_{0},\theta\right)\right)\right.\\\nonumber
	&\left.+\int\limits _{\sqrt{\frac{\omega}{D}}d}^{\infty}dr\left(\frac{j_{l}\left(e^{i\frac{\pi}{4}}\sqrt{\frac{\omega}{D}}d\right)}{\sqrt{\frac{\omega}{D}}d}h_{l}^{\left(1\right)}\left(e^{i\frac{\pi}{4}}r\right)f_{l}^{m_{1}}\left(\theta,\theta_{0}\right)+\frac{h_{l}^{\left(1\right)}\left(e^{i\frac{\pi}{4}}r\right)}{r}j_{l}\left(e^{i\frac{\pi}{4}}\sqrt{\frac{\omega}{D}}d\right)f_{l}^{m_{1}}\left(\theta_{0},\theta\right)\right)\right]
\end{align}
\begin{align}
	&=-\frac{d}{\sqrt{D}}\frac{\pi^{2}v}{D^{2}}\sum_{m_{1}=-2}^{2}\sum_{l=\min\ m}^{\infty}\left(-1\right)^{m_1+l}A_{m,m_{1},m_{1}\pm1}\int\limits _{0}^{1}d\cos\theta\int\limits _{0}^{1}d\cos\theta_{0}\times\\\nonumber
	&\int\limits _{\sqrt{\frac{\omega}{D}}d}^{\infty}dr\left(\frac{h_{l}^{\left(1\right)}\left(e^{i\frac{\pi}{4}}r\right)}{r}j_{l}\left(e^{i\frac{\pi}{4}}\sqrt{\frac{\omega}{D}}d\right)+h_{l}^{\left(1\right)}\left(e^{i\frac{\pi}{4}}r\right)\frac{j_{l}\left(e^{i\frac{\pi}{4}}\sqrt{\frac{\omega}{D}}d\right)}{\sqrt{\frac{\omega}{D}}d}\right)\left(f_{l}^{m_{1}}\left(\theta_{0},\theta\right)+f_{l}^{m_{1}}\left(\theta,\theta_{0}\right)\right) 
\end{align}
\begin{align}\label{rotatedim9}
	&=-2\frac{d}{\sqrt{D}}\frac{\pi^{2}v}{D^{2}}\sum_{m_{1}=-2}^{2}\sum_{l=\min\ m}^{\infty}\left(-1\right)^{m_1+l}A_{m,m_{1},m_{1}\pm1}\int\limits _{0}^{1}d\cos\theta\int\limits _{0}^{1}d\cos\theta_{0}\times\\\nonumber
	&\int\limits _{\sqrt{\frac{\omega}{D}}d}^{\infty}dr\left(\frac{h_{l}^{\left(1\right)}\left(e^{i\frac{\pi}{4}}r\right)}{r}j_{l}\left(e^{i\frac{\pi}{4}}\sqrt{\frac{\omega}{D}}d\right)+h_{l}^{\left(1\right)}\left(e^{i\frac{\pi}{4}}r\right)\frac{j_{l}\left(e^{i\frac{\pi}{4}}\sqrt{\frac{\omega}{D}}d\right)}{\sqrt{\frac{\omega}{D}}d}\right)f_{l}^{m_{1}}\left(\theta,\theta_{0}\right) 
\end{align}
For $l=m_{1}=0$ the angular integration of \eqref{rotatedim9} will be:
\begin{align}
	&\int\limits _{0}^{1}d\cos\theta\int\limits _{0}^{1}d\cos\theta_{0}f_{0}^{0}\left(\theta,\theta_{0}\right)=\int\limits _{0}^{1}d\cos\theta\int\limits _{0}^{1}d\cos\theta_{0}\left(3\cos^{2}\theta_{0}-1\right)\sin^{2}\theta\cos\theta\\
	&=\left(\int\limits _{0}^{1}d\cos\theta_{0}\left(3\cos^{2}\theta_{0}-1\right)\right)\left(\int\limits _{0}^{1}d\cos\theta\sin^{2}\theta\cos\theta\right)=0
\end{align}

For larger $l$ values there are no contributions which go as $\sim\frac{1}{\sqrt{\omega}}$	

\section{EVALUATION OF THE EFFECT OF VARYING NV DEPTH IN AN ENSEMBLE MEASUREMENT}

In Eq. 2 of the main text we show that in the Lorentzian model the uncertainty is
\beq\label{Last1}
\Delta \tilde{v} = \frac{1}{\tilde{v}}\frac{1}{\gamma_e B_{RMS} \tau_D^{1/2}  \sqrt{T}}.
\eeq
We use it to show that for parameters fitting water, the accuracy is $\frac{\Delta{v}}{v}\approx600 \frac{\sqrt{\textrm{s}}}{\sqrt{T}}\frac{A}{(\mu\textrm{m})^2}$. 
We would now like to show that variations in the NV's distance from the diamond surface, which arise naturally in ensemble measurements, result in a negligible effect. 

Lets assume that the NV's depth $d$ is a random variable uniformly distributed on $[d_1,d_2]$. The power spectrum can be estimated by $\left<S(\omega)\right>=\frac{1}{d_2-d_1}\int\limits_{d_1}^{d_2} S(\omega,d) d(d)$.
Following the derivation of \eqref{Last1} for $\left<S(\omega)\right>$ is challenging analytically, therefore we result to numeric evaluation of the accuracy at $\omega=0$ with the parameters for water. Our analysis yields the result 
\beq
\frac{\Delta{v}}{v}\approx 870 \frac{\sqrt{\textrm{s}}}{\sqrt{T}}\frac{A}{(\mu\textrm{m})^2}
\eeq
for $d_1=5 \ \text{nm},\ d_2=30 \ \text{nm}$. This deviates from result of a constant distance, presented in the main text by a negligible factor of 1.45.